\documentclass{aa}  

\usepackage{graphicx}
\usepackage{txfonts}
\usepackage{txfonts}
\usepackage{amsmath,mathtools}
\usepackage{here}
\usepackage[usenames,dvipsnames]{xcolor}
\usepackage{booktabs}
\usepackage{amssymb}
\usepackage{amsmath}
\usepackage{rotating}
\usepackage[normalem]{ulem}

\newcommand{\Rsun}{\ensuremath{\,\mathrm{R_\odot}}\xspace}
\newcommand{\Msun}{\ensuremath{\,\mathrm{M_\odot}}\xspace}
\newcommand{\Lsun}{\ensuremath{\,\mathrm{L_\odot}}\xspace}

\definecolor{blue700}{HTML}{536DFE}

\definecolor{deeppurple}{HTML}{6200EA}
\usepackage[breaklinks=true,colorlinks,linkcolor=blue,citecolor=deeppurple]{hyperref} 

\begin{document} 

\title{Different to the core: The pre-supernova structures of massive single and binary-stripped stars}

   \author{E. Laplace
          \inst{1}
                  \and
                   S. Justham
                   \inst{1, 2}
          \and
          M. Renzo
          \inst{3, 4}
          \and
          Y. G\"{o}tberg
          \inst{5}
          \and 
          R. Farmer
          \inst{1, 7}
                   \and
                  D. Vartanyan
                  \inst{6}
                   \and
                  S. E. de Mink
                  \inst{7, 1, 8}
          }

   \institute{Anton Pannekoek Institute of Astronomy and GRAPPA, Science Park 904, University of Amsterdam, 1098XH Amsterdam, The Netherlands
        \and 
        School of Astronomy and Space Science, University of the Chinese Academy of Sciences, Beijing 100012, China
        \and
        Department of Physics, Columbia University, New York, NY 10027, USA
        \and
        Center for Computational Astrophysics, Flatiron Institute, New York, NY 10010, USA
        \and
        The Observatories of the Carnegie Institution for Science, 813 Santa Barbara Street, Pasadena, CA 91101, USA
        \and
        Department of Physics and Astronomy, University of California, Berkeley, CA 94720 
        \and
        Max Planck Institute for Astrophysics, Karl-Schwarzschild-Str. 1, 85748 Garching, Germany
        \and
        Center for Astrophysics, Harvard-Smithsonian, 60 Garden Street, Cambridge, MA 02138, USA
        \\
        $^{*}$\email{e.c.laplace@uva.nl}
}

 \abstract
 {
 	The majority of massive stars live in binary or multiple systems and will interact with a companion during their lifetimes, which helps to explain the observed diversity of core-collapse supernovae. Donor stars in binary systems can lose most of their hydrogen-rich envelopes through mass transfer. As a result, not only are the surface properties affected, but so is the core structure. However, most calculations of the core-collapse properties of massive stars rely on single-star models. We present a systematic study of the difference between the pre-supernova structures of single stars and stars of the same initial mass (11 -- 21\Msun) that have been stripped due to stable post-main-sequence mass transfer at solar metallicity. We present the pre-supernova core composition with novel diagrams that give an intuitive representation of the isotope distribution. As shown in previous studies, at the edge of the carbon-oxygen core, the binary-stripped star models contain an extended gradient of carbon, oxygen, and neon. This layer remains until core collapse and is more extended in mass for higher initial stellar masses. It originates from the receding of the convective helium core during core helium burning in binary-stripped stars, which does not occur in single-star models. We find that this same evolutionary phase leads to systematic differences in the final density and nuclear energy generation profiles. Binary-stripped star models have systematically higher total masses of carbon at the moment of core collapse compared to single-star models, which likely results in systematically different supernova yields. In about half of our models, the silicon-burning and oxygen-rich layers merge after core silicon burning. We discuss the implications of our findings for the ``explodability,'' supernova observations, and nucleosynthesis of these stars. Our models are publicly available and can be readily used as input for detailed supernova simulations.}
 
   \keywords{ stars: massive -- supernovae: general -- binaries: close -- stars: evolution -- nuclear reactions, nucleosynthesis, abundances -- stars: neutron}

   \maketitle
%
%-------------------------------------------------------------------

\section{Introduction}
The question of how massive stars end their lives is one of the most important in stellar astrophysics. Recent developments in supernova simulations, through the inclusion of more sophisticated physics and advancements in computational capabilities, have produced the first successful three-dimensional explosions of stars by independent groups \citep[e.g.,][]{takiwaki_three-dimensional_2012,lentz_three-dimensional_2015,melson_neutrino-driven_2015,muller_dynamics_2015,roberts_general-relativistic_2016,kuroda_full_2018,ott_progenitor_2018,summa_rotation-supported_2018,muller_three-dimensional_2019,vartanyan_successful_2019,burrows_overarching_2020}, though the debate is still open regarding which components are essential \citep[for a recent review, see][]{burrows_core-collapse_2021}. On the observational side, the rise of robotic transient surveys, which are revealing an unprecedented number and diversity of supernovae, is giving us exceptional samples to compare with theoretical predictions. Examples of these facilities include the Zwicky Transient Facility \citep[ZTF,][]{bellm_zwicky_2019}, the Vera Rubin Observatory \citep{ivezic_large_2008}, the DLT40 survey telescope PROMPT \citep{tartaglia_early_2018}, and the All-Sky Automated Survey for Supernovae \citep[ASAS-SN,][]{kochanek_all-sky_2017}. For both stellar explosion models and the interpretation of supernova data, robust stellar models that accurately reflect our current understanding of massive star evolution are required.

Most massive stars live in multiple systems and will interact with a companion during their lifetime \citep[e.g.,][]{sana_binary_2012}. As a result of these interactions, stars can transfer their hydrogen-rich envelope to a companion star before core collapse, leading to supernovae that appear different from those that would be produced if all massive stars were single \citep{Wheeler+Levreault1985,podsiadlowski_presupernova_1992}.  Transient observations have revealed a diverse population of explosions that resemble the supernovae expected from such binary-stripped stars \citep[for a review see, e.g.,][]{modjaz_new_2019}.

Stars that are stripped by binary interactions before they reach core collapse are also important in forming the observed population of compact-object binaries from isolated binaries \citep[see, e.g.,][]{bhattacharya_formation_1991, dewi_late_2003, podsiadlowski_effects_2004, dewi_double-core_2006}.   Therefore, understanding stripped-envelope stars, and the outcomes of their supernovae, is crucial for understanding the formation of stellar-mass gravitational-wave merger sources \citep[see, e.g.,][]{abbott_observation_2016,abbott_gw170817_2017}.    

Pioneering work revealed differences between single and binary-stripped stellar structures, namely, systematically less massive final cores \citep{kippenhahn_entwicklung_1967,habets_evolution_1986}, except for the mass range in which a second dredge-up decreases the mass of the helium core in single stars \citep{podsiadlowski_effects_2004,poelarends_supernova_2008}. \citet{langer_standard_1989,langer_wolf-rayet_1991} and \citet{woosley_evolution_1993} found that wind mass loss in pure helium stars leads to a shrinking convective core that affects the final core mass and composition. Subsequent work investigated the conditions for which envelope loss alters the final core structure enough to change whether a massive star of a given initial mass would form a neutron star or black hole at core collapse (for early studies see, e.g., \citealt{brown_formation_1996}, \citealt{brown_formation_2001}, \citealt{Wellstein+Langer1999}, and \citealt{pols_helium-star_2002}).

However, despite the importance of massive stripped-envelope stars and the potential effects of envelope loss on the evolution and structure of the core, the majority of detailed stellar structures at the onset of core collapse are computed for single stars.   Studies that follow the binary interaction in detail are rare.  A common assumption is that the structures of binary progenitors can be adequately approximated with pure helium star models \citep[e.g.,][]{woosley_evolution_2019}. An alternative simplifying approximation is that the outcomes following mass transfer in a binary system are equivalent to mass loss over an assumed timescale until a certain surface composition has been reached \citep[e.g.,][]{schneider_pre-supernova_2021}.

Few calculations of binary stellar models self-consistently capture changes in the composition and interior structure through the final hydrostatic burning phases of massive stars. Instead, stellar evolution calculations of binaries are commonly stopped at earlier stages, such as the end of core carbon burning \citep[e.g.,][]{eldridge_implications_2006,yoon_type_2010, yoon_type_2017,eldridge_supernova_2018,gilkis_effects_2019}. Models of core-collapse progenitors often use a small nuclear reaction network that only approximately captures the late phases of nuclear burning (\citealt{timmes_integration_1999,timmes_inexpensive_2000}; recent examples include \citealt{aguilera-dena_pre-collapse_2020} and \citealt{schneider_pre-supernova_2021}). In reality, after a silicon-rich core has been formed in massive stars, leptonic losses due to electron-capture processes significantly change the interior stellar structure and composition \citep{hix_silicon_1996}. Such electron-capture processes can also be important in earlier phases, notably for massive stars that develop a partially degenerate core before the end of oxygen burning \citep{Thielemann+Arnett1985,Jones+2013}. \citet{farmer_variations_2016} demonstrated the potential impact of the size of the nuclear reaction network and of the numerical resolution on the structure of pre-core-collapse models. However, appropriately extensive nuclear reaction networks are time- and memory-intensive, and thus relatively computationally expensive. Therefore, few suitably detailed models are available \citep{woosley_evolution_2002,woosley_nucleosynthesis_2007,renzo_systematic_2017}. This lack of detailed progenitor models matters because even small differences in stellar structure can have large consequences for the outcomes of simulations of stellar core collapse (e.g.,\,the location of the silicon-oxygen interface; see \citealt{vartanyan_revival_2018}). 

To better understand the impact of binary interaction on the stellar structure at core collapse, a systematic comparison between the pre-supernova core structures of single stars and of stars stripped in binary systems is needed. Recent studies have presented pre-core-collapse models of stars that have lost their hydrogen-rich envelopes \citep{marchant_pulsational_2019,woosley_evolution_2019,schneider_pre-supernova_2021} and even their helium-rich layers \citep{tauris_ultra-stripped_2015, kruckow_progenitors_2018}.  

Independent groups have also explored the ``explodability'' of stars and the distribution of remnant compact object masses, using the evolution of carbon-oxygen cores with varying carbon to oxygen mass fractions \citep[e.g.][]{patton_towards_2020} or prescriptions based on the masses of carbon-oxygen cores and helium shells \citep{fryer_compact_2012,ertl_two-parameter_2016,ertl_explosion_2020,mandel_simple_2020,mandel_binary_2021}. A limitation of these studies is their assumed homogeneous composition distributions, at either the start or end of core helium burning, which may not accurately capture the complex structure revealed by more detailed stellar evolution models.

In this study we systematically compare the evolution and pre-supernova structures of single stars and donor stars in binary systems. We present models of stars at solar metallicity with initial masses of 11 to 21\Msun that are representative of neutron star progenitors. After core oxygen depletion, we solve the stellar structure simultaneously with a nuclear network of 128 isotopes, so as to self-consistently model the burning, including the evolution of the electron fraction during that phase. Following \citet{farmer_variations_2016}, we employ a sufficiently high spatial and temporal resolution to ensure a converged final helium core mass. We compare the composition structures using novel diagrams that represent the stellar structure on a two-dimensional surface and enable the visualization of the full isotope distribution. We investigate the origin of the systematic differences in structure and composition by studying details of the late-time evolution. We first review the general effect of stable mass transfer in a binary system and the differences that arise compared to the evolution of single-star models of the same initial mass in Sect. \ref{sec:evolution}. We present our main findings on the systematic differences in the pre-supernova density and composition structure of single and binary-stripped star models with the same core mass in Sect. \ref{sec:endcomp}. In Sect. \ref{sec:origins} we investigate the origin of these differences. We discuss the implications and limitations of our findings in Sect. \ref{sec:discussion} and conclude in Sect. \ref{sec:conclusion}.

%--------------------------------------------------------------------
\section{Method}
\label{sec:method}
We employed the MESA stellar evolution code \citep[version 10398;][]{paxton_modules_2011,paxton_modules_2013,paxton_modules_2015,paxton_modules_2018,paxton_modules_2019} to compute the stellar structure of massive stars from the beginning of core hydrogen burning until the onset of core collapse. We calculated two sets of 11 stellar models with the same initial masses of 11 -- 21\Msun. The first set follows the evolution of single massive stars, while the second models the evolution of stars with the same initial masses but in a close binary system.

For the binary models, we used the same setup as in \citet{laplace_expansion_2020}, in which we simplified the computation of the binary interaction by approximating the initially less massive companion star (the secondary) as a point mass with initially 80\% of the mass of the primary star. We assumed mass transfer occurs conservatively such that no mass is lost during Roche-lobe overflow. We followed the time-dependent mass transfer evolution, which depends on the radial and orbital evolution of our models, computed self-consistently. We focused on binary systems undergoing the most common type of mass transfer, known as stable case~B mass transfer \citep{de_mink_binaries_2008}. This term designates mass exchange initiated by the primary star filling its Roche lobe after expanding during the hydrogen-shell burning phase that follows the end of core-hydrogen burning \citep{kippenhahn_entwicklung_1967, tutukov_evolution_1973}. We chose initial orbital periods between $25$ and $35$~d. This is not done systematically, but we believe the evolutionary stage at the beginning of mass transfer is sufficiently similar so as to not significantly affect our results \citep{gotberg_ionizing_2017,laplace_expansion_2020}. For these choices of binary parameters at solar metallicity, stars of 11\Msun and above do not interact again with their companion after the first phase of mass transfer for these choices of physics and binary parameters \citep[][]{yoon_type_2017,laplace_expansion_2020}. Thus, after the primary star reached core helium depletion (defined as the moment when the central helium mass fraction in the core decreases below $10^{-4}$), we only followed the evolution of the primary star.

For both the single stars and the primary stars in binary systems, we used the same choice of physical assumptions, explained in detail below.

\subsection{Starting and stopping conditions}
We defined the starting point of the evolution as the moment when the abundance of helium in the center increases by 5\%. From this point on, we evolved the models until the onset of core collapse, which we defined as the moment when the in-fall velocity of any point within the boundary of the iron-rich core  reaches $1000\,\rm{km\,s}^{-1}$ \citep{woosley_evolution_2002}. Here, "iron" includes all species for which the mass number is higher than 46. Throughout this work, we define the core boundaries as the mass coordinate where the mass fraction of the depleted element (for example $^{28}$Si in the case of the iron core) decreases below 0.01 and the mass fraction of the most abundant element (for example iron) increases above 0.1. We did not take rotation into account for the evolution of single stars nor the binary stars. 

\subsection{Metallicity and opacities} 
The models were computed at solar metallicity \citep[Z = 0.0142, where Z is the mass fraction of elements heavier than helium;][]{asplund_chemical_2009}, and we employed the opacity tables from \citet{ferguson_low-temperature_2005} and from OPAL \citep{iglesias_radiative_1993,iglesias_updated_1996}. 
We assumed an initial helium mass fraction of $Y = 2Z\,+\, 0.24$ and an initial hydrogen mass fraction of $X = 1 - Y - Z$, following \citet{tout_zero-age_1996} and \citet{pols_stellar_1998}. 

\subsection{Nuclear reaction network} 
To obtain accurate information for the interior composition profile at the onset of core collapse, a large nuclear reaction network is needed. It allows all electron-capture processes that become significant after core-oxygen depletion and that affect the core structure through lepton losses to be followed. \citet{farmer_variations_2016} showed that only models computed with nuclear reaction networks containing at least 127 isotopes do not exhibit significant variations in their pre-supernova structure (e.g., the mass of the iron core) compared to models with larger networks. Therefore, after core-oxygen depletion, we employed a nuclear reaction network consisting of 128 isotopes \citep[\texttt{mesa128};][]{timmes_integration_1999,timmes_inexpensive_2000,paxton_modules_2011}. 

\textsc{MESA} solves the fully coupled stellar structure and composition equations simultaneously using a single reaction network \citep{paxton_modules_2011,paxton_modules_2015}. This enables a self-consistent calculation of all quantities, including but not limited to the energy generation rate, the electron fraction, and the composition. 
We computed nuclear reactions in the stellar interior until the end of core oxygen burning with an alpha-chain network containing the 21 most important isotopes for these evolutionary phases \citep[\texttt{approx21};][]{timmes_inexpensive_2000, paxton_modules_2011}.
We note two imperfections introduced by our use of the approximate alpha-chain network in these early burning phases. The first is a consequence of \texttt{approx21} not containing any isotopes of neon aside from $^{20}$Ne. To approximate the burning of $^{14}$N to $^{22}$Ne, the network adopts the construction $^{14}$N($\frac{3}{2}\alpha,\gamma$)$^{20}$Ne (for which, see \citealt{pols_approximate_1995}). This inevitably leads to a small systematic error in the electron fraction ($Y_{e}$) after helium burning, and also affects the later isotope distribution. This applies to all our models, but we consider that the eventual overall effect is small. We demonstrate the size of the effect in Appendix \ref{sec:appendix:nuclear_network}, in which we compare representative models that use our standard method to more computationally expensive models that use the full network for the whole evolution (see also, e.g., the comparisons in \citealt{sukhbold_high-resolution_2018}). We have indicated in figure captions where regions labeled as $^{20}$Ne are substituting for $^{22}$Ne, which occurs when the neon formed in helium-rich zones. The second issue is potentially more significant, but only for the lower-mass models in our grid. The alpha-chain network \texttt{approx21} neglects some electron-capture reactions that become important at high densities, including before the end of oxygen burning (see, e.g., \citealt{Thielemann+Arnett1985} and \citealt{Jones+2013}). This is a common approximation for core-collapse models with similarly low masses as those we study (see, e.g., \citealt{schneider_pre-supernova_2021}), but is a source of systematic error.  As a guide to the range in which this may be important for our results, we note that \citet{woosley_evolution_2019} adopted a full nuclear network for helium stars with initial masses below $4.5 M_{\odot}$ for this reason, which roughly corresponds to our three (two) lowest-mass stripped (single) stars. We consider that this would not significantly affect the main direct paired comparisons we present, nor our main conclusions.

We chose to perform a single switch of nuclear network, rather than gradually increasing the number of isotopes taken into account for distinct evolutionary steps. This allowed us to minimize numerical artifacts that can be introduced by these switches \citep{renzo:thesis,renzo_systematic_2017}.
We visually inspected our models to confirm that switching networks did not introduce obvious discontinuities in the evolution or artificial features in the structures. We present examples of this with Kippenhahn diagrams in Appendix \ref{sec:appendix:kipp} (see also Appendix \ref{sec:appendix:nuclear_network}).
We employed values from \citet{angulo_compilation_1999} for the $^{12}\mathrm{C}(\alpha,\gamma)^{16}\mathrm{O}$ rate.

\subsection{Mixing}
For convective mixing we used the mixing length theory (MLT) approximation \citep{bohm-vitense_uber_1958} with a mixing length parameter $\alpha_{\rm{MLT}} = 1.5$. We adopt the Ledoux criterion for evaluating stability against convective mixing and assume efficient semi-convection, with a semi-convection parameter $\alpha_{\mathrm{sc}} = 1.0$ \citep{Langer+1991}. Due to numerical issues in the outer layers of the most massive stellar models, we treated convection using the \texttt{MLT++} scheme of MESA \citep{paxton_modules_2013} before core oxygen depletion. This method artificially increases the energy flux in radiation-dominated convectively inefficient layers \citep[cf.][]{jiang_local_2015,jiang_outbursts_2018}. 
We took into account convective overshooting above the core and shells by using a step overshooting parameter of 0.335 pressure scale heights, appropriate for stars in our mass range \citep[see][]{brott_rotating_2011}. We did not assume any undershooting.

\subsection{Winds} 
For the stellar winds, we used the "Dutch" scheme of \textsc{MESA} \citep{de_jager_mass_1988,nugis_mass-loss_2000,vink_mass-loss_2001}. For the wind mass loss of binary-stripped stars, we employed the extrapolated empirical prescription of \citet{nugis_mass-loss_2000} \citep[for more details, see][]{laplace_expansion_2020}.
The timing and amount of wind mass loss has a significant impact on the pre-collapse core structure of massive stars \citep{renzo_systematic_2017,gilkis_effects_2019}, but we expect the binary interactions to produce a larger effect (because of the shorter timescale and higher mass-loss rates). We investigate the effect of varying the wind mass-loss rate in Appendix \ref{sec:appendix:winds}.

\subsection{Resolution} 
Numerical spatial resolution, that is, the choice of the number and minimum step between mass shells, can affect the pre-supernova structure of stellar models \citep{farmer_variations_2016}. Testing showed that converged values of stellar parameters (e.g., the helium core mass) could be obtained by choosing at least 1000 mass cells in each model (with an average of 5000 throughout the evolution) and ensuring that about one thousandth of the total mass be contained in each cell (\texttt{max\_dq} = $10^{-3}$). Our models typically contain 40,000 time steps until core oxygen depletion, and 200,000 more until core collapse. Further details can be found in our MESA inlists and models available online\footnote{\href{https://doi.org/10.5281/zenodo.4506803}{https://doi.org/10.5281/zenodo.4506803}}.

The analysis was performed with the following open-source codes: \texttt{mesaPlot} \citep{farmer_mesaPlot_2019}, \texttt{matplotlib} \citep{Hunter_matplotlib_2007}, \texttt{numpy} \citep{van_der_walt_numpy_2011}, \texttt{ipython/jupyter}, \citep{perez_ipython_2007} and \texttt{TULIPS} \citep{laplace_tulips_2021}.

% --------------------------------------------------
\begin{sidewaysfigure*} % ----- Evolution of the composition
        \centering
        %       \vspace{-1.3cm}
        \begin{minipage}{\textwidth}
                \centering
                \includegraphics[width=\textwidth]{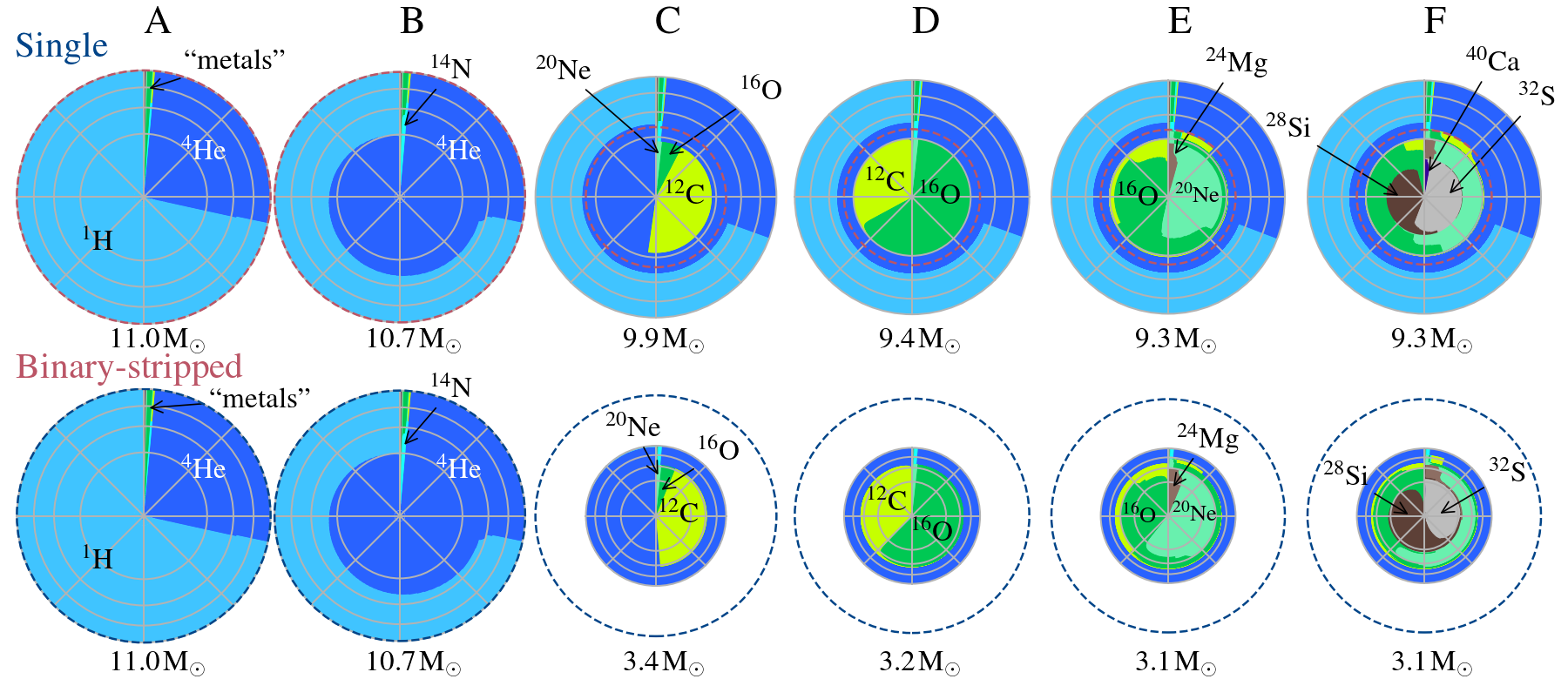}
        \end{minipage}
        \begin{minipage}{\textheight}
                \includegraphics[width=0.33\textwidth]{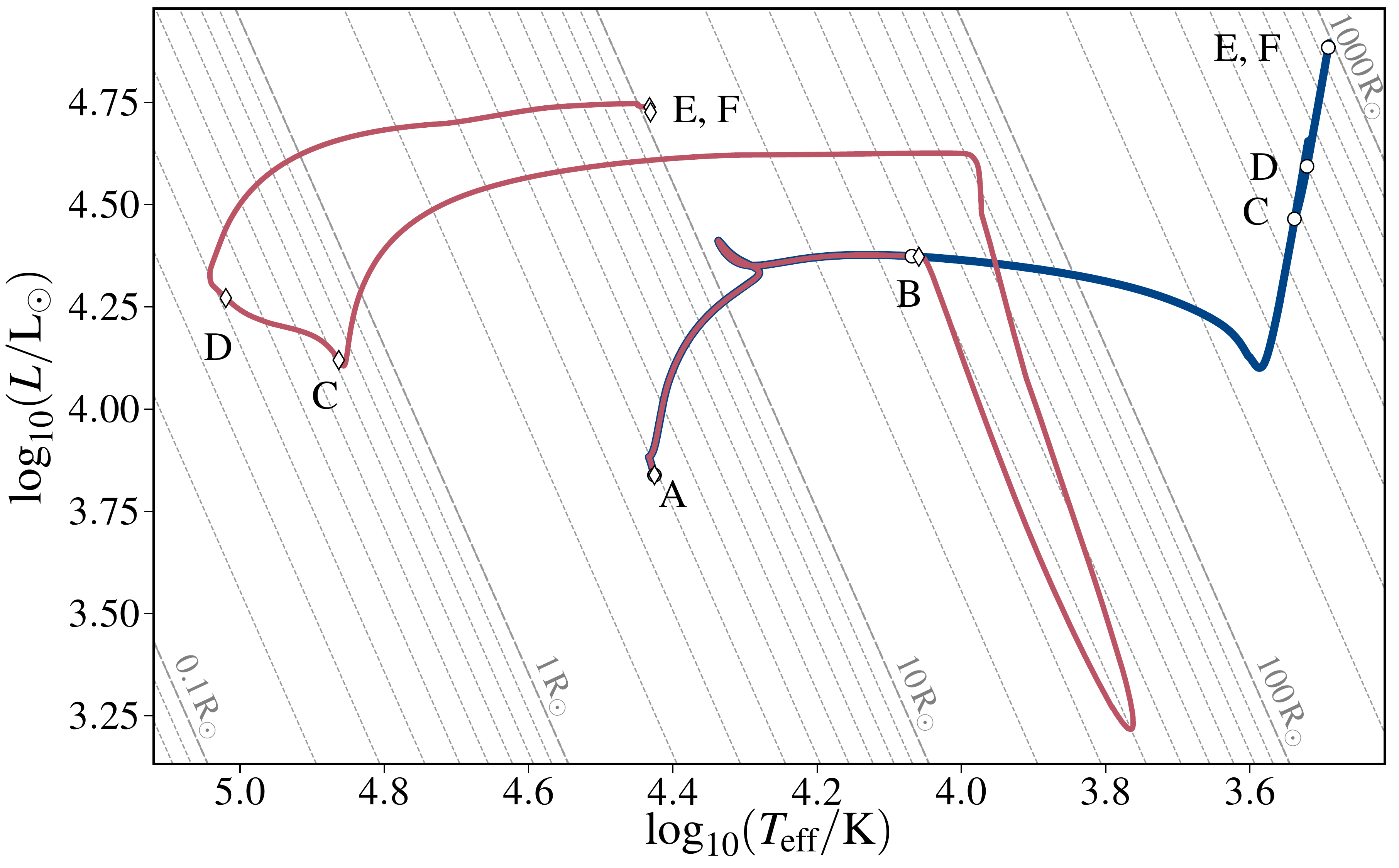}
                \includegraphics[width=0.33\textwidth]{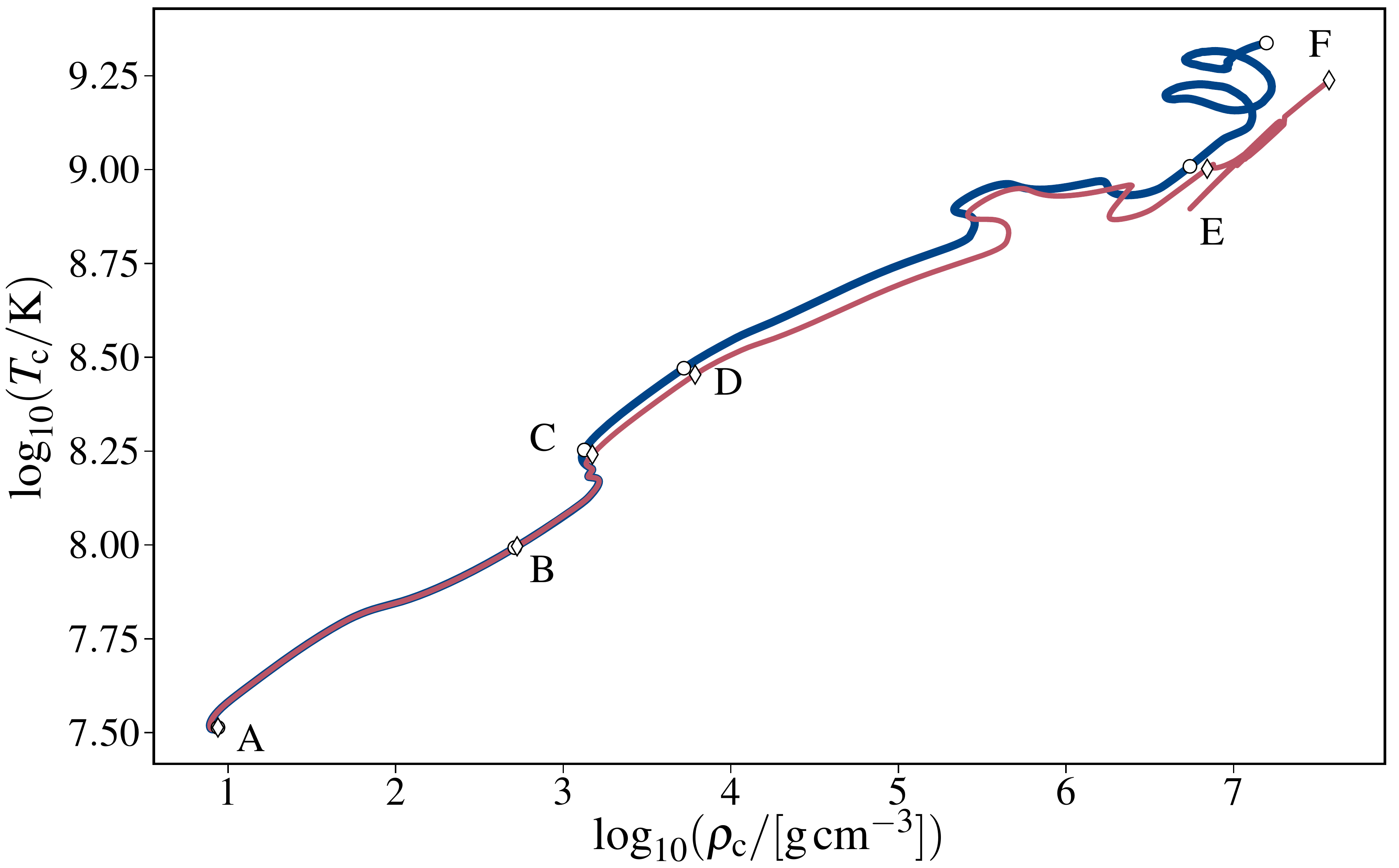}
                \includegraphics[width=0.33\textwidth]{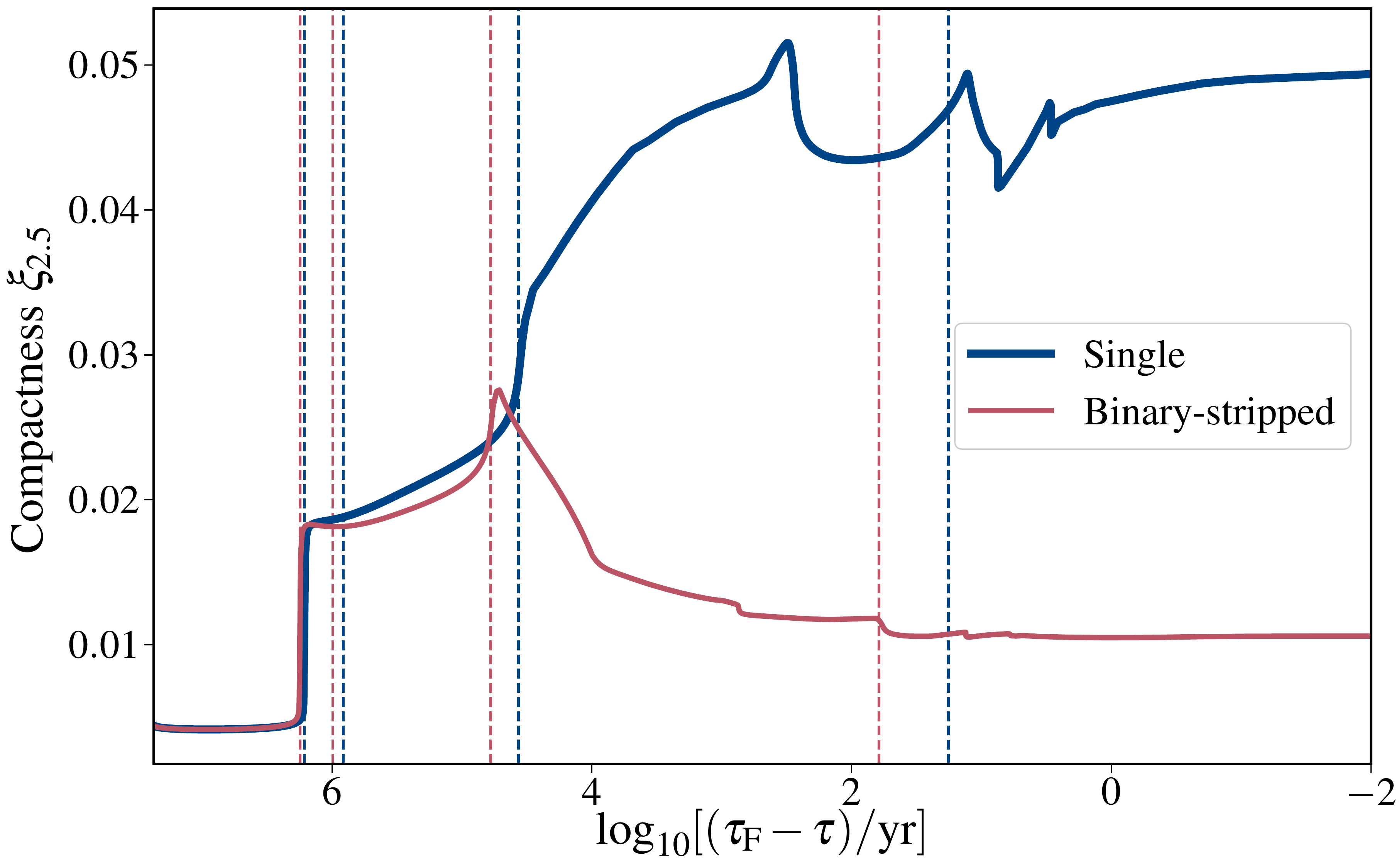}
        \end{minipage}
        \caption{Evolution of example single (blue) and binary-stripped star (red) models with the same initial mass of 11\Msun until core oxygen depletion. Key moments of the evolution, discussed in Sect. \ref{sec:evolution}, are marked with letters A to F.
                \textbf{Upper panels} - Chemical structure of the single star (\textit{top}) and the donor star in a binary system (binary-stripped star, \textit{bottom}) with the same initial mass of 11\Msun at key moments of the evolution. The radial direction is proportional to the square root of the total mass of the model (see text). The total mass is given below each model. Colors indicate the local mass fraction of each isotope. The surface area spanned by each element is proportional to its total mass. The diagrams are divided into concentric rings. Each ring is a pie chart of the most abundant isotopes in the models. From the outside moving inward, gray circles mark regions containing 100, 75, 50, and 25 percent of the total mass, respectively. Lines are placed at equal intervals of one-eighth of the total fraction of isotopes. To aid the comparison, dashed blue or red circles indicate the total mass of the alternate model (single or binary-stripped) at the same stage. The neon isotope labeled in panel C would be $^{22}$Ne in reality and is only $^{20}$Ne as an artifact of the \texttt{approx21} network. Likewise, for these relatively low-mass models, the isotope distribution at the end of core oxygen burning (panel F) is significantly affected by the simplifications of \texttt{approx21}.
                \textbf{Bottom panels} - Evolution of the single (\textit{blue}) and binary-stripped (\textit{red}) models in the HR diagram (\textit{left}) and the central density - central temperature (\textit{center}) plane. Circles and diamonds mark important evolutionary steps for the single and binary-stripped star model, respectively. We also show the evolution of the compactness parameter (\textit{right}) as a function of the time until core oxygen depletion. Vertical dashed lines mark important evolutionary steps.
        }
        \label{fig:evolution}
\end{sidewaysfigure*}
\section{Comparison of an example single and binary evolutionary model}
\label{sec:evolution}
In this section we review generic differences in evolution between single-star models and binary models of the same initial mass. To this end, we compare the evolution of two representative models that start their evolution with the same initial mass of 11\Msun.
In Fig. \ref{fig:evolution}, we present the evolution of these models. In the bottom panels, we display evolutionary tracks on the Hertzsprung-Russell (HR) diagram and on the central density -- central temperature ($\log\rho_{\rm{c}}$- $\log T_{\rm{c}}$) plane. We also show the evolution of the compactness parameter $\xi_{M}$, commonly used to characterize the core structure of stars (see also Sect. \ref{sec:origins:compactness}), defined as follows \citep{oconnor_black_2011}:
\begin{equation}
\xi_{M} = \frac{M / \Msun}{R(M) / 1000 \mathrm{~km}}\,.
\end{equation}
Here, we evaluate the compactness parameter at $M = 2.5\Msun$, which approximately corresponds to the boundary between black hole and neutron star masses \citep{ugliano_progenitor-explosion_2012}.
We further show composition structure diagrams at specific times of the evolution. Since these diagrams\footnote{The composition diagrams are constructed using the python package \texttt{TULIPS} that is being prepared for release as open-source software, \citep{laplace_tulips_2021}.} differ from standard representations of the stellar composition found in the literature, we briefly describe them here. 
Each color in the composition diagrams represents a different isotope. The center and edge of the shaded circles in each of the composition diagrams correspond to the center and surface of each star, respectively, and rings about the center represent intermediate mass coordinates. Specifically, the radius of each ring about the center is proportional to the square root of the Lagrangian mass coordinate. The fraction of each ring that is shaded in a particular color indicates the mass fraction of the corresponding isotope at that mass coordinate. This combination means that the total area of any color in the plot is proportional to the total mass of the corresponding isotope in the stellar model. At each mass coordinate, the isotopes are ordered counterclockwise, by increasing atomic number (and, for isotopes with identical atomic number, by increasing mass number). Nuclear fusion typically causes the mass fraction of the dominant, low atomic number, species to decrease. Hence in regions of nuclear fusion, evolutionary composition changes typically cause color patterns to move clockwise on the composition diagram\footnote{A movie showing the changes in composition can be found at \href{https://doi.org/10.5281/zenodo.4506803}{https://doi.org/10.5281/zenodo.4506803}.}. To aid the comparison between different diagrams, we show dashed red (blue) circles that indicate the total mass of the corresponding binary (single)-star model.

In the sections that follow we describe the example single and binary models at key phases of the evolution, referring to the phases labeled A-F in Fig. \ref{fig:evolution}.

\subsection{Early evolution until mass transfer in binary models (A -- B)}
\label{sec:evolution:zams_to_rlof}
The first part of the evolution, starting from the zero-age main-sequence (labeled point A in Fig. \ref{fig:evolution}) is identical for the single-star model and the primary star in the binary model. Tides and rotational effects due to the binary evolution have a negligible impact at this stage. The stars begin their evolution with a solar metallicity composition, that is, abundances of 0.7174, 0.2684, and 0.0142 for hydrogen, helium, and heavier elements, respectively \citep{asplund_chemical_2009}, as shown in the composition diagrams.

After the end of core hydrogen burning, the stars burn hydrogen in a shell and expand, as can be observed on the HR diagram in Fig. \ref{fig:evolution}, in which the diagonal lines show loci at constant radii. At point B, the binary star fills its Roche lobe and starts to transfer matter to its companion, leading to a divergence of the evolutionary tracks on the HR diagram. At this point, the stars still have an identical chemical structure (see point B in the composition diagrams of Fig. \ref{fig:evolution}), with a pristine composition in the outer layers and a core that is composed mainly of helium, with small mass fractions (less than 0.01) of nitrogen, which has been produced by the CNO cycle. The spiral structure visible in the center of the composition diagram (point B in Fig. \ref{fig:evolution}) reflects the chemical gradient developed above the helium core. This is the result of the recession in mass coordinate of the convective core during core hydrogen burning. This can also be seen in the evolutionary tracks on the $\log\rho_{\rm{c}}$- $\log T_{\rm{c}}$ plane in Fig. \ref{fig:evolution}, where the tracks of the single and binary star are indistinguishable between point A and B. The same is true for the compactness parameter.

\subsection{Development of key differences until the end of core helium burning (B -- D)}
\label{sec:evolution:rlof_to_hedepl}
The donor star in the binary transfers matter to its companion and loses nearly all its outer hydrogen envelope (B -- C), becoming a "binary-stripped" star and leading to a dramatic change in chemical structure and surface properties \citep[for a detailed description, see, e.g.,][]{gotberg_ionizing_2017,laplace_expansion_2020}. This change is apparent on the composition diagrams at point C of Fig. \ref{fig:evolution} (see also the dashed colored circles giving the total mass of the alternate model). 
Meanwhile, the single star continues to expand and cool while burning hydrogen in a shell. At point C, both models have fused roughly half of the helium inside their cores. This is the moment when the central temperature and density conditions of the stars start to diverge (see the $\log\rho_{\rm{c}}$- $\log T_{\rm{c}}$ diagram and compactness parameter evolution in Fig. \ref{fig:evolution}). At the same evolutionary stage (C), the binary-stripped star has a slightly denser and cooler core than the single star, because from this point on, the binary-stripped star behaves, to a first approximation, like the core of a star with a lower initial mass \citep[cf.][]{kippenhahn_entwicklung_1967}.

At the end of core-helium burning (D, defined as the moment when the central helium mass fraction drops below $10^{-4}$), the helium core mass of the single-star model is larger than that of the binary-stripped star (see the dashed circle on the composition diagram at point D in Fig. \ref{fig:evolution}). This can be explained by two effects: (1) the helium core mass of single stars increases due to the creation of helium by the hydrogen-burning shell \citep[e.g.,][]{woosley_evolution_2019}; and (2) the binary-stripped star loses mass due to winds, leading to a decrease in the helium core mass (see also Appendix \ref{sec:appendix:helium_core_mass}). 
At core helium depletion, the core composition differ significantly between the models (see the composition diagrams at point D in Fig. \ref{fig:evolution}). While the cores of both the single and the binary-stripped star model are composed of the same products of core helium burning (namely carbon, oxygen, and neon), the relative ratios of these elements are different. The mass fraction of carbon is larger in the binary-stripped star with an abundance of 0.38 compared to 0.33 for the single star. In contrast, the mass fraction of oxygen is smaller for the binary-stripped star model, 0.60 compared to 0.65 for the single star. This is caused by differences in the mass and density of their cores and by the distinct behavior of their convective cores during core helium burning \citep[cf.][]{langer_standard_1989,woosley_evolution_1993, brown_formation_2001}. Higher core masses and lower densities, together with a growth of the convective helium-burning core, favor a more efficient destruction of carbon through alpha captures in single-star models and leads to the observed differences in central carbon and oxygen mass fractions \citep[][see Sect. \ref{sec:origins:co_gradient} for more details]{woosley_evolution_1993,brown_formation_2001}. 

The binary-stripped star model develops an extended carbon-oxygen gradient at the edge of the core (visible on the composition diagram as the lime-colored outer "arm" at the bottom of the dark green region at point D in Fig. \ref{fig:evolution}). This is due to the convective core shrinking during core helium burning as a result of wind mass loss \citep[][for more details, see Sect. \ref{sec:origins:co_gradient} and Appendix \ref{sec:appendix:winds}]{langer_standard_1989,woosley_evolution_1993}.
At core helium depletion, the compactness parameter reaches a value of 0.03 for both the single and binary-stripped star models.
 
\subsection{Evolution after core helium depletion (D -- F)}
\label{sec:evolution:after_hedepl}
After the end of core helium burning, both stars expand again while burning helium in a shell (labeled D -- E in Fig.~\ref{fig:evolution}) and reach their final location on the HR diagram\footnote{Assuming no dynamical transient happens shortly before core collapse \citep[e.g.,][]{shiode_stability_2012,khazov_flash_2016,fuller_pre-supernova_2017,fuller_pre-supernova_2018}}. The stars enter the final burning stages of heavier elements, starting with core carbon burning, and the evolution accelerates due to neutrino losses \citep[e.g.,][]{fraley_supernovae_1968}. During this phase, a large difference in the evolution of the compactness parameter can be observed in Fig. \ref{fig:evolution}. The compactness value of the binary-stripped star model decreases, while it increases for the single-star model. This reflects the change in radius evaluated at the same mass of 2.5\Msun. 

The previously built-up chemical composition differences remain until the end of core carbon burning (labeled E in Fig.~\ref{fig:evolution}, which marks the moment when the central mass fraction of carbon drops below $10^{-4}$). The binary-stripped star model is less abundant in oxygen and more abundant in neon and magnesium than the single star. This can be attributed to the different composition at the beginning of core carbon burning and to the different burning conditions during core carbon burning as shown in $\log\rho_{\rm{c}}$- $\log T_{\rm{c}}$ diagram in Fig \ref{fig:evolution}. The compactness parameter of the single star shows two prominent maxima, after each of which the compactness parameter decreases significantly. These maxima are coincident with instances of off-center carbon ignition (as shown in Appendix \ref{sec:appendix:kipp}). Such off-center burning has previously been found to be important for compactness evolution (\citealp[cf. Fig. 7 of][]{renzo_systematic_2017}, see also \citealp{sukhbold_compactness_2014}).

The composition profiles become increasingly different during this phase of the evolution (E -- F, where F we define as the moment when the central oxygen mass fraction drops below $10^{-4}$). During this stage, the stars develop a central core mainly composed of silicon, sulfur, and calcium. In the single-star model, the mass fraction of silicon is lower than in the binary-stripped model. In contrast, the mass fraction of sulfur and calcium is larger in the single-star model compared to the binary-stripped star model. 

When massive stars reach oxygen depletion (labeled F in Fig. \ref{fig:evolution}), they have less than a few days left to live \citep[e.g.,][]{woosley_evolution_2002}. This is the moment when oxygen-shell burning followed by core and then shell silicon burning, occur. Immediately after, the iron-rich core begins to collapse. At this point, we stop our models, since the high density and extremely neutron-rich material requires a nuclear equation of state and detailed treatments of the neutrino physics. Kippenhahn diagrams that show the entire evolution of the stellar structure for these two example models can be found in Fig.~\ref{fig:kipp} of the Appendix.

\begin{figure*}[h!] % ----- single / binary composition profile at core collapse ----
        \centering
        \includegraphics[width=\textwidth]{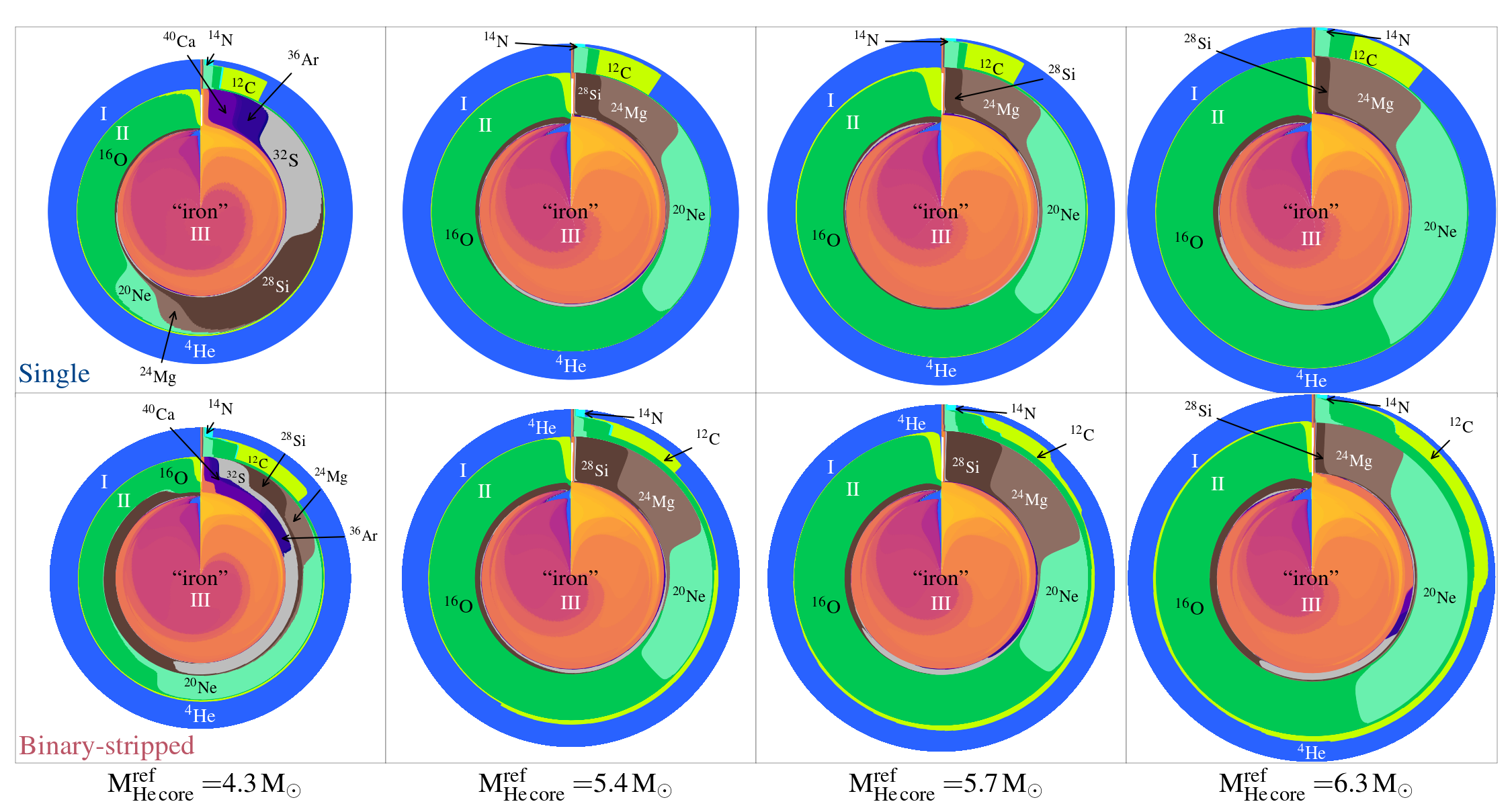}
        \caption{Final composition profiles of selected single (top row) and binary-stripped (bottom row) star models at the onset of core collapse. We show four example models with similar reference core masses, which are indicated below each column. The diagrams are constructed in a similar fashion as Fig. \ref{fig:evolution}, with each color representing an isotope and the surface area spanned by this color being proportional to the mass of this isotope in the star. The radius of each diagram is proportional to the final helium core mass. The hydrogen-rich envelope of single stellar models is not shown. Three prominent regions are marked with roman numerals: I) a helium-rich layer, II) an oxygen-rich layer, and III) an inner iron-rich zone. The lowest-mass example models (leftmost column) both show enhanced mass fractions of heavy elements in the oxygen-rich region (II) due to shell merger events. Binary-stripped star models contain an extended carbon-oxygen gradient at the edge of the oxygen-rich region (II) that is absent in single-star models.}
        \label{fig:end_comp_sb}
\end{figure*}
% --- Main results ---------------------------------------
\section{Differences between single and binary-stripped models at core collapse}
\label{sec:endcomp}
In the previous section we compared single and binary-stripped star models with the same initial mass. In this section we compare the properties of models with similar core masses. This is because the explosion properties are mainly determined by the mass of their carbon-oxygen cores \citep[e.g.,][]{woosley_evolution_2002,farmer_mind_2019}. We define the reference core mass as the mass of the helium core at the end of central helium burning. The boundary of the helium core is set as the mass at which the mass fraction of hydrogen decreases below 0.01 and the mass fraction of helium is larger than 0.1. We discuss the robustness of this definition in Appendix \ref{sec:appendix:helium_core_mass}.

\subsection{Composition at core collapse}
\label{sec:endcomp:composition}

In Fig. \ref{fig:end_comp_sb} we show the interior composition at core collapse of selected single-star models (top row) and binary-stripped star models (bottom row). We select four pairs of models with similar reference core masses of $\mathrm{M}_{\mathrm{He\,core}}^{\mathrm{ref}} \simeq 4.3, 5.4, 5.7, 6.3\Msun$ (see also Table~\ref{tab:general}). We focus on the composition inside the helium core. The radius of each diagram is proportional to the total helium core mass. The hydrogen-rich outer layers in the single-star models are not shown for clarity. These diagrams naturally bring into focus three distinct regions that contain a significant fraction of the total mass of the star, given below.

The first region is the helium-rich layer (I). This region is mainly composed of $^{4}$He. At the outer edge, the ashes of the CNO cycle can be observed with mass fractions of 0.99 for $^{4}$He, and up to 0.01 for $^{14}$N. Moving inward, $^{4}$He still makes up the largest mass fraction of the region, up to 0.75. The rest of the mass is contained in the products of helium burning, namely $^{12}$C, $^{16}$O, and $^{20}$Ne, in order of decreasing mass fraction. At the inner edge of this region, binary-stripped stars exhibit a composition gradient that is not present in single stars due to the shrinking of the convective core during helium burning (see Sect. \ref{sec:origins:co_gradient}).
        
The second is the oxygen-rich layer (II). This region is mainly composed of $^{16}$O, but also $^{20}$Ne, $^{24}$Mg, and $^{28}$Si with typical mass fractions of about 0.6, 0.2, 0.06, and 0.02, respectively. The exact distribution differs between single and binary-stripped star models with similar reference core masses. In the models with reference core masses of $4.3$\Msun (leftmost column in Fig. \ref{fig:end_comp_sb}), isotopes such as $^{36}$Ar and $^{40}$Ca have been mixed outward into this region (due to shell mergers, see Sect. \ref{sec:endcomp:shellmergers}).

The third is the iron-rich region (III). This is the center-most region, mainly composed of iron-group isotopes (with atomic mass numbers from 52 to 62). The composition diagrams show a clear, smooth, spiral pattern in the center (Fig. \ref{fig:end_comp_sb}). This is because isotopes with higher mass numbers (for example $^{58}$Fe) are more abundant in the inner-most layers, while light iron-group isotopes (for example $^{54}$Fe) are more abundant at the outer edge of the core. We also note the presence of $^{4}$He (in blue) produced by late photodisintegration. At the edge of this region, a small fraction of $^{28}$Si and products of incomplete $^{28}$Si burning such as $^{32}$S, $^{36}$Ar, and $^{40}$Ca is present. It can be identified as a narrow ring around the iron-rich region in the composition diagrams. This ring is more extended in mass for binary-stripped star models than for single stars, indicating binary-stripped stars have higher mass fractions of $^{28}$Si and its burning products at the edge of the iron-rich core than their single star counterparts.
%\end{enumerate}

Not all the material shown at the moment captured in Fig. \ref{fig:end_comp_sb} will be ejected during the supernova. Almost all of region III becomes enclosed in the compact object that forms in the center. The layers above the iron-rich core are mixed and reprocessed by the supernova shock, creating new isotopes through supernova nucleosynthesis, in particular almost all iron-group elements ejected.

In Table \ref{tab:general}, we give detailed values of core masses and masses of the most important isotopes present at the onset of core collapse. Composition diagrams for the full set of models at the moment of core collapse are shown in Appendix \ref{sec:appendix:full_set}.

\subsection{Shell mergers}
\label{sec:endcomp:shellmergers}
About half of our models experience "shell merger" events, where a convective burning shell merges with the layers above. In our models, we find that the silicon-burning shell merges with the oxygen-rich layers above in the final day before core collapse. This has been found in some three-dimensional core-collapse calculations \citep[e.g.,][for a discussion of the uncertainties, see Sect. \ref{sec:discussion:shell_mergers}]{couch_revival_2013,collins_properties_2018,yoshida_three-dimensional_2020,andrassy_3d_2020,yadav_large-scale_2020,fields_development_2020,mcneill_stochastic_2020}. The result is that a high-temperature layer containing silicon and its burning products is mixed into a lower-temperature region mainly composed of un-burned oxygen, neon, and magnesium isotopes. This produces alpha particles that, in turn, lead to enhanced alpha-capture reactions that result in high abundances of $^{28}$Si, $^{32}$S, $^{36}$Ar, and $^{40}$Ca in the oxygen-rich layer (region II of Fig. \ref{fig:end_comp_sb}), at the expense of $^{16}$O, $^{20}$Ne, and $^{24}$Mg.

\begin{figure} % ----- Mg24 / Si28 --- 
        \centering
        \includegraphics[width=0.5\textwidth]{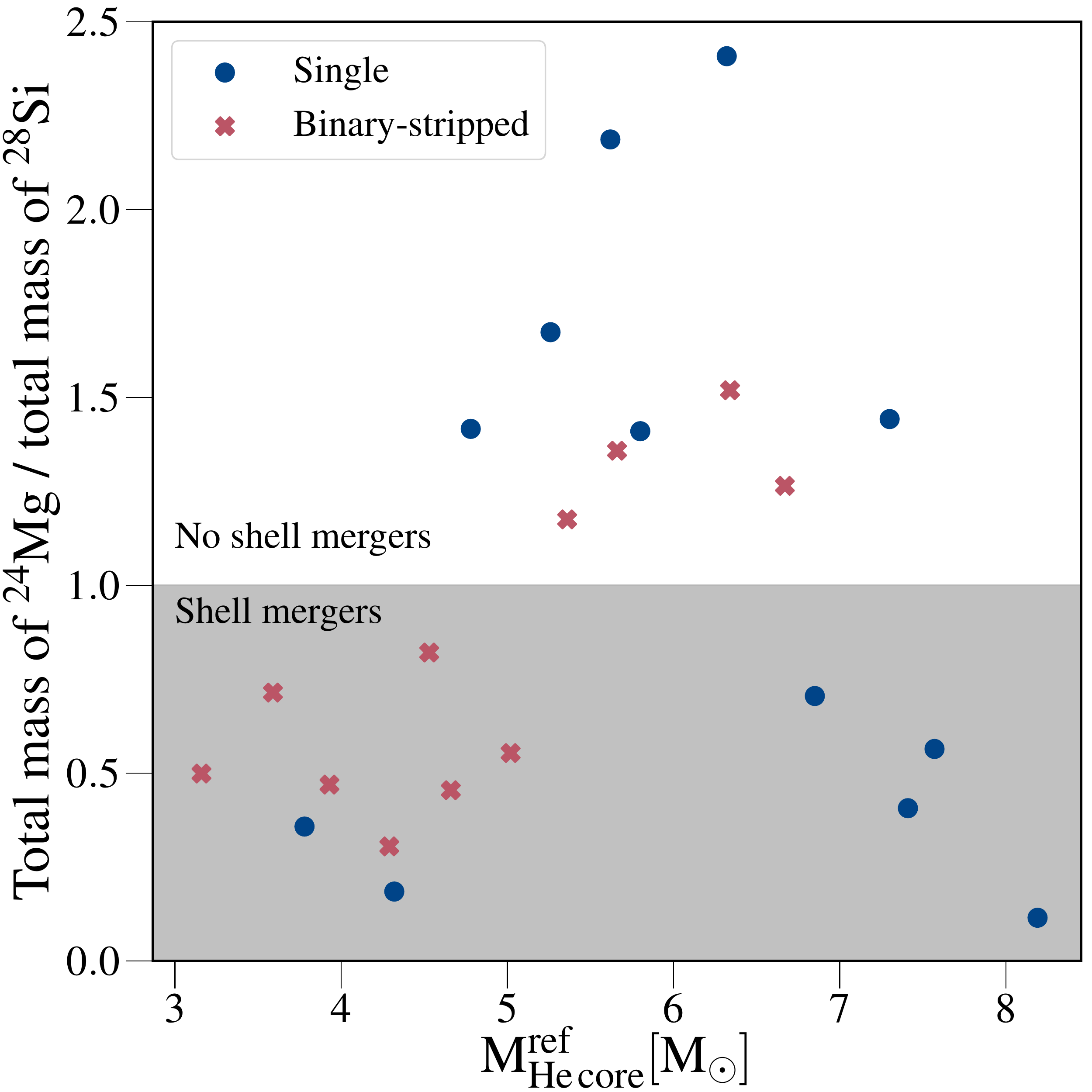}
        \caption{Ratio of the total mass of $^{28}$Si and  $^{24}$Mg at the onset of core collapse as a function of the reference core mass for single (blue) and binary-stripped (red) progenitors. Models with a ratio below one experience shell mergers (gray region). In our models, the occurrence of shell mergers is related to the helium core mass after helium burning.}
        \label{fig:mg24_si28}
\end{figure}

In Fig.~\ref{fig:mg24_si28} we show the ratio of the total mass of $^{24}$Mg and $^{28}$Si at the onset of core collapse as a function of the reference core mass. Models for which this ratio is lower than one (more $^{28}$Si than $^{24}$Mg) have experienced shell mergers (see also Appendix \ref{sec:appendix:shell_merger}). 
We find that shell mergers occur mainly for the lowest- and highest-mass models in our grid, although the robustness of this trend is not clear. Most models with reference core masses lower than about 5\Msun or higher than 6.8\Msun experience shell mergers.

\begin{figure} % ----- Ca40 distribution ----
        \centering
        \includegraphics[width=0.5\textwidth]{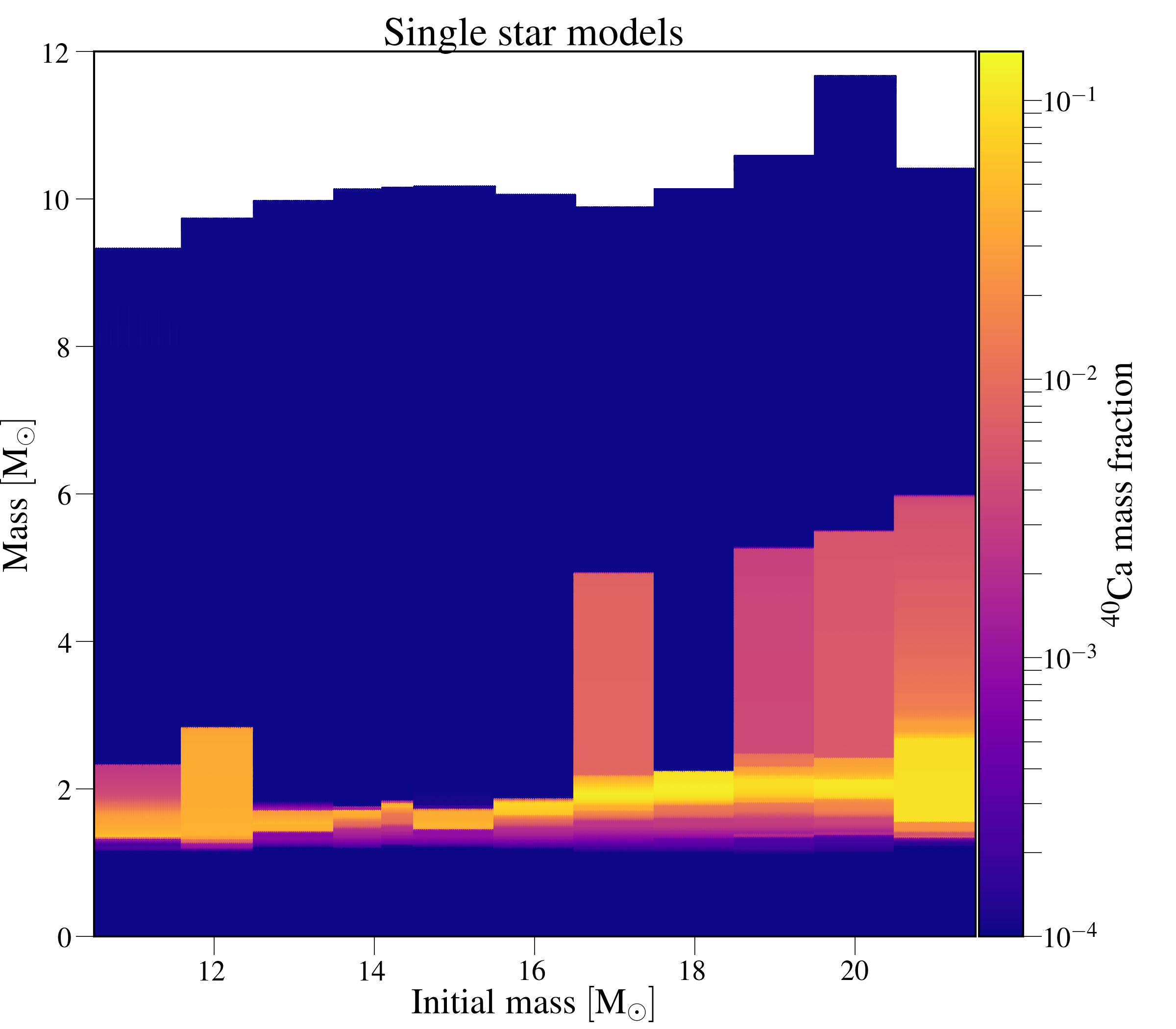}
        \includegraphics[width=0.5\textwidth]{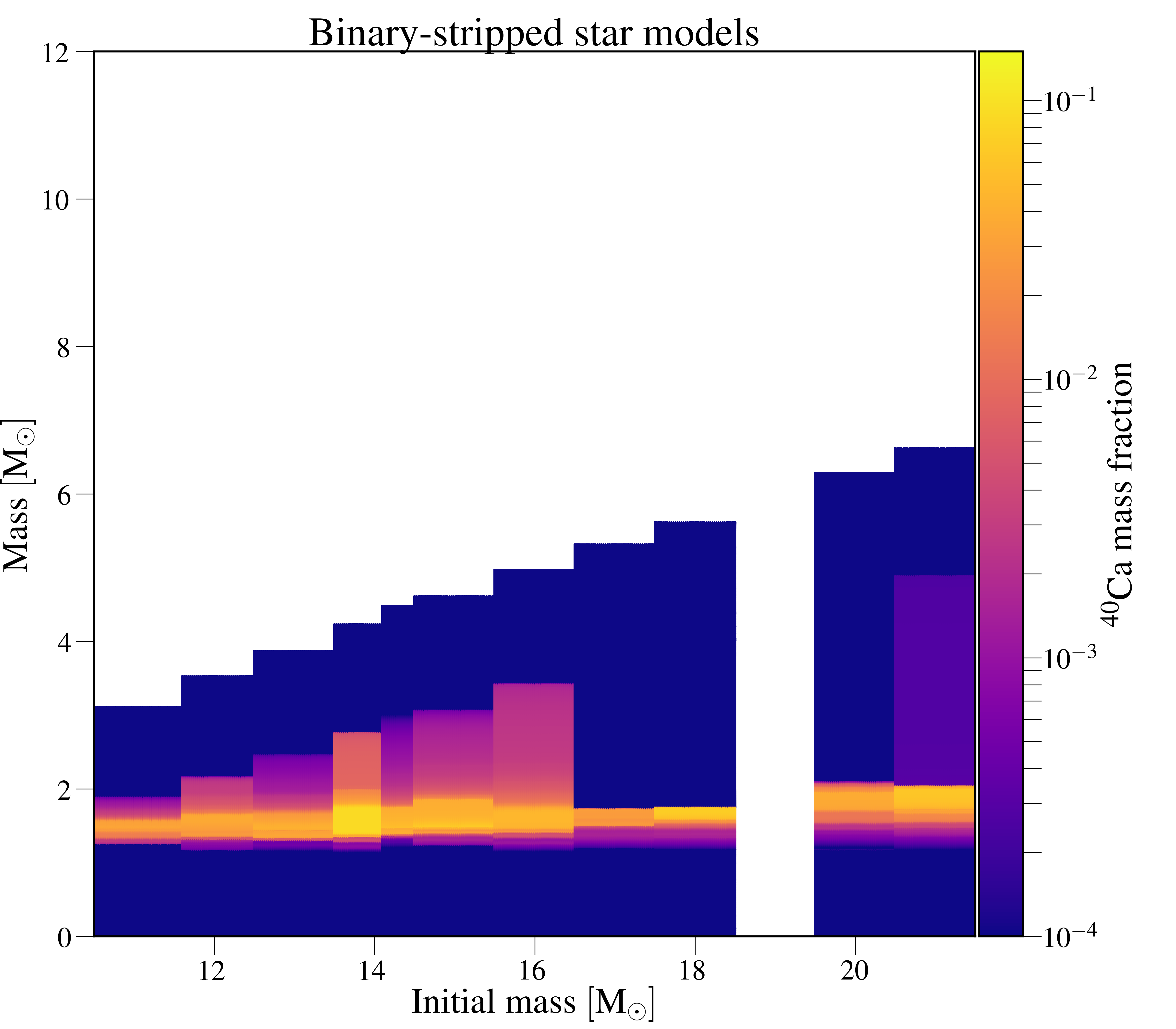}
        \caption{Distribution of $^{40}$Ca in the interior of all stellar models as a function of their mass (top: single stellar models; bottom: binary-stripped stars). All models, represented as bars, are shown as a function of their initial mass. The length of each bar gives the mass of this model at the onset of core collapse. The binary-stripped star model with an initial mass of 19\Msun is not shown because it did not reach core collapse due to numerical issues.}
        \label{fig:ca40_distr}
\end{figure}

In Fig. \ref{fig:ca40_distr} we show the distribution of $^{40}$Ca in the interior of all models at the moment of core collapse as a function of initial mass. We show the profiles of single and binary-stripped star models in the top and bottom panels, respectively. Models with shell mergers have a higher $^{40}$Ca abundance outside the silicon core. The presence of $^{40}$Ca in the oxygen-rich region can have important observational consequences, which might allow for observational tests of the physical nature of shell mergers (see the discussion in Sect. \ref{sec:discussion:nucleosynthesis}).

\begin{sidewaystable*}
        \caption{Core masses and total masses of the most abundant isotopes at the onset of core collapse for stars stripped in binaries and single stars with the same initial masses. These models are for solar metallicity ($Z = 0.014$).}
       \begin{tabular}{cccccccccccccccccc}
\toprule\midrule
$M_{\mathrm{init}}$ & $M_{\mathrm{CC}}$ & $M_{\mathrm{He\, core}}$ & $M_{\mathrm{CO\, core}}$ & $M_{\mathrm{Si\, core}}$ & $M_{\mathrm{Fe\, core}}$ & $M_{^{1}\mathrm{H}}$ & $M_{^{4}\mathrm{He}}$ & $M_{^{12}\mathrm{C}}$ & $M_{^{14}\mathrm{N}}$ & $M_{^{16}\mathrm{O}}$ & $M_{^{20}\mathrm{Ne}}$ & $M_{^{24}\mathrm{Mg}}$ & $M_{^{28}\mathrm{Si}}$ & $M_{^{32}\mathrm{S}}$ & $M_{^{36}\mathrm{Ar}}$ & $M_{^{40}\mathrm{Ca}}$ & $M_{\mathrm{Fe}}$\\ 
$[M_{\odot}]$ & $[M_{\odot}]$ & $[M_{\odot}]$ & $[M_{\odot}]$ & $[M_{\odot}]$ & $[M_{\odot}]$ & $[M_{\odot}]$ & $[M_{\odot}]$ & $[M_{\odot}]$ & $[M_{\odot}]$ & $[M_{\odot}]$ & $[M_{\odot}]$ & $[M_{\odot}]$ & $[M_{\odot}]$ & $[M_{\odot}]$ & $[M_{\odot}]$ & $[M_{\odot}]$ & $[M_{\odot}]$\\ 
\midrule
\multicolumn{8}{l}{\textbf{Binary stripped stars}}\\ 
11.0 & 3.11 & 3.11 & 1.88 & 1.43 & 1.32 & 0.00018 & 1.13 & 0.062 & 0.0060 & 0.20 & 0.05 & 0.05 & 0.11 & 0.05 & 0.008 & 0.0074 & 0.54 \\
12.1 & 3.52 & 3.52 & 2.09 & 1.49 & 1.33 & 0.00046 & 1.21 & 0.095 & 0.0045 & 0.27 & 0.11 & 0.07 & 0.10 & 0.06 & 0.013 & 0.0129 & 0.62 \\
13.0 & 3.87 & 3.87 & 2.46 & 1.42 & 1.32 & 0.00034 & 1.25 & 0.104 & 0.0064 & 0.49 & 0.18 & 0.1 & 0.20 & 0.103 & 0.019 & 0.0137 & 0.46 \\
14.0 & 4.23 & 4.23 & 2.57 & 1.38 & 1.27 & 0.00020 & 1.28 & 0.128 & 0.0064 & 0.49 & 0.33 & 0.08 & 0.26 & 0.201 & 0.044 & 0.0479 & 0.46 \\
14.6 & 4.48 & 4.48 & 2.99 & 1.47 & 1.37 & 0.00020 & 1.29 & 0.14 & 0.0044 & 0.73 & 0.33 & 0.16 & 0.19 & 0.103 & 0.021 & 0.0191 & 0.55 \\
15.0 & 4.61 & 4.61 & 2.44 & 1.47 & 1.36 & 0.00026 & 1.28 & 0.167 & 0.0033 & 0.72 & 0.47 & 0.09 & 0.2 & 0.137 & 0.029 & 0.0247 & 0.55 \\
16.0 & 4.97 & 4.97 & 3.42 & 1.47 & 1.33 & 0.00027 & 1.28 & 0.175 & 0.0052 & 1.04 & 0.27 & 0.19 & 0.35 & 0.154 & 0.026 & 0.0222 & 0.55 \\
17.0 & 5.31 & 5.31 & 3.73 & 1.71 & 1.47 & 0.00063 & 1.29 & 0.193 & 0.0063 & 1.34 & 0.36 & 0.27 & 0.23 & 0.051 & 0.009 & 0.0069 & 0.67 \\
18.0 & 5.61 & 5.61 & 2.49 & 1.65 & 1.46 & 0.00058 & 1.25 & 0.226 & 0.0051 & 1.56 & 0.35 & 0.32 & 0.24 & 0.059 & 0.014 & 0.0134 & 0.70 \\
20.0 & 6.28 & 6.28 & 4.60 & 1.80 & 1.52 & 0.00064 & 1.16 & 0.345 & 0.0032 & 1.68 & 0.82 & 0.26 & 0.17 & 0.08 & 0.016 & 0.0134 & 0.82 \\
21.0 & 6.62 & 6.62 & 4.07 & 1.82 & 1.54 & 0.00085 & 1.07 & 0.430 & 0.0007 & 1.98 & 0.68 & 0.34 & 0.27 & 0.125 & 0.027 & 0.023 & 0.72 \\
\midrule 
\multicolumn{8}{l}{\textbf{Single stars}}\\ 
11.0 & 9.32 & 3.78 & 2.31 & 1.44 & 1.32 & 3.79 & 2.95 & 0.113 & 0.0211 & 0.42 & 0.11 & 0.09 & 0.25 & 0.127 & 0.021 & 0.0173 & 0.49 \\
12.1 & 9.73 & 4.33 & 2.81 & 1.31 & 1.26 & 3.68 & 3.01 & 0.121 & 0.0185 & 0.57 & 0.12 & 0.07 & 0.37 & 0.273 & 0.06 & 0.0596 & 0.44 \\
13.0 & 9.97 & 4.79 & 3.19 & 1.53 & 1.41 & 3.52 & 3.04 & 0.123 & 0.0167 & 1.01 & 0.29 & 0.20 & 0.14 & 0.047 & 0.011 & 0.0114 & 0.58 \\
14.0 & 10.12 & 5.28 & 3.60 & 1.67 & 1.47 & 3.27 & 3.03 & 0.137 & 0.0145 & 1.23 & 0.42 & 0.23 & 0.14 & 0.037 & 0.008 & 0.0066 & 0.71 \\
14.6 & 10.15 & 5.63 & 3.90 & 1.70 & 1.51 & 3.03 & 2.98 & 0.169 & 0.0147 & 1.36 & 0.50 & 0.24 & 0.11 & 0.035 & 0.008 & 0.0071 & 0.8 \\
15.0 & 10.16 & 5.82 & 4.07 & 1.69 & 1.44 & 2.9 & 2.96 & 0.158 & 0.014 & 1.61 & 0.37 & 0.29 & 0.20 & 0.046 & 0.011 & 0.0113 & 0.62 \\
16.0 & 10.05 & 6.34 & 4.47 & 1.71 & 1.5 & 2.44 & 2.84 & 0.147 & 0.0124 & 1.68 & 0.74 & 0.31 & 0.13 & 0.061 & 0.016 & 0.0162 & 0.75 \\
17.0 & 9.88 & 6.87 & 4.97 & 1.85 & 1.58 & 1.94 & 2.72 & 0.188 & 0.0106 & 1.82 & 0.6 & 0.22 & 0.31 & 0.234 & 0.081 & 0.0607 & 0.77 \\
18.0 & 10.13 & 7.32 & 5.42 & 1.89 & 1.62 & 1.78 & 2.71 & 0.178 & 0.0102 & 2.10 & 0.78 & 0.33 & 0.23 & 0.154 & 0.042 & 0.0492 & 0.81 \\
19.0 & 10.58 & 7.42 & 5.49 & 1.97 & 1.65 & 1.84 & 3.04 & 0.209 & 0.0143 & 2.04 & 0.72 & 0.14 & 0.35 & 0.262 & 0.086 & 0.048 & 0.83 \\
20.0 & 11.66 & 7.59 & 5.67 & 1.96 & 1.63 & 2.50 & 3.26 & 0.204 & 0.0148 & 2.06 & 0.67 & 0.23 & 0.41 & 0.308 & 0.089 & 0.0617 & 0.84 \\
21.0 & 10.40 & 8.21 & 6.20 & 1.54 & 1.40 & 1.00 & 3.00 & 0.228 & 0.0146 & 2.31 & 0.72 & 0.07 & 0.65 & 0.547 & 0.142 & 0.1503 & 0.58 \\
\bottomrule
\end{tabular}

        {\textbf{Notes.} We define the core boundaries as the mass coordinate where the mass fraction of the depleted element (for example, $^{1}$H in the case of the helium core) decreases below 0.01 and the mass fraction of the most abundant element (for example, $^{4}$He) increases above 0.1. In the case of the CO core, we calculate the maximum between the values of the $^{12}$C and $^{12}$O core boundary values. For the mass of iron ($M_{\mathrm{Fe}}$), we have added the mass of $^{52}$Fe, $^{53}$Fe, $^{54}$Fe, $^{55}$Fe, $^{56}$Fe, $^{57}$Fe, and $^{58}$Fe. Of these isotopes, $^{54}$Fe, $^{56}$Fe, and $^{58}$Fe dominate and contribute with similar amounts.}
        \label{tab:general}
\end{sidewaystable*}
\subsection{Final total masses of isotopes for single and binary-stripped stars}
\label{sec:endcomp:comparison}

\begin{figure}[!h]% ---------Overview figures of CC parameters as a function of initial mass -------------
        \centering
        \includegraphics[width=0.5\textwidth]{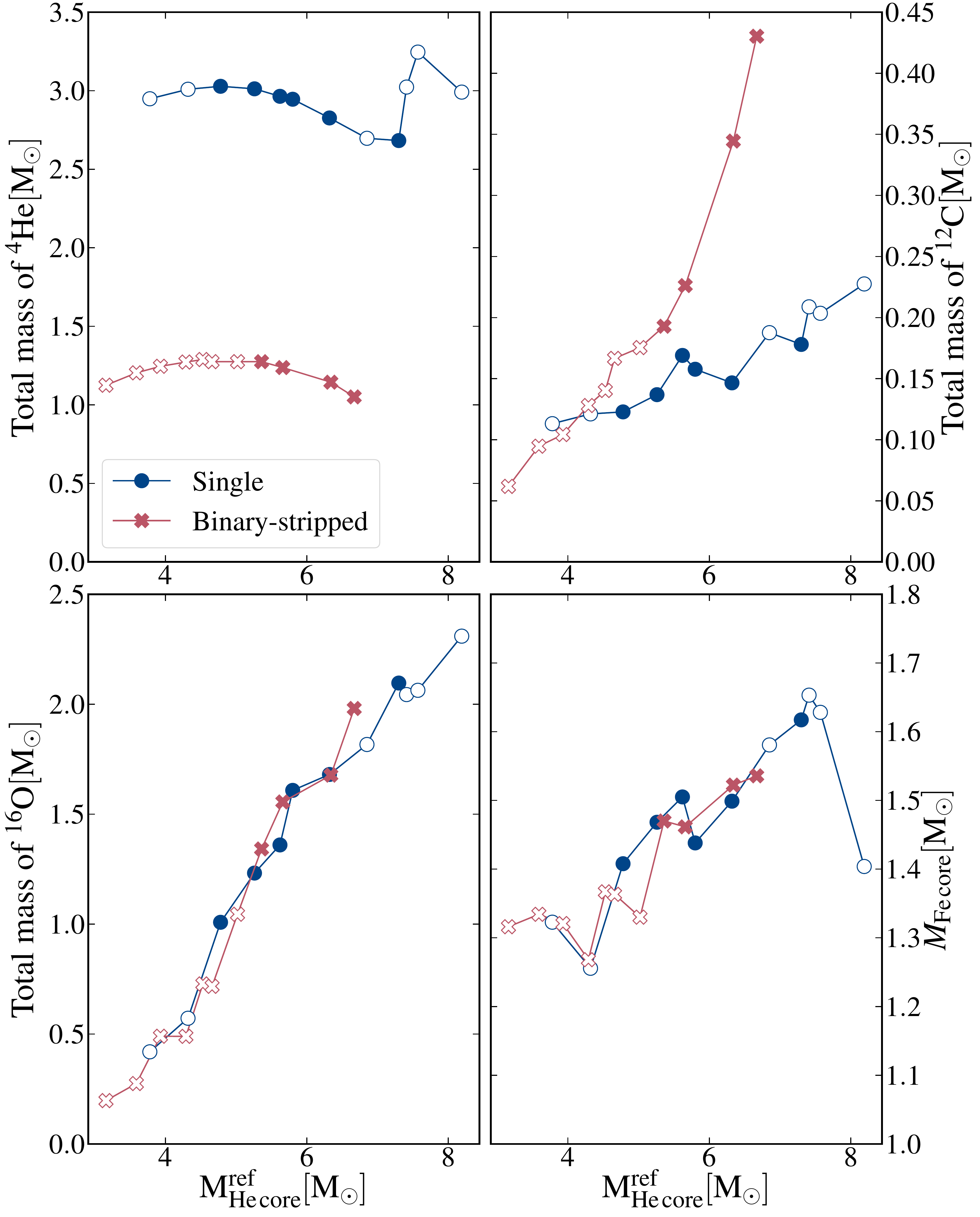}
        \caption{Composition and core properties of single (blue circles) and binary-stripped (red crosses) star models at the moment of core collapse as a function of the reference core mass. Open symbols indicate models that experienced a shell merger event. From top to bottom and left to right:  total masses of $^{4}$He, $^{12}$C, and $^{16}$O outside the iron core. The final panel shows the iron core mass.}
        \label{fig:core_overview}
\end{figure}
Differences in composition at the onset of core collapse are important indicators for differences in chemical yields. In Fig. \ref{fig:core_overview}, we show the total mass of selected isotopes at the end of the evolution integrated throughout the star as a function of the reference core mass. After the explosion, some of the isotopes will be reprocessed by the supernova shock or become part of the compact object that will form in the center, but the exterior layers (where, $\rho\lesssim 10^6 \mathrm{g}\,\mathrm{cm}^{-3}$) will leave the star relatively unaffected by the explosion.

\subsubsection{Final total mass of helium outside the iron-rich core}The amount of $^{4}$He outside the iron-rich core is systematically lower in binary stripped stars compared to single stars (on average, 3\Msun for single stars and compared to 1\Msun for binary-stripped star models). This is the combined effect of the quenched H-burning shell, which does not produce as many ashes in the binary-stripped models, and the stellar winds, which tap directly into the helium-rich material for the binary-stripped stars with core mass larger than 7\Msun.

\subsubsection{Final total mass of carbon outside the iron core}Binary-stripped star models in our grid tend to have higher masses of $^{12}$C at the end of their life than the single-star models (see top right panel of Fig. \ref{fig:core_overview}). Below reference core masses of 4.5\Msun, binary-stripped stars and single stars contain similar masses of $^{12}$C above the iron core. For higher masses, the total amount of $^{12}$C in binary-stripped star models is significantly higher than their single star counterparts, and the difference increases for more massive models, where the binary models have more than twice as much $^{12}$C. We discuss the origin of this difference, the retreat of the convective helium-burning core due to wind-mass loss in binary-stripped stars, in Sect. \ref{sec:origins:c_mass}.

\subsubsection{Final total mass of oxygen outside the iron core}The final total mass of $^{16}$O is similar for the single and binary-stripped star models (see bottom left panel of Fig. \ref{fig:core_overview}). This is because, even though the reactions involved in the creation and destruction of $^{16}$O are fractionally different in single and binary-stripped stars, they compensate each other in a similar way. We discuss this further in Sect. \ref{sec:origins:c_mass}. The total mass of $^{16}$O increases linearly with the reference core mass and does not depend on the occurrence of shell mergers.

\subsubsection{Final iron core mass}
Overall, the iron core mass is similar for all models, with masses from 1.25 to 1.65\Msun. Here, "iron" includes all species for which the mass number is higher than 46, and the core boundary is computed with respect to $^{28}$Si. The iron core mass increases slightly with an increasing reference core mass, except for the most massive single-star model. Binary-stripped and single stellar models have similar iron core masses that range from 1.3 and 1.63\Msun.

\begin{figure*}[ht]% ---------Density profiles at CC -------------
        \centering
        \includegraphics[width=\textwidth]{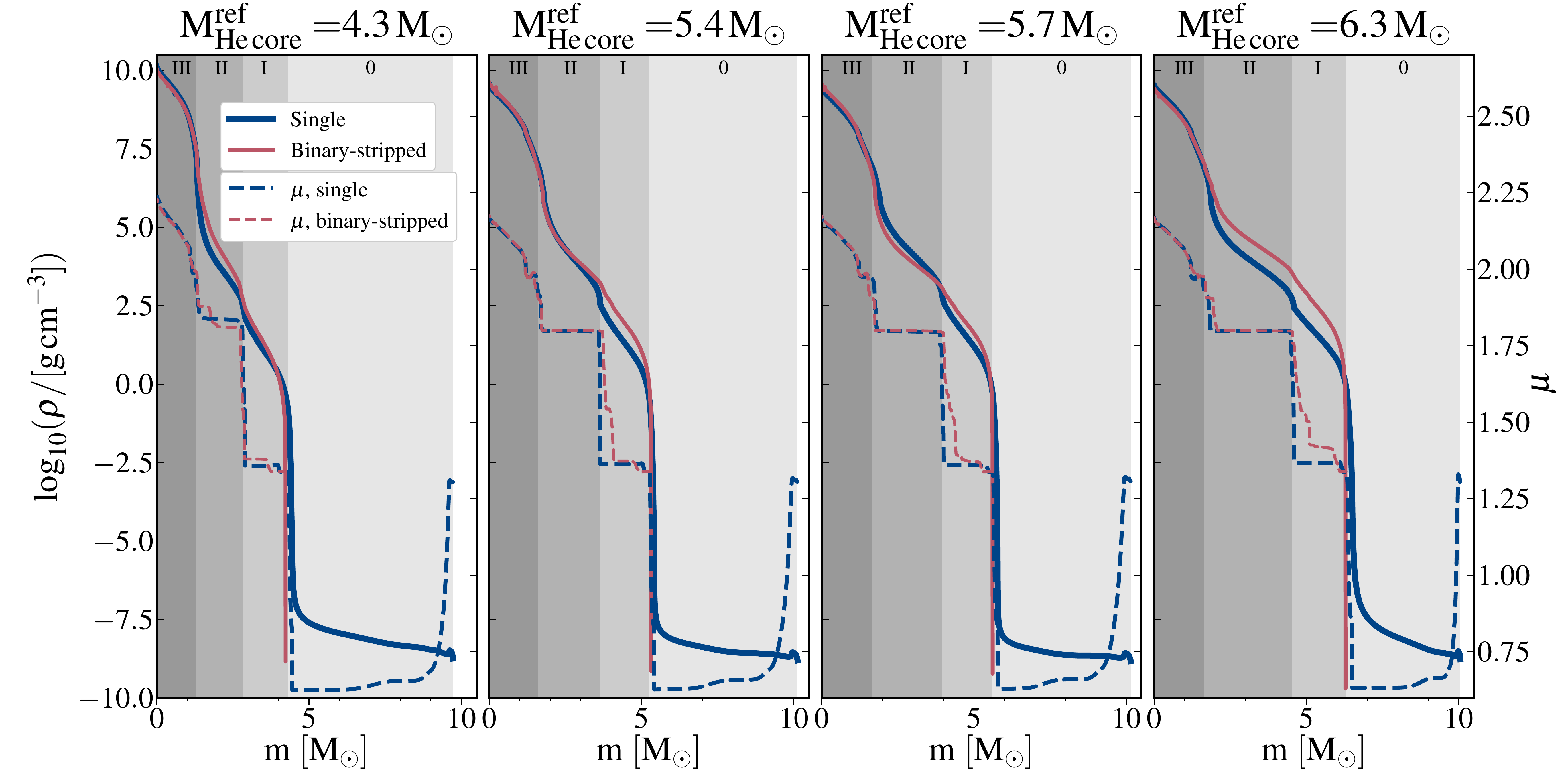}
        \caption{Density profiles at the moment of core collapse for single (blue) and binary-stripped star models (red) with similar reference core masses. The mean molecular weight profiles are indicated with dashed lines. From left to right, shaded regions give the approximate locations of the iron-rich, oxygen-rich, helium-rich, and hydrogen-rich regions, labeled with Roman numerals as in Fig. \ref{fig:end_comp_sb}. The leftmost models, with a reference core mass of 4.3\Msun, both experience shell mergers during their evolution. The density of the binary-stripped star models in the helium-rich region (I) is systematically higher than that of the single-star models with a similar reference core mass.}
        \label{fig:profiles_cc}
\end{figure*}

\subsection{Density profiles}
\label{sec:endcomp:density_profiles}
We find systematic differences in the final density and mean molecular weight profiles of single and binary-stripped star models with similar reference core masses. These differences are of particular importance because even small differences in the density profile have been shown to have a large impact on the explodability of stars \citep[e.g.,][]{vartanyan_successful_2019}. In Fig~\ref{fig:profiles_cc}, we show the density and mean molecular weight profiles at the onset of core collapse as a function of the mass coordinate for the same example models as in Fig.~\ref{fig:end_comp_sb}.

Overall, the density profiles span similar values for single and binary-stripped star models with the same reference core mass, with central densities of about $10^{10} \,\mathrm{g}\,\mathrm{cm}^{-3}$, dropping by almost 20 orders of magnitude throughout the star.
However, binary-stripped stars have steeper density profiles at the inner edge of the helium-rich region (at the boundary of regions 0 and I) compared to single-star models due to the absence of a hydrogen-rich layer in the binary-stripped stars. 
In addition, we find that binary-stripped star models have a systematically higher density in the helium-rich region (I) compared to the single-star models. At the same mass coordinates, we also find large differences in the mean molecular weight profiles: binary-stripped star models display a shallow drop of the mean molecular weight, whereas single-star models contain sharp drops in the mean molecular weight profile. The difference in the mean molecular weight profiles can be attributed to the presence of a carbon-oxygen gradient at the edge of the oxygen-rich region (see Sect. \ref{sec:origins:co_gradient}). 

In the oxygen-dominated layers (II), no such trends can be found. Instead, we find that single and binary-stripped star models have large differences in density. At the inner edge of this region, the mean molecular weight reaches a small peak that is linked to the presence of a small silicon-rich layer at this location (see also Fig. \ref{fig:end_comp_sb}).
In the inner-most iron-rich region (III), the density profiles show similar trends. At the surface of the single-star models, the mean molecular weight increases due to recombination of elements.

\subsection{Properties of the helium core}
\label{sec:endcomp:helium_core}

\begin{figure}% -------- Properties of the helium core -------------
        \centering
        \includegraphics[width=0.5\textwidth]{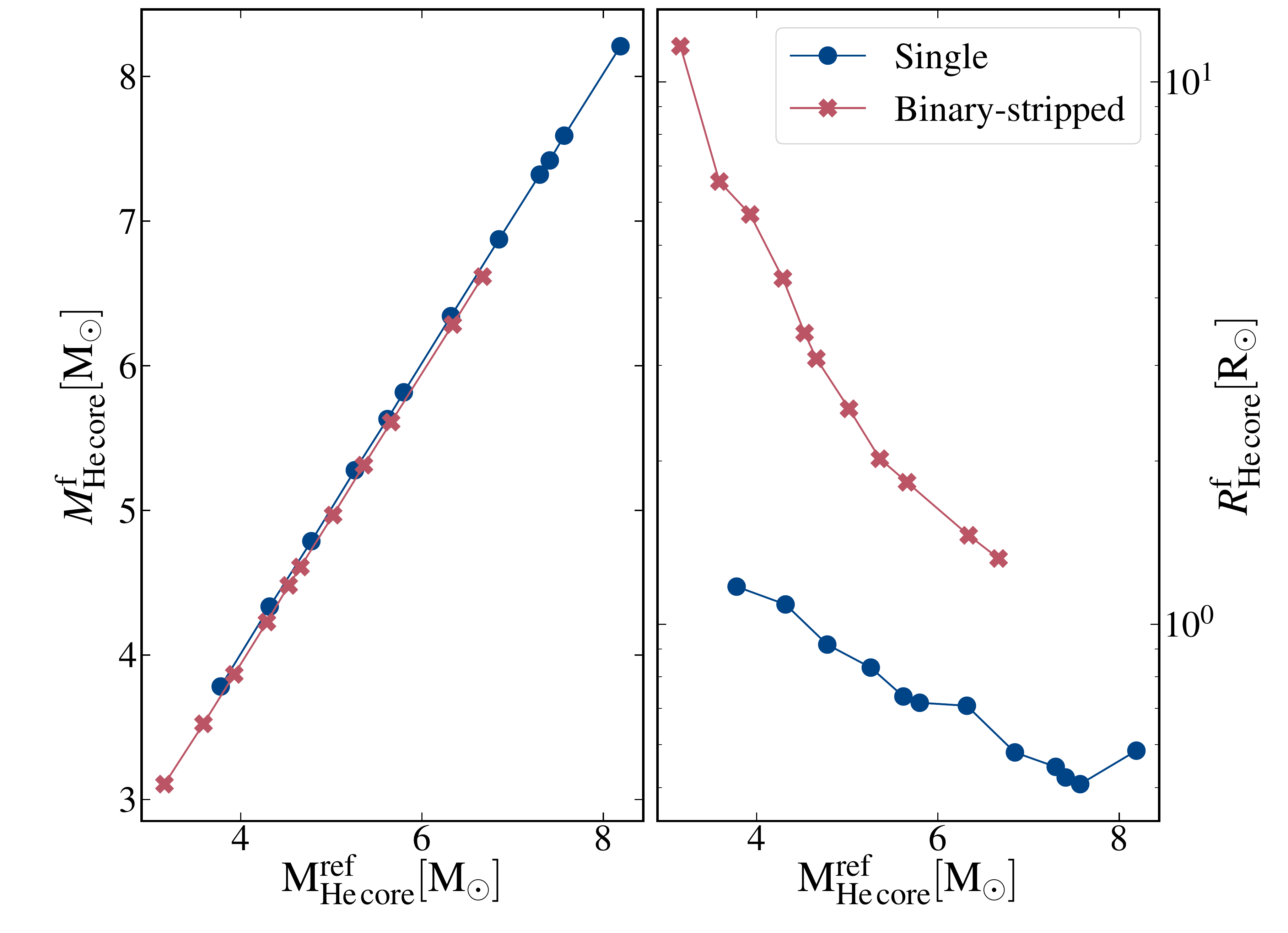}
        \caption{Final helium core mass and radius of single-star models (blue circles) and models of stars stripped in binaries (red crosses) at the onset of core collapse as a function of the reference core mass.}
        \label{fig:he_core}
\end{figure}

We present final properties of the helium core in Fig. \ref{fig:he_core}.
For the same reference core masses, the radius of the helium core is systematically larger for the binary-stripped star models (1.3 to 11.6 \Rsun) compared to the single-star models (0.6 to 1.2\Rsun). This trend in radius remains for different definitions of the helium core mass (see Appendix \ref{sec:appendix:helium_core_mass}). For both the single and binary-stripped star models, the helium core radii decrease with increasing reference core mass. For binary-stripped stars, this trend in radius at solar metallicity is well known \citep{yoon_type_2010,yoon_type_2017}.

The final mass of the helium core increases linearly with the reference core mass (i.e., the mass of the helium core at the end of core helium burning) for both the single and binary-stripped star models. A small offset can be observed between the highest-mass binary-stripped star models and the single-star models. These are due to the effect of wind mass loss for the binary-stripped stars and to the growth of the helium core after core helium depletion for the single stars (see also Sect. \ref{sec:evolution:rlof_to_hedepl} and the evolution of the helium core mass shown in appendix \ref{sec:appendix:helium_core_mass:ev}).

% --------------------------- Origins ----------------------------------------------------
\section{Origins of differences between single and binary-stripped star progenitors}
\label{sec:origins}
In the previous sections we showed that binary-stripped star models develop systematically smaller helium core masses than single stars with the same initial mass. Since interior properties of stars strongly depend on the core mass, internal differences between these models are expected. What can appear surprising, however, is that the final core properties of single and binary-stripped stars are also different when comparing models with similar helium core masses. The origin of these differences is mainly linked to the rate and timing of mass loss, which cannot easily be modeled starting from a naked helium core. We explore these origins mainly with two models, an initially 16\Msun single star and a star with an initial mass of 20\Msun that is stripped in a binary system before core helium depletion. Both develop a reference core mass of 6.3\Msun at the end of core helium burning, and as such, are well suited for a comparison. 

\subsection{The chemical gradient around the helium-depleted core}
\label{sec:origins:co_gradient}
\begin{figure}[h]% ----- Evolution of the convective core ----
        \centering
        \includegraphics[width=0.5\textwidth]{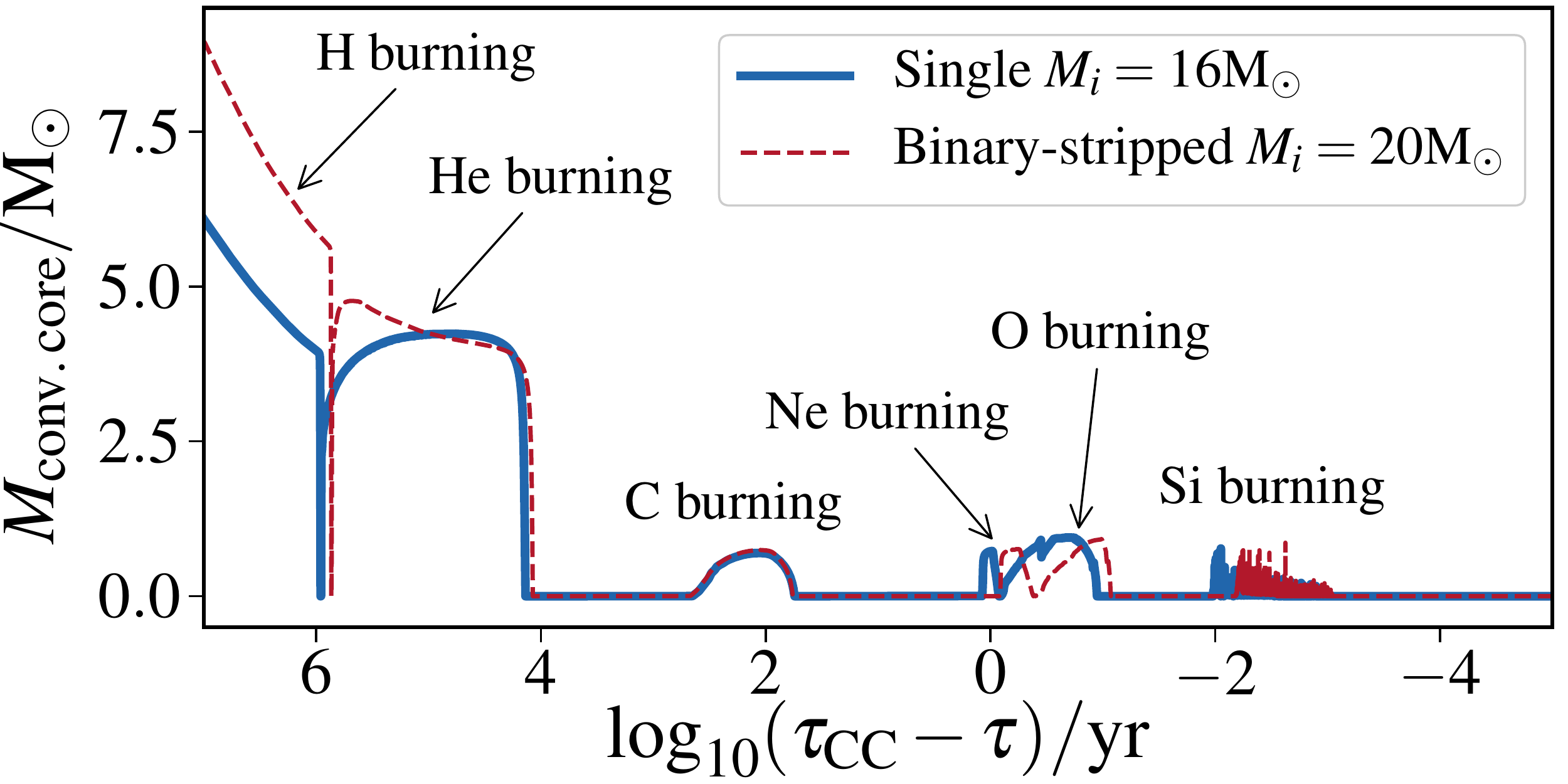}%
        \caption{Evolution of the convective core mass of a single and a binary-stripped star model with the same reference core mass of 6.3\Msun. During core helium burning, the convective core grows in the single-star model, while it decreases in the binary-stripped star model due to the effect of wind mass loss.}
        \label{fig:conv_core}
\end{figure}

Helium burning proceeds differently in the cores of binary-stripped stars and those of single stars, even in cores of similar mass \citep[cf.][]{woosley_presupernova_1995}. Helium is mainly destroyed through two channels: (1) the triple-alpha process (2) alpha captures onto carbon that create oxygen. The reaction rates of both reactions have a different dependence on the density $\rho$ and on the abundance of $^4$He nuclei, $X_{\mathrm{He}}$. For the triple-alpha process, the reaction rate $r_{\mathrm{3\alpha}}$ scales with the density cubed:
\begin{equation}
r_{\mathrm{3\alpha}} \propto X_{\mathrm{He}}^{3}\rho^{3},
\label{eq:3alpha}
\end{equation}
while the rate of alpha captures onto carbon, $r_{\mathrm{C\alpha}}$, is proportional to the density squared:
\begin{equation}
r_{\mathrm{C\alpha}} \propto X_{\rm{C}}X_{\mathrm{He}}\rho^{2},
\label{eq:Calpha}
\end{equation}
where $X_{\mathrm{C}}$ is the abundance of $^{12}$C \citep[e.g.,][]{burbidge_synthesis_1957}.
As we show in Fig.~\ref{fig:conv_core}, convective helium-burning cores grow in mass during core helium burning for single stars due to hydrogen-shell burning (see Sect. \ref{sec:evolution:rlof_to_hedepl}), while they decrease in mass for binary-stripped stars due to wind mass loss \citep{langer_standard_1989,langer_wolf-rayet_1991,woosley_evolution_1993}. For single stars, the growth of the convective helium-burning core brings an additional supply of helium that favors the destruction of carbon through alpha captures (see Eq. \ref{eq:Calpha}). As a result, the mass fraction of oxygen is higher in the cores of single stars compared to the cores of binary-stripped stars, at the expense of carbon, even for the same reference core mass. The exact fraction of carbon and oxygen is subject to the still uncertain rate of the alpha-capture reaction onto carbon \citep{weaver_nucleosynthesis_1993,brown_formation_1996,farmer_constraints_2020}.
\begin{figure*} % ----- Composition at core helium depletion ----
        %       \hspace{-4cm}
        \begin{minipage}{\textwidth}
                \includegraphics[width=\textwidth]{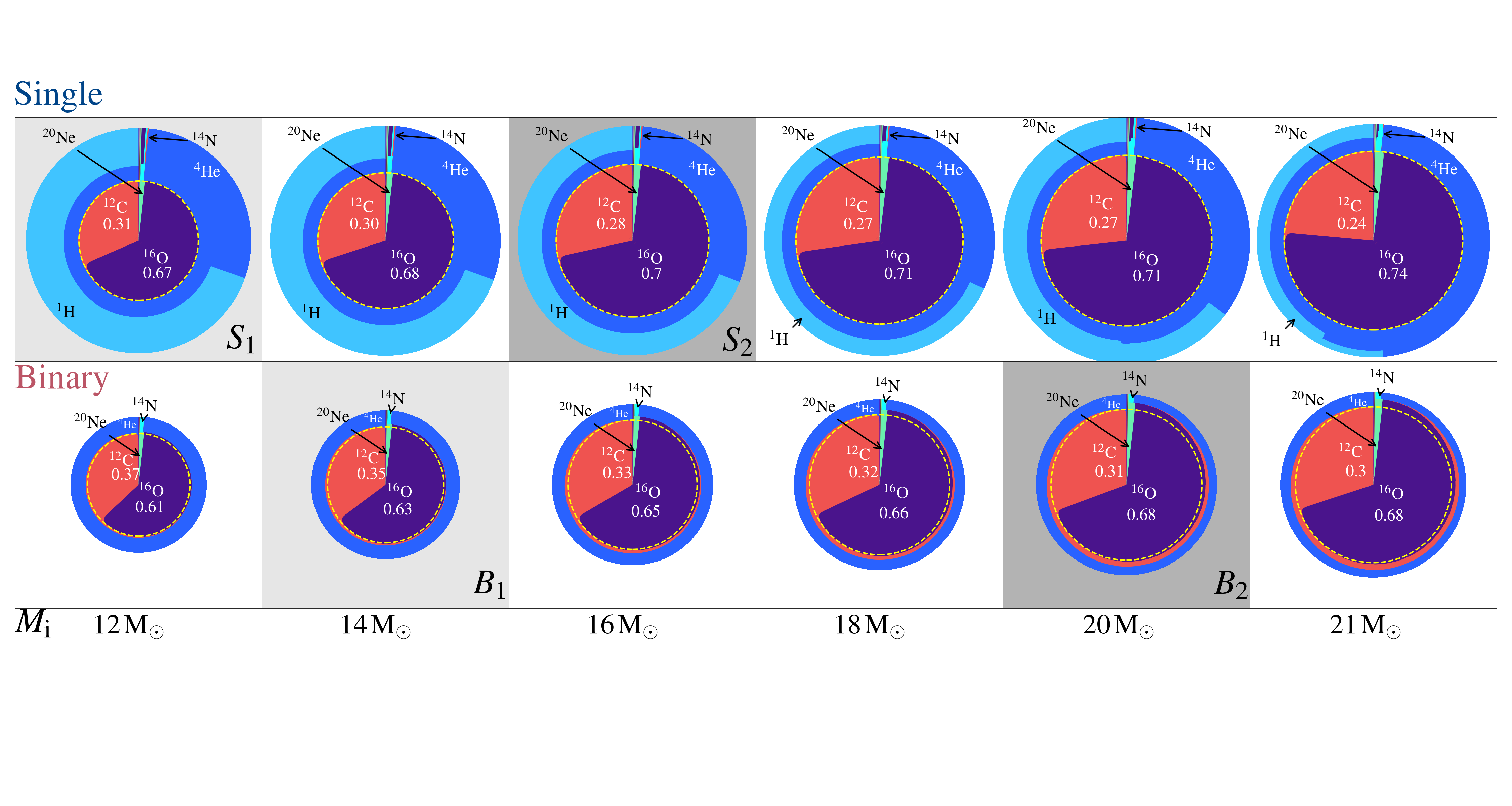}
        \end{minipage}
        \vspace{-2cm}
        \caption{Composition structure at core helium depletion for a subset of the single (top) and binary-stripped star (bottom) grids. Below each set, we indicate the initial masses of these models. Single ($S_{1}$ and $S_{2}$) and binary-stripped star ($B_{1}$ and $B_{2}$) models with similar reference core masses are highlighted with the same gray background color. Dashed yellow circles indicate the edge of the helium-depleted core. Outside the helium-depleted core, binary-stripped star models contain a layer that consists of a gradient of $^{12}$C, $^{16}$O, and $^{20}$Ne, whose extent increases with an increasing helium core mass. This layer is not present in single-star models. The neon isotope would be $^{22}$Ne in reality and is only $^{20}$Ne as an artifact of the \texttt{approx21} network.}
        \label{fig:hedepl_composition}
\end{figure*}

This is illustrated in Fig. \ref{fig:hedepl_composition}, where we show a subset of composition diagrams at the moment of core helium depletion (when the central mass fraction of helium decreases below $10^{-4}$) for single and binary-stripped star models in our grid. We highlight the differences in the central mass fractions of carbon and oxygen, which are indicated on each composition diagram. Models with the same initial mass are shown in each column. We indicate two sets of single and binary-stripped models, $S_{1,2}$ and $B_{1,2}$, respectively, that reach similar reference core masses at the end of core helium burning (marked with the same background color). For both sets, the central carbon mass fractions are smaller in the single-star models (0.31 and 0.28, for $S_{1}$ and $S_{2}$, respectively) compared to the binary-stripped star models (0.35 and 0.31, for $B_{1}$ and $B_{2}$, respectively) for the same reference core mass. The opposite occurs for the oxygen mass fractions (0.67 and 0.7 for the single-star models compared to 0.63 and 0.68 for the binary-stripped star models with a similar reference core mass). This confirms the finding of systematic differences in the chemical structure of single and binary-stripped stars from independent groups (see also Sect. \ref{sec:discussion:literature}).

The composition diagrams in Fig. \ref{fig:hedepl_composition} emphasize the trend in central carbon and oxygen mass fractions with increasing initial mass. For both the single (top row) and binary-stripped star models (bottom row), the central carbon mass fraction decreases with increasing initial mass (from 0.31 to 0.24 for the single-star models, and from 0.37 to 0.30 in the models of stars stripped in binaries, for the same initial masses of 12 to 21\Msun) at the expense of the central oxygen fraction (from 0.67 to 0.74 for the single-star models, and from 0.61 to 0.68 in the binary-stripped star models).

The mass dependence of the central mass fractions can be naturally understood as a consequence of the two main nuclear reactions involved in the burning of helium having rates with different density dependences (see Eqs. \ref{eq:3alpha} and \ref{eq:Calpha}). Stars with higher reference core masses have lower core densities and higher temperatures, and this favors alpha-captures onto carbon. Thus, carbon is destroyed more efficiently in the cores of more massive stars during core helium burning \citep[cf.][]{woosley_presupernova_1995,brown_formation_2001}.

Figure~\ref{fig:hedepl_composition} also shows the presence of a composition gradient of carbon and oxygen around the helium-depleted core. The composition gradient is left behind by the convective helium-burning core, which recedes in the binary-stripped star models due to the effect of wind mass loss. This is the red region just outside the helium-depleted core (indicated with a yellow dashed circle in the composition diagrams of Fig. \ref{fig:hedepl_composition}). The carbon-oxygen gradient around the helium-depleted core is not visible in the cores of single stars because the helium-burning core grows in mass instead of receding, leading to a steep composition gradient at the edge. The composition gradient of binary-stripped stars becomes more pronounced and more extended in mass as the total mass of the model increases. This can be seen by the growth of the red and purple rings (highlighting $^{12}$C and $^{16}$O, respectively) just outside the helium-depleted core in Fig. \ref{fig:hedepl_composition}. Higher-mass binary-stripped star models experience stronger wind mass loss, which leads to a faster and more pronounced retreat of the convective helium-burning core and to the observed effect on the carbon-oxygen gradient (see also Appendix \ref{sec:appendix:winds}). The chemical gradient at the edge of the helium core brings a different chemical environment at the location helium-burning shell in binary-stripped stars, ultimately leading to different total masses of carbon compared to single stars (see also Sect. \ref{sec:origins:c_mass}).

\begin{figure} % ----- Density profile at core helium depletion ---
        \centering
        \includegraphics[width=0.5\textwidth]{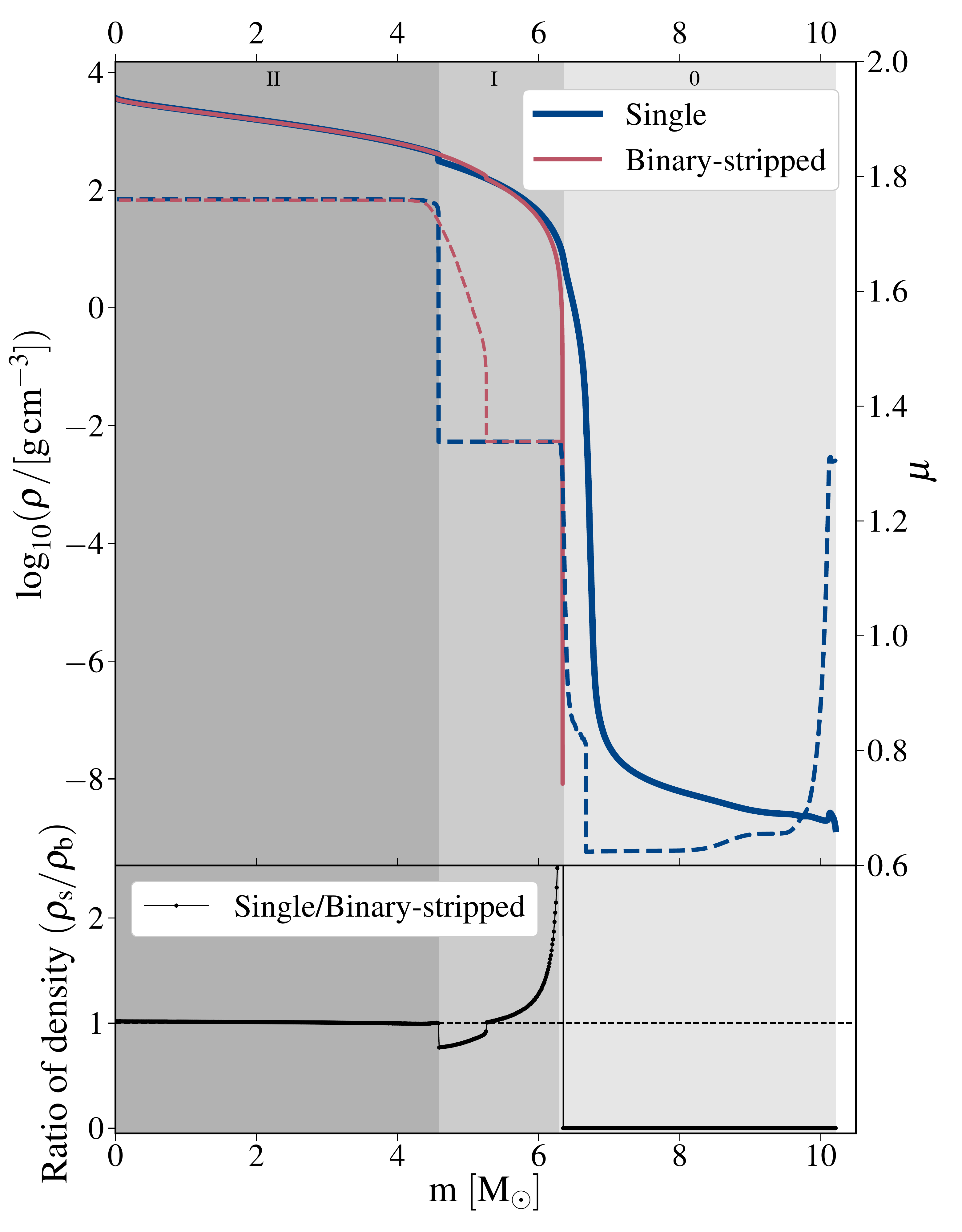}
        \caption{(top) Density and mean molecular weight profile of a single and a binary-stripped star model with the same reference core mass of 6.3\Msun at the moment of core helium depletion (D). Gray background colors indicate the approximate location of the oxygen-rich (II), helium-rich (I), and hydrogen-rich (0) regions. (bottom) Ratio of the density of the single and binary-stripped star models as a function of mass.}
        \label{fig:profile_hedepl}
\end{figure}

The differences in chemical structure also have consequences for the interior density structure. In Fig. \ref{fig:profile_hedepl} we compare the density structure and mean molecular weight at core helium depletion of two models with the same reference core mass. To highlight the differences, we show the interpolated ratio of the density in binary-stripped and single-star models in the bottom panels. Approximate locations of the oxygen-rich, helium-rich, and hydrogen-rich regions are indicated with colors as in Fig. \ref{fig:profiles_cc}. The density of the innermost, oxygen-rich core (II, up to $m = 4.5\Msun$) is the same for the single and binary-stripped star model. At the boundary between the oxygen-rich and helium-rich layers (shaded regions II and I in Fig. \ref{fig:profile_hedepl}, respectively), the density drops in both models, though this decrease is slower for the binary-stripped star model. At the same mass coordinate, we observe a sharp drop in mean molecular weight for the single-star model, while the binary-stripped star model shows a shallower profile. This difference is due to the extended carbon-oxygen gradient at the edge of the helium-depleted core in the binary-stripped star model. Because the gradient contains some $^{22}$Ne, the value of the mean molecular weight is different from the value of 1.33 expected for a layer entirely composed of $^{4}$He, $^{12}$C, $^{16}$O, or $^{20}$Ne.

At the outer edge of the CO-enhanced region ($m\approx5\Msun$ in Fig.~\ref{fig:profile_hedepl}), both models reach the same values for the density and mean molecular weight, 1.34. However, at the outer edge of the helium-rich layer, differences are visible. The single-star model has a higher density than the binary-stripped star model, as can be observed in the lower panel. This is because the single star contains a hydrogen-burning shell at the outer edge of the helium-rich region and a hydrogen-rich envelope. In contrast, the binary-stripped star model has lost its hydrogen envelope at this point, and the outer edge of the helium-rich layer (shaded region I in Fig.~\ref{fig:profile_hedepl}) corresponds to the surface of the star.

\subsection{Origin of distinct final total isotopic masses: The importance of the composition gradient}
\label{sec:origins:c_mass}
\begin{figure} % ----- Evolution of C and O mass inside and outside C/O core 6.3 Msun -----
        \centering
        \includegraphics[width=0.5\textwidth]{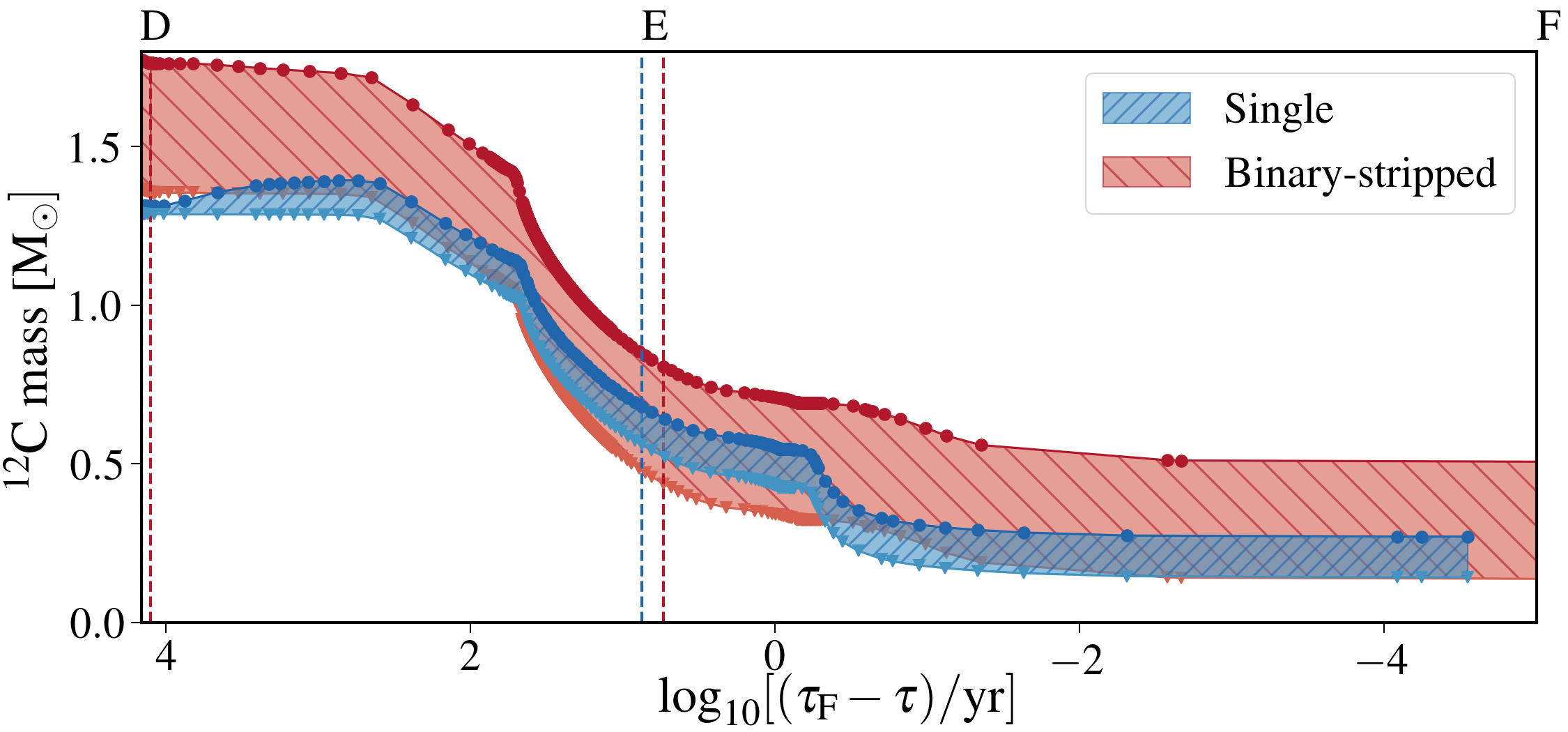}
        \includegraphics[width=0.5\textwidth]{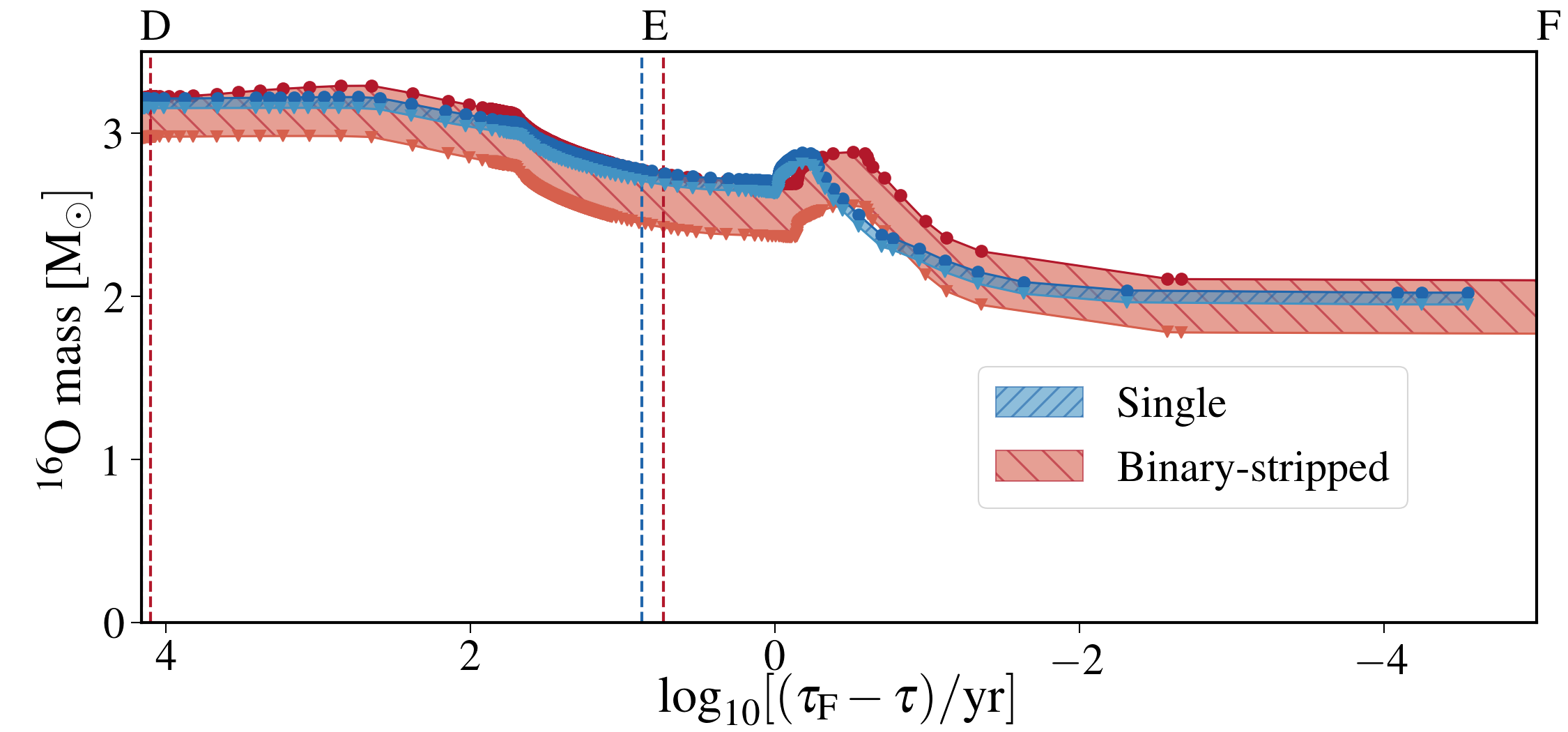}
        \includegraphics[width=0.5\textwidth]{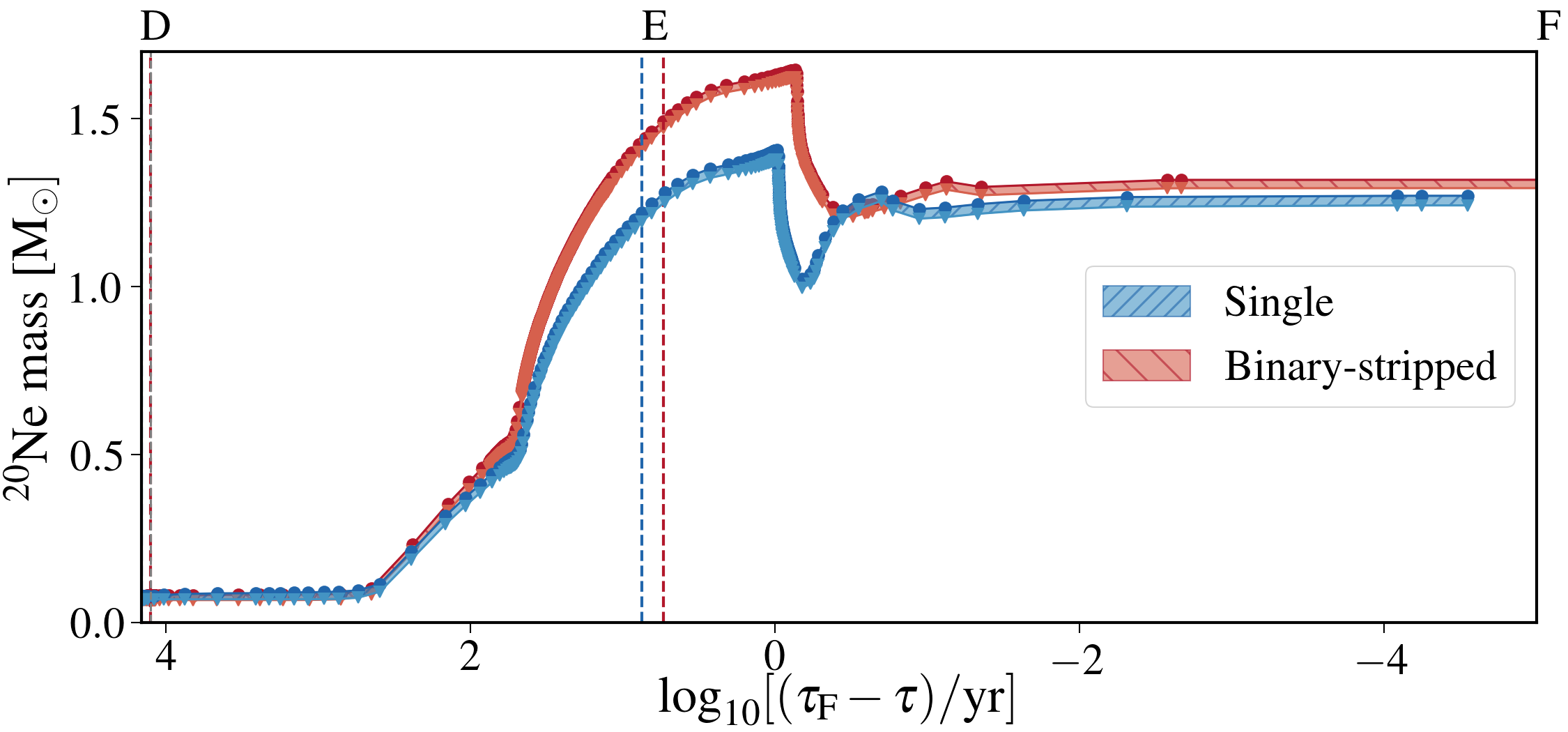}
        \caption{Evolution of the total mass of $^{12}$C (\textit{top}), $^{16}$O (\textit{middle}), $^{20}$Ne (\textit{bottom}) for a single (blue) and a binary-stripped star model (red) with the same reference core mass of 6.3\Msun. The evolution is shown from the moment of core helium depletion as a function of the time until core oxygen depletion. The curve marked by inverted triangles (i.e., the lower boundary of the colored bands in each plot) indicates the total mass of each isotope inside the helium-depleted core. The upper boundary marked by circles shows the total mass of this isotope in the star. The red and blue shaded regions thus highlight the mass of this isotope in the shell above the core.}
        \label{fig:c12_o16_single_vs_stripped}
\end{figure}

In Sect. \ref{sec:endcomp:comparison} we showed that binary-stripped star models end their lives with higher total masses of $^{12}$C compared to single-star models (see Fig. \ref{fig:core_overview}), while the total masses of $^{16}$O remain similar. Here we discuss how these differences in composition arise.
In Fig. \ref{fig:c12_o16_single_vs_stripped}, we present the evolution of the total mass of $^{12}$C, $^{16}$O, and $^{20}$Ne as a function of time until core oxygen depletion for the example binary-stripped and single-star models with a reference core mass of 6.3\Msun (with initial masses of 20\Msun and 16\Msun, respectively). During the helium-shell burning phase (first half of D-E), the total mass of carbon decreases slightly in the binary-stripped star, while it increases in the single-star model. In contrast, the total mass of oxygen increases in the binary-stripped star model, while it remains unchanged in the single-star model. The shaded bands highlight the mass of each isotope outside the helium-depleted core (label D in Fig. \ref{fig:c12_o16_single_vs_stripped}). Meanwhile, the total masses of carbon and oxygen inside the helium-depleted core remain constant. 
The binary-stripped star model retains a more massive layer of $^{12}$C (0.2\Msun) and $^{16}$O (0.25 \Msun) outside the helium depleted core than the single-star model (about 0.07\Msun and 0.02\Msun for $^{12}$C and $^{16}$O, respectively) throughout the evolution.
This is because of differences in the relative rates of helium-burning channels: alpha captures onto carbon dominate in the binary-stripped model. In contrast, the single-star model burns helium primarily through the triple-alpha process. This is because of the differences in composition at the location of the helium-burning shell in the single stars compared to the binary-stripped stars.

At the end of core oxygen burning (label F in Fig. \ref{fig:c12_o16_single_vs_stripped}), the interior $^{12}$C mass of both models (triangles) is similar. However, the total $^{12}$C mass, shown with circles, is higher for the binary-stripped star model due to the mass of $^{12}$C outside the helium-depleted core. The interior $^{16}$O mass is lower for the binary-stripped star model than for the single-star model since helium burning through the triple-alpha process is favored over alpha-captures onto $^{12}$C (see Sect. \ref{sec:origins:co_gradient}). Despite the differences, the total mass of $^{16}$O is similar in the single and binary-stripped star models. Although $^{16}$O is created during helium-shell burning and more actively during neon burning in the binary-stripped star model compared to the single-star model, oxygen is also destroyed more rapidly during carbon core and shell burning. These effects compensate each other and create a similar final total mass of $^{16}$O for the single and binary-stripped star model.

\subsection{Differences in shell burning}
\label{sec:origins:shell_burning}
In Sect. \ref{sec:endcomp:density_profiles} we found a systematic difference between the density profiles of single and binary-stripped stars with same reference core mass. Specifically, the He-rich layer of binary-stripped stars is systematically denser than that of single star. This impacts the subsequent shell burning phases and thus the pre-collapse structure.

\begin{figure*}[ht!] % ----- Density profile evolution ---
        \centering
        \includegraphics[width=0.5\textwidth]{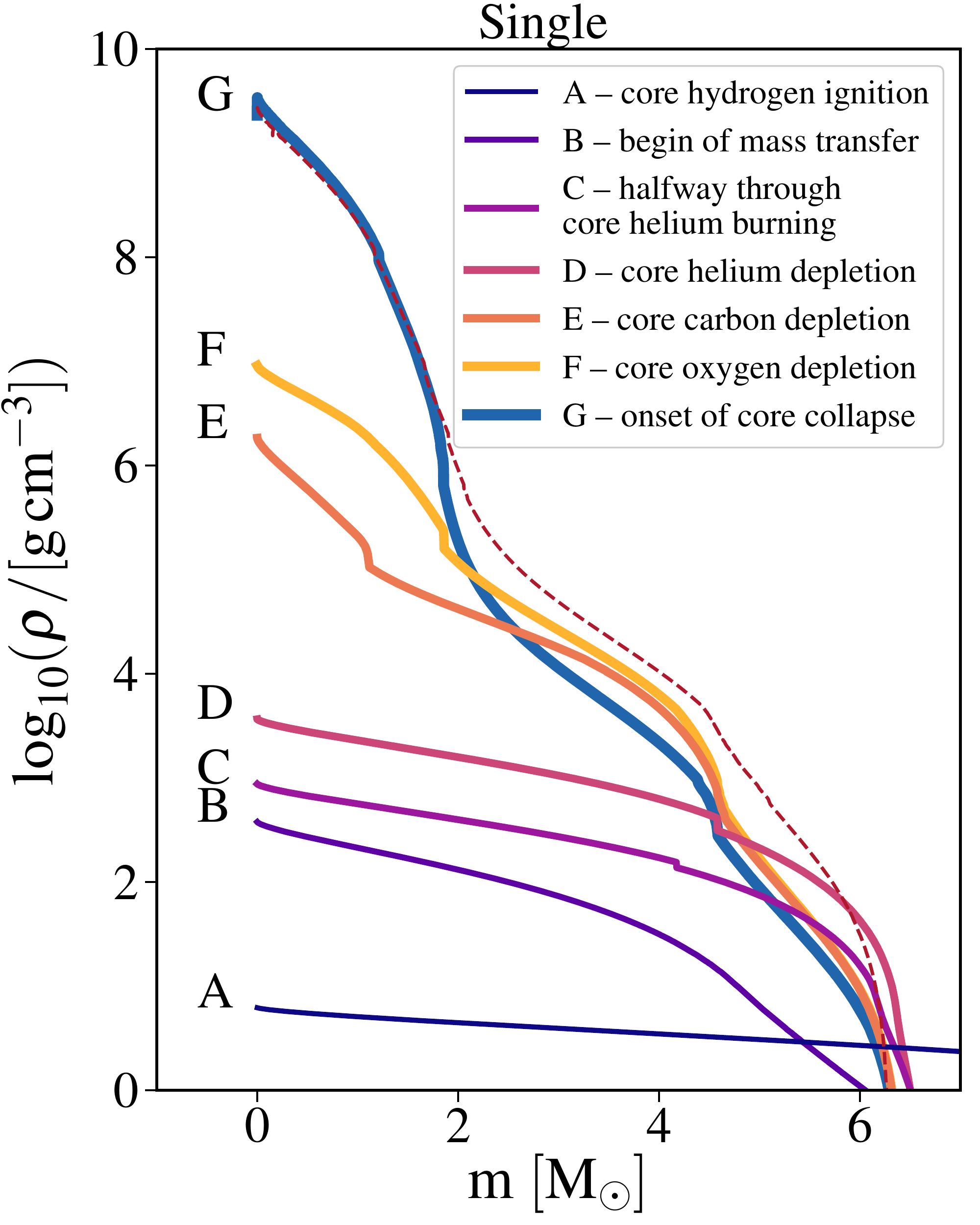}%
        \includegraphics[width=0.5\textwidth]{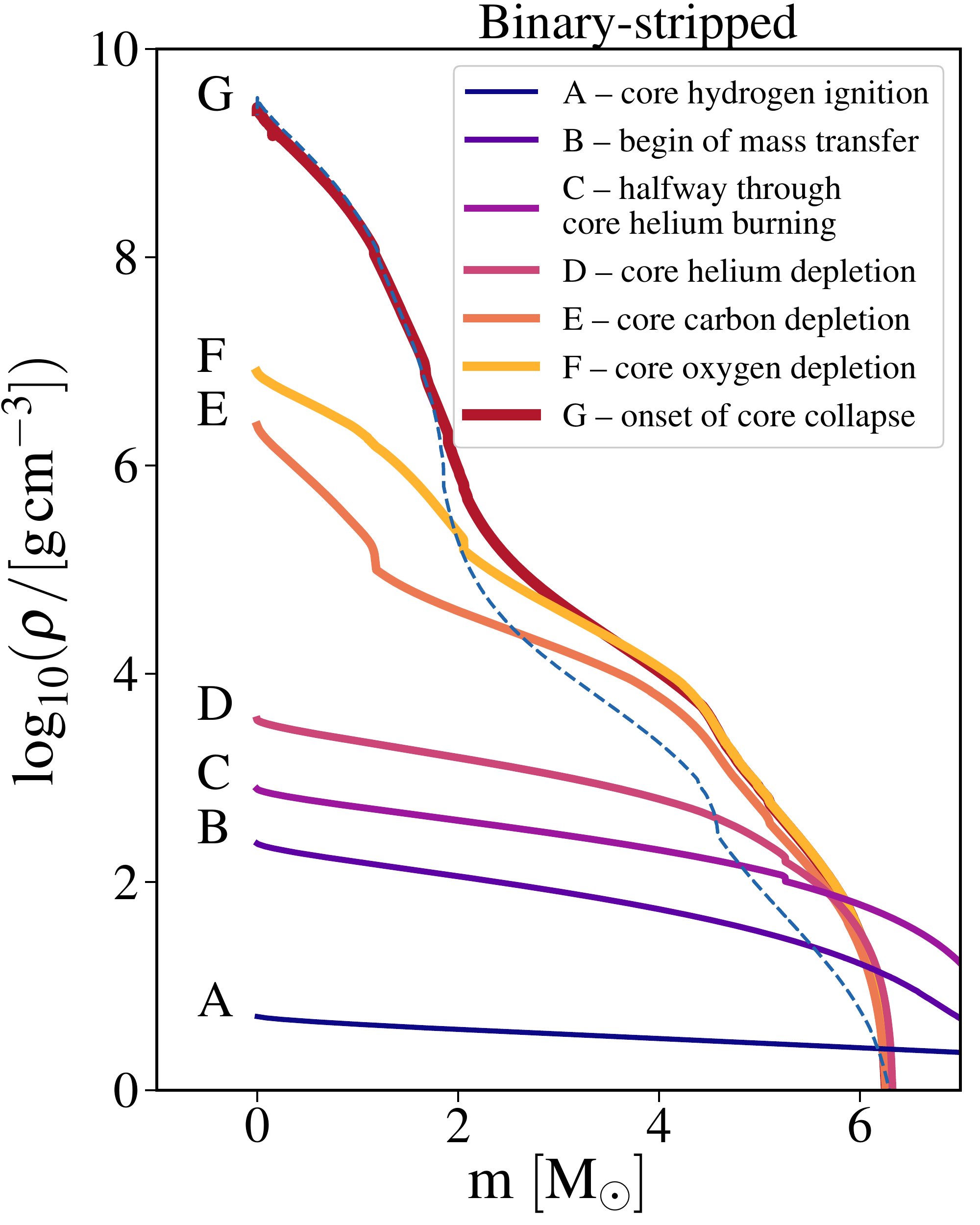}
        \caption{Density profile evolution of a single (left) and stripped (right) star model with the same reference core mass of 6.3\Msun from zero age main sequence to core collapse. Dashed lines show the corresponding profiles at the onset of core collapse.}
        \label{fig:profile_ev}
\end{figure*}

In Fig. \ref{fig:profile_ev} we show the evolution of density profiles at notable moments of the evolution (marked with the same labels as in Fig. \ref{fig:evolution}) for our example pair composed of a single and a binary-stripped star model that both reach a similar reference core mass of 6.3\Msun at the end of core helium burning. We show the profiles from the onset of core hydrogen burning (A) to core collapse (G). We focus on the inner 7\Msun of the stellar structures. The evolution of the density profile is similar for both models until core helium depletion (label D). As the star evolves, the density of the inner core (up to 5\Msun) increases monotonically. At core helium depletion (point D), the density profile of the binary-stripped star shows a sharp drop at 6.3\Msun, corresponding to the surface of the star.

The change in density between the single and binary-stripped star in the helium-rich layer ($5\Msun \lesssim m \lesssim 6$\Msun) happens between core helium depletion and the end of core carbon burning (D-E). While the density of this layer increases for the binary-stripped star model, it decreases for the single-star model.
This phase of the evolution is marked by the ignition of the helium shell and the start of core carbon burning. 

\begin{figure*}[ht!] % ----- Burning profile evolution ---
        \centering
        \includegraphics[width=\textwidth]{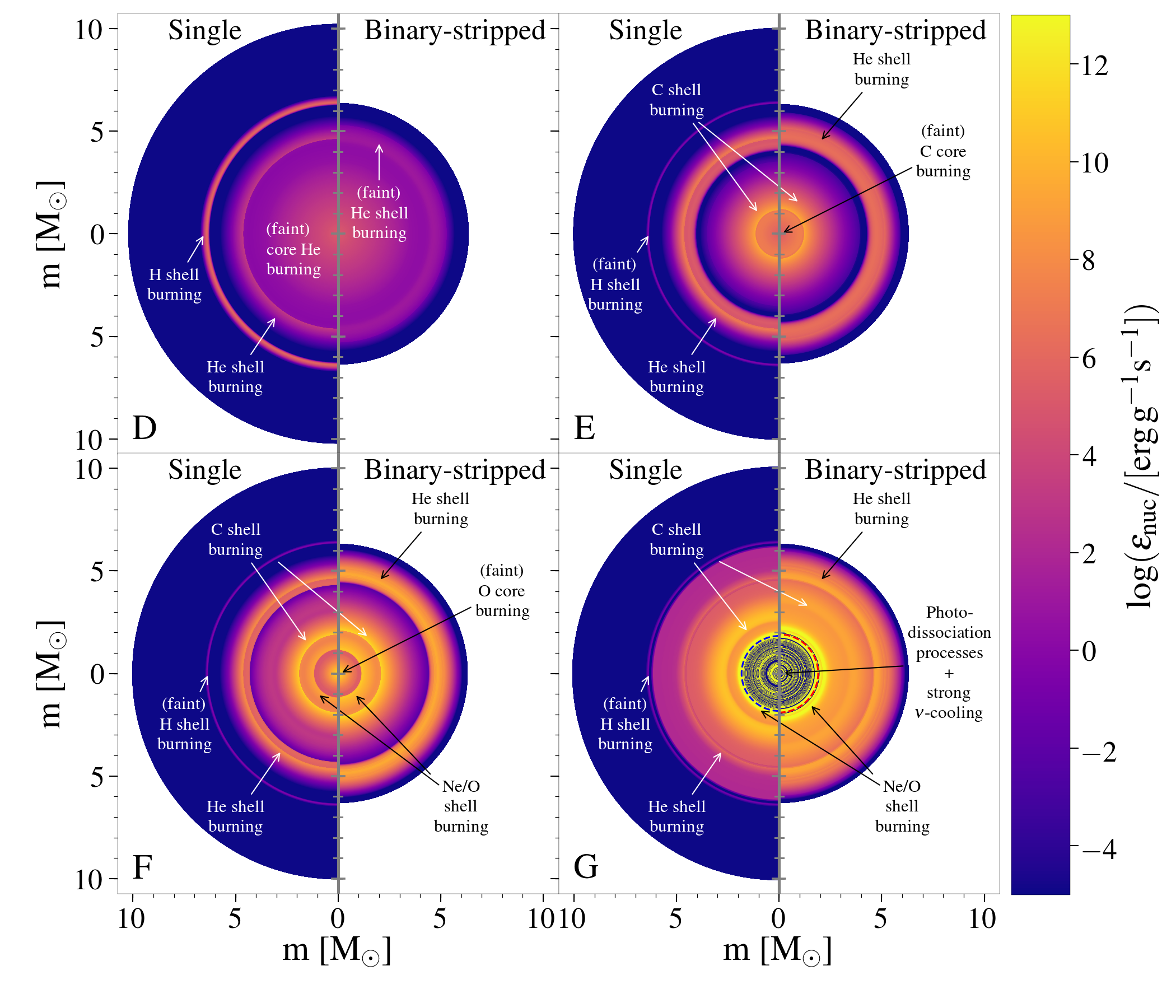}
        \vspace{-1cm}
        \caption{Comparison of the nuclear burning structures of a single (left half-circle) and binary-stripped (right half-circle) star model with the same reference core mass of 6.3\Msun. Each stellar model is represented as a multitude of two-dimensional half-circles, where the radius of each circle is directly proportional to the mass coordinate. The nuclear energy generation rate is indicated with colors. All models are compared for the same evolutionary steps (D: core helium depletion; E: core carbon depletion; F: core oxygen depletion; G: onset of core collapse). In panel G (onset of core collapse), we give the locations of the mass boundary above the compact object that forms in the center (defined as the mass coordinate where the specific entropy drops below a value of $4 \,\mathrm{erg}\,\mathrm{g}^{-1}\,\mathrm{K}^{-1}$; \citealp{ertl_two-parameter_2016}) with a dashed blue line (single) and red line (binary-stripped). The binary-stripped star model forms a more massive compact object in its core than the single-star model (1.90\Msun compared to 1.83\Msun).}
        \label{fig:burn_ev}
\end{figure*}

We show the nuclear burning profiles for the specific points of the evolution in Fig. \ref{fig:burn_ev}. We represent the single and binary-stripped stellar model pair as half circles (left: single-star models, right: binary-stripped star models), where the size of the circle is directly proportional to the mass coordinate. The nuclear energy generation rate throughout the stellar structures is shown with colors.

At the moment of core helium depletion (panel D in Fig. \ref{fig:burn_ev}), we find various differences in the burning structure of the example single and binary-stripped star model. The single-star model contains a hydrogen-burning shell, which is absent in the binary-stripped star. In addition, it has a helium-burning shell ($m=4.5\Msun$ in Fig. \ref{fig:burn_ev}). In contrast, at the same mass coordinate, the binary-stripped star model has a specific energy generation rate that is two orders of magnitude smaller.

Between core helium depletion and core carbon depletion (points D and E in Fig. \ref{fig:burn_ev}), a large change occurs in the burning structures. In the single star, the specific energy generation rate in the hydrogen-burning shell decreases significantly, while it increases in the helium-burning shell. This decline of the output from the hydrogen-burning shell is linked to the change in density in the helium-rich region.

We also note large differences between the helium-shell burning of single and binary-stripped star model from this point on until the end of the evolution (points E to G in Fig. \ref{fig:burn_ev}). In the binary-stripped star model, the helium-shell burning region is more extended ($4.2\Msun < m < 5.5$ \Msun). The single-star model has a higher maximum energy generation rate (up to $\epsilon_{\rm{nuc}} = 10^{8}\rm{erg}\,\rm{g}^{-1}\rm{s}^{-1}$, compared to $10^{7}\rm{erg}\,\rm{g}^{-1}\rm{s}^{-1}$ for the binary-stripped star model). However, for the binary-stripped star model, a high specific energy generation rate is found in the entire helium-burning region, while for the single-star model, it drops quickly outside of the region with the maximum value.
These differences can be understood by the differences in composition at the location of the helium-burning shell at the end of core helium burning. Because of the extended chemical gradient at the edge of the helium-depleted core, the binary-stripped star model contains a significantly higher mass fraction of carbon next to the helium-burning shell, while the single-star model contains very little carbon near the burning shell at the beginning of helium-shell burning. This means, just as in the case of central helium burning (see Sect. \ref{sec:origins:co_gradient}), that there are differences in the relative contributions of nuclear reactions involved in helium-shell burning. For helium-shell burning, the situation is reversed from core helium burning: more alpha captures onto carbon occur in the binary-stripped star model than in the single-star model (see also the change in total carbon mass and total oxygen mass between point D and E in Fig \ref{fig:c12_o16_single_vs_stripped}).

After the end of core carbon burning (panel F in Fig. \ref{fig:burn_ev}), additional changes arise in the density and nuclear burning structure. The stars ignite carbon in a shell around the core. However, the location of this shell is significantly different between the single and binary-stripped star models. In the single-star model, the carbon shell ignites at $m\approx 1.7\Msun$, while in the binary-stripped star, it ignites further out, at $m\approx 2.1\Msun$. This can be explained by the differences in carbon mass fractions at the end of core helium burning \citep{chieffi_presupernova_2020}. The binary-stripped star has a higher core carbon abundance and the carbon shell ignites further out from the center than in the single-star model with the same reference core mass. However, the extent (in mass) of convective core oxygen burning is similar (up to $m\approx 1.2\Msun$). 
We also give the location of the mass boundary expected for the compact object created during the explosion. We set this boundary at the mass coordinate where the specific entropy drops below a value of $4 \,\mathrm{erg}\,\mathrm{g}^{-1}\,\mathrm{K}^{-1}$ \citep{ertl_two-parameter_2016}. For the same reference helium core mass, the binary-stripped star model creates a compact object that is more massive than for the single-star model (1.90\Msun compared to 1.83\Msun).

\subsection{Compactness evolution}
\label{sec:origins:compactness}

\begin{figure*}[h]% ----- Compactness evolution ----
        \centering
        \includegraphics[width=0.47\textwidth]{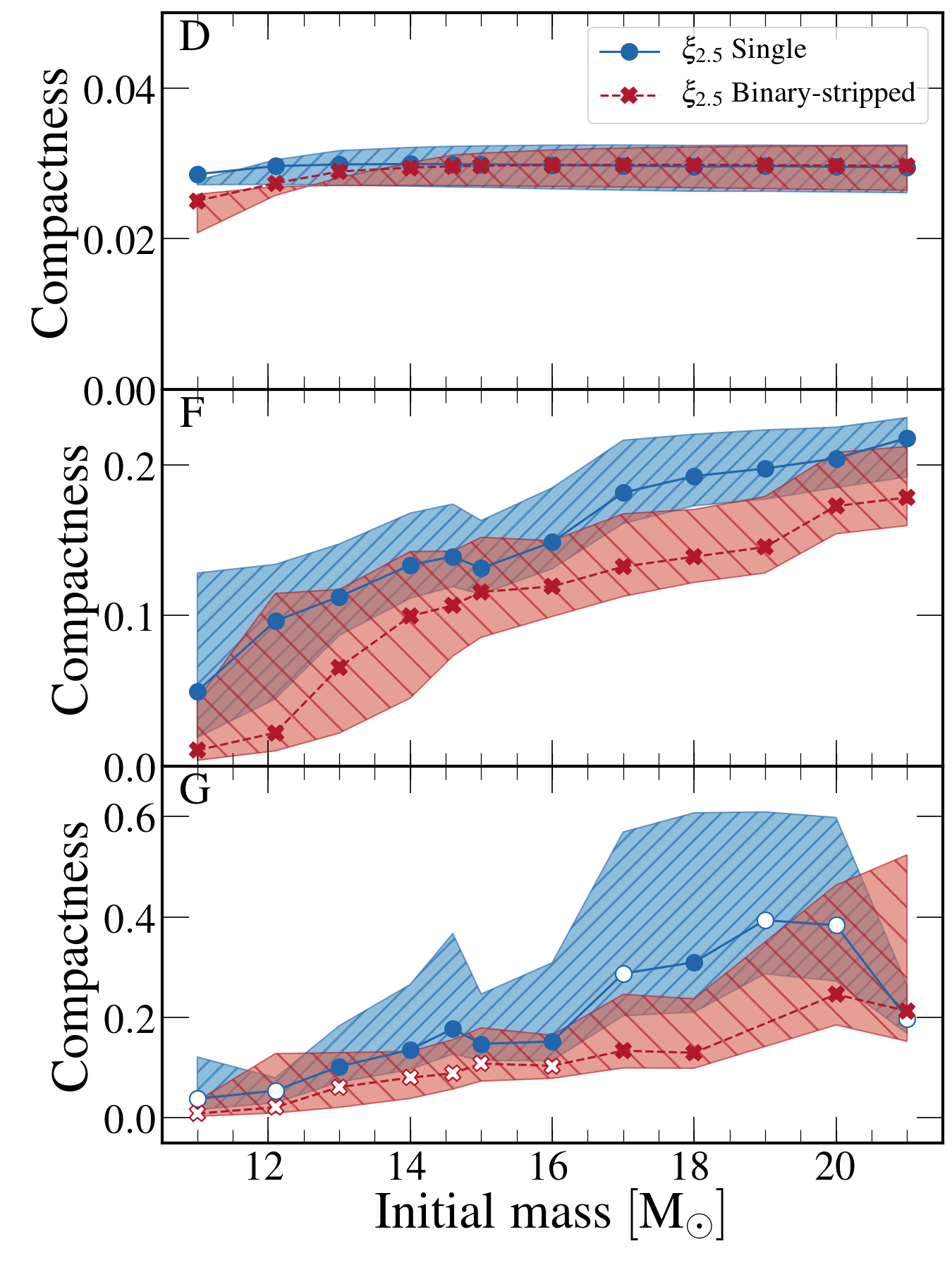}%
        \includegraphics[width=0.47\textwidth]{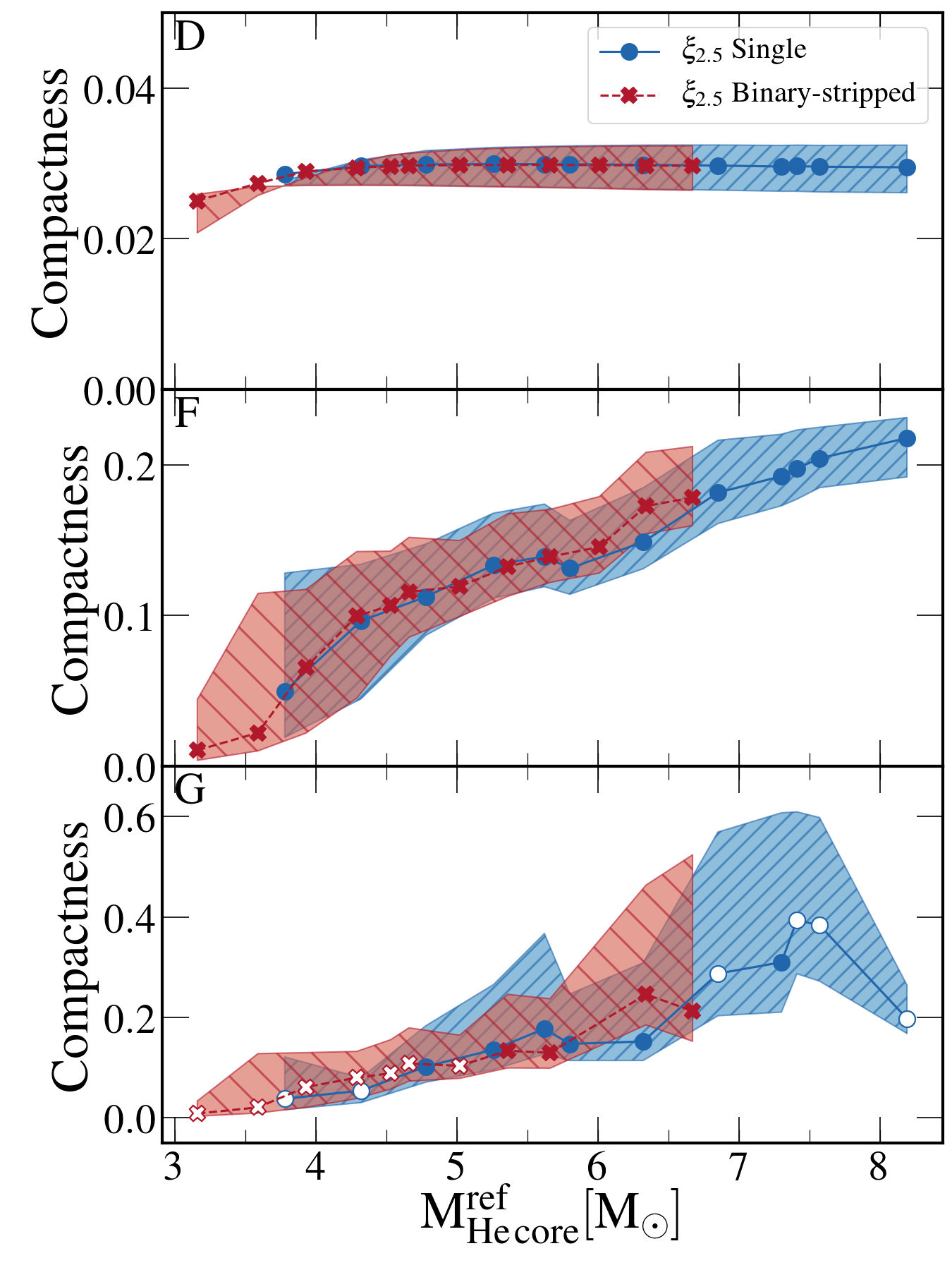}
        \vspace{-0.4cm}
        \caption{Evolution of the compactness parameter for the full grid of single (blue) and binary binary-stripped (red) models as a function of initial mass (\textit{left}) and reference core mass (\textit{right}). From top to bottom, we show the compactness at the moment of core helium depletion (D, \textit{top}), core oxygen depletion (F, \textit{middle}), and the onset of core collapse (G, \textit{bottom}). For the models at the onset of core collapse, open symbols indicate models that have experienced shell mergers.
                For every phase of the evolution, we indicate the range of changes in the compactness parameter with blue and red shaded regions. The upper bound of this region corresponds to the compactness evaluated at $m=2\Msun$ and the lower bound to the compactness evaluated at $m=3\Msun$. Trends in the compactness parameter as a function of mass are consistent between different definitions of the compactness for the range considered. The reference core mass is defined as the mass of the helium core at the moment of core helium depletion.}
        \label{fig:compactness_evolution}
\end{figure*}

The compactness parameter is often used to characterize the core structure of stars. It has been shown to correlate with the explodability with one-dimensional parametric supernova explosion models \citep{oconnor_black_2011,ugliano_progenitor-explosion_2012}, although studies of three-dimensional explosion of stars have questioned this \citep[e.g.,][]{ott_progenitor_2018,kuroda_full_2018}. Nevertheless, it can be used to predict the type of compact object resulting from a supernova explosion (typically, neutron stars for low values of the compactness and black holes for higher values). Other measures of explodability have been proposed from one-dimensional parametric simulations \citep[for a recent example, see][]{ertl_explosion_2020} or multidimensional simulations.

We show the evolution of the compactness parameter for the full grid of models in Fig. \ref{fig:compactness_evolution} at three moments of the evolution: D core helium depletion, F core oxygen depletion, and G the onset of core collapse. In addition, we give general properties of the models at the moment of core collapse in Table~\ref{table:cc_general}.
We also evaluate the compactness at mass coordinates 2 and 3\Msun, shown as a blue (red) band in Fig. \ref{fig:compactness_evolution} for the single star (binary-stripped) models. We find that trends in compactness are robust against variations in the mass coordinate at which the compactness parameter is evaluated for all moments of the evolution presented here \citep{ugliano_progenitor-explosion_2012}.
In the left panels, we show the compactness as a function of initial mass, and in the right panels, as a function of the reference core mass.

At core helium depletion, the compactness parameter is similar for models of all initial masses, with a value of about 0.03. Only the three lowest-mass models show differences between single and binary-stripped models. At that epoch, binary-stripped star models with initial masses smaller than 13.5\Msun have a slightly smaller compactness than their single star counterparts. This is because the mass coordinate at which the compactness is evaluated lies outside of the oxygen core in these models, which means the compactness parameter captures the relative expansion of the helium-rich envelope compared to the single stars (see also Table~\ref{tab:general}).

By the time of core oxygen depletion (panels marked F in Fig. \ref{fig:compactness_evolution}), the value of the compactness parameter changes significantly between all models \citep{renzo_systematic_2017}. Higher-mass models have higher compactness. Binary-stripped star models have systematically lower values of the compactness (0.01 to 0.18) than single-star models of the same initial mass (0.05 to 0.21). However, when comparing models with the same reference core mass (defined as the helium core mass at core helium depletion) the compactness is similar between single and binary-stripped star models, except for the two most massive binary-stripped star models. The difference for the highest-mass models could be due to the transition to radiative core carbon burning around this mass range for binary-stripped stars that changes the compactness of the core \citep[][]{brown_formation_1996,brown_formation_1999,brown_formation_2001,sukhbold_compactness_2014,woosley_evolution_2019,schneider_pre-supernova_2021}.

At the onset of core collapse (panel G in Fig. \ref{fig:compactness_evolution}), the differences between the compactness of single and binary-stripped star models of the same initial mass increases even further. The compactness of binary-stripped star models remains systematically smaller than that of single-star models. For the same reference core mass, we find similar values between single and binary-stripped star models. Changes in the compactness are qualitatively similar to those from similar recent studies \citep[e.g.,][]{woosley_evolution_2019,chieffi_presupernova_2020, patton_towards_2020,schneider_pre-supernova_2021}. 
We do not find clear differences in compactness for models that experience shell mergers before core collapse (marked with open symbols in the lower panels of Fig. \ref{fig:compactness_evolution}). This is because the Si/O shell mergers observed in this study typically occur at mass coordinates of $m\simeq1.3\Msun$, below the mass range we chose to evaluate the compactness and the merged convective shells do not extend beyond the mass cuts chosen for the compactness.

When comparing the electron fraction $Y_{e}$  in Table~\ref{table:cc_general}, we also find different values for single and binary-stripped star models. Different values of the electron fraction imply that the late nuclear burning conditions in the cores of single and binary-stripped stars are different.
\begin{table}
        \caption{Core properties of single and binary-stripped star models at the onset of core collapse.}
       \begin{tabular}{cccccc}
\toprule\midrule
$M_{\mathrm{init}}$ & $M_{\mathrm{CC}}$ & $M_{\mathrm{He\, core}}^{\mathrm{ref}}$ & $\xi_{2.5}$ & Y$_e$ & Shell merger\\ 
$[M_{\odot}]$ & $[M_{\odot}]$ & $[M_{\odot}]$ &  & $(m < 1\, M_{\odot})$ & \\ 
\midrule
\multicolumn{5}{l}{\textbf{Binary stripped stars}}\\ 
11.0 & 3.11 & 3.16 & 0.01 & 0.446 & True\\ 
12.1 & 3.52 & 3.59 & 0.02 & 0.446 & True\\ 
13.0 & 3.87 & 3.93 & 0.06 & 0.446 & True\\ 
14.0 & 4.23 & 4.29 & 0.08 & 0.446 & True\\ 
14.6 & 4.48 & 4.53 & 0.09 & 0.447 & True\\ 
15.0 & 4.61 & 4.66 & 0.11 & 0.447 & True\\ 
16.0 & 4.97 & 5.02 & 0.10 & 0.448 & True\\ 
17.0 & 5.31 & 5.36 & 0.13 & 0.452 & False\\ 
18.0 & 5.61 & 5.66 & 0.13 & 0.452 & False\\ 
20.0 & 6.28 & 6.34 & 0.25 & 0.453 & False\\ 
21.0 & 6.62 & 6.67 & 0.21 & 0.453 & False\\ 
\midrule 
\multicolumn{5}{l}{\textbf{Single stars}}\\ 
11.0 &   9.32 & 3.78 & 0.04 & 0.446 & True\\ 
12.1 &   9.73 & 4.32 & 0.05 & 0.444 & True\\ 
13.0 &   9.97 & 4.78 & 0.10 & 0.448 & False\\ 
14.0 & 10.12 & 5.26 & 0.14 & 0.452 & False\\ 
14.6 & 10.15 & 5.62 & 0.18 & 0.453 & False\\ 
15.0 & 10.16 & 5.80 & 0.15 & 0.449 & False\\ 
16.0 & 10.05 & 6.32 & 0.15 & 0.452 & False\\ 
17.0 &  9.88 & 6.85 & 0.29 & 0.454 & True\\ 
18.0 & 10.13 & 7.30 & 0.31 & 0.455 & False\\ 
19.0 & 10.58 & 7.41 & 0.39 & 0.455 & True\\ 
20.0 & 11.66 & 7.57 & 0.38 & 0.455 & True\\ 
21.0 & 10.40 & 8.19 & 0.20 & 0.448 & True\\ 
\bottomrule
\end{tabular}

        \label{table:cc_general}
\end{table}

\section{Discussion}
\label{sec:discussion}
\subsection{Implications for explodability}
\label{sec:discussion:explodability}
For the same initial mass, binary-stripped star models have a smaller reference core mass, a smaller compactness, and a different distribution of isotopes compared to single stars. Both the smaller mass and compactness suggest that binary-stripped stars will explode more easily than if they had not interacted with a companion \citep[though explodability might be better described with more sophisticated means; e.g.,][]{ertl_two-parameter_2016,ertl_explosion_2020}. Multidimensional calculations indicate that the explosion depends on details of the stellar structure, especially the density profile \citep[e.g.,][]{vartanyan_revival_2018}. \citet{muller_three-dimensional_2019} did not find significant differences between the explosions of low-mass single and binary-stripped stars \citep[based on naked helium core models from][]{tauris_ultra-stripped_2015} of the same core mass. However, even for the same reference core mass, we find systematic differences in the density structure of single and binary-stripped stars we computed. In the helium-rich region binary-stripped stars have higher densities than single stars, and at the edge of the helium-rich region the density drops more sharply in the binary-stripped star models. In addition, we find distinct electron fractions in the cores of single and binary-stripped stars, which imply differences in the late nuclear burning conditions. These differences in structure can affect the accretion rates during core collapse and the explosion dynamics. This could have a significant impact for population synthesis models of compact objects. We explore detailed explosion models and implications for
the supernova explodability of single and binary-stripped stars in a companion paper \citep{vartanyan_binary-stripped_2021}.

\subsection{Implications for nucleosynthesis and nebular lines}
\label{sec:discussion:nucleosynthesis}
Our models reveal that the composition structures of stellar cores at the moment of core collapse is complex. We identified three main regions, the helium-rich, oxygen-rich, and iron-rich layers.

The distribution of isotopes in the cores of binary-stripped stars is different from the homogeneous distribution often assumed for the evolution of naked carbon-oxygen cores, \citep[e.g.,][]{patton_towards_2020}. For instance, stars stripped in binaries contain an extended gradient of carbon, oxygen, and neon above the helium-depleted core that is not present in single stellar structures. This extended composition gradient leads to changes in the final density, mean molecular weight, and composition profiles. This difference is first induced by changes during the central helium burning phase and is further accentuated in the subsequent nuclear burning stages.

These differences imply that the supernova nucleosynthesis expected from single and binary-stripped stars might be systematically different. In particular, we expect $^{12}\rm{C}$ and $^{16}\rm{O}$ yields to remain different after the supernova explosion. This is because during supernova nucleosynthesis, part of the remaining carbon will be synthesized into oxygen. Large consequences can be expected by this difference in yields because it would affect nebular oxygen lines, which are currently widely used to infer the progenitor masses of core-collapse supernova explosions. In addition, a systematic difference in the nucleosynthesis from binary-stripped and single stars would have implications for the interpretation of galaxy chemical evolution models. We explore the differences in nucleosynthesis for $^{12}\rm{C}$ in \citet{farmer_cosmic_2021}.

\subsection{Comparison with models in the literature}
\label{sec:discussion:literature}
\begin{figure}[h]% ----- Compoarison to other work ----
        \centering
        \includegraphics[width=0.5\textwidth]{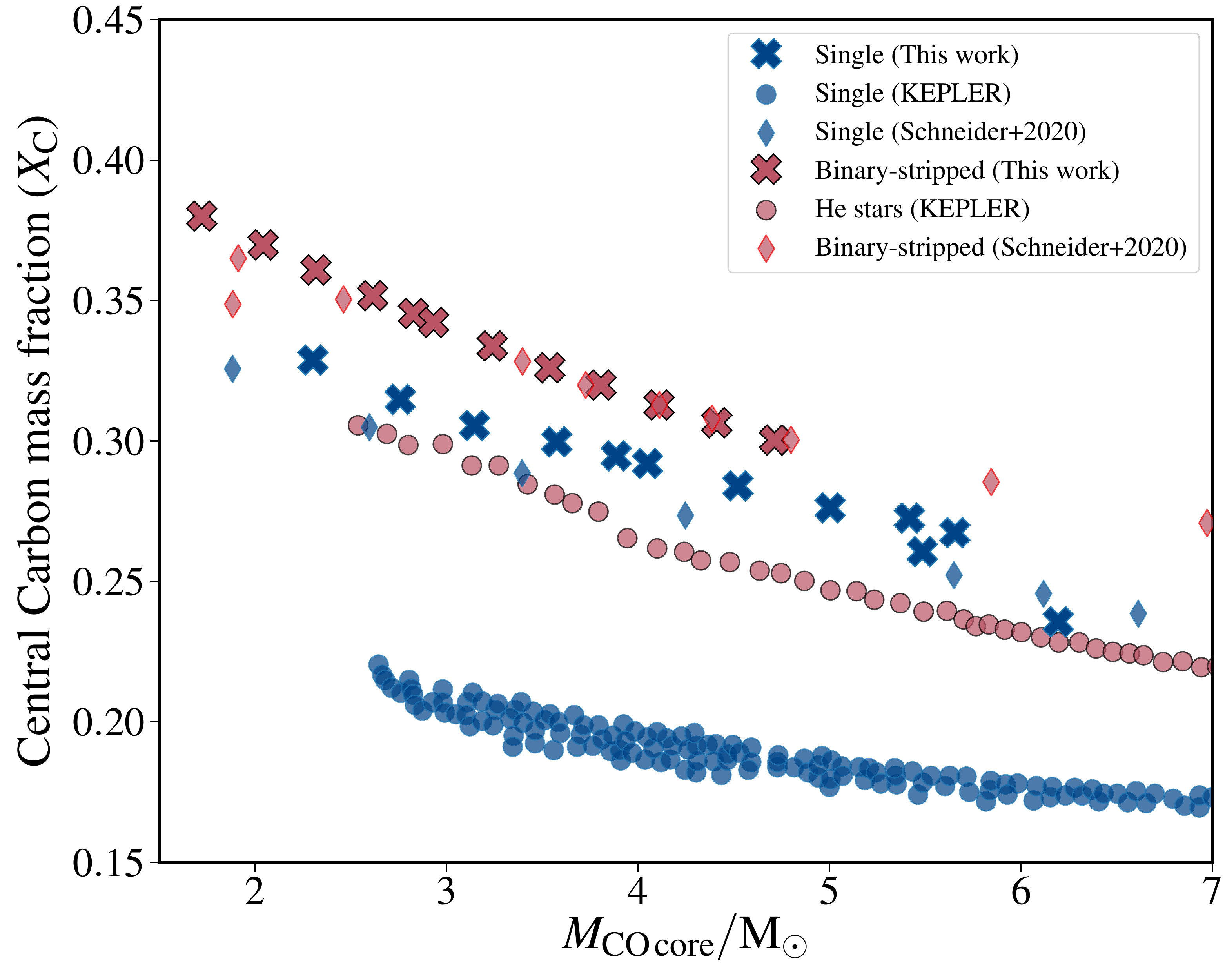}%
        \caption{Comparison of our models to models from the literature on the central carbon mass fraction to CO-core mass plane. These values are evaluated at the end of core helium depletion. All models show higher central carbon mass fractions for binary-stripped star models compared to single-star models.}
        \label{fig:comp_other_work}
\end{figure} 
Our models show a consistent trend of systematically higher central carbon mass fraction for binary-stripped star models compared to single-star models, as do other models in the literature. We show the central carbon mass fraction as a function of the mass of the carbon-oxygen core evaluated at the moment of core helium depletion in Fig. \ref{fig:comp_other_work} and compare our models to single-star models from \citet{sukhbold_high-resolution_2018} and pure helium star models from \citet{woosley_evolution_2019} (as shown in Fig. 6 of \citealt{patton_towards_2020}). These models were computed with the KEPLER stellar evolution code with similar physical assumptions to each other. We also compare our models to single star and case B mass transfer binary models from \cite{schneider_pre-supernova_2021}, computed with MESA. The physical assumptions are similar.

There is a good agreement between our models and the values reported in \citet{schneider_pre-supernova_2021} for both the single and binary-stripped star models. Especially for the binary-stripped star models, we barely find any differences in central mass fraction, except at the lowest-mass end. This agreement originates from the similar physical assumptions regarding the choice of wind mass loss, overshooting, and of the $^{12}\mathrm{C}(\alpha,\gamma)^{16}\mathrm{O}$ rate. We expect that the small differences at the lowest-mass end are linked to the treatment of mass transfer. \citet{schneider_pre-supernova_2021} parameterize mass loss on a thermal timescale until an arbitrary surface composition is reached, while in our models the binary mass-transfer phase is computed self-consistently. Our single-star models have systematically higher core carbon abundances than the models from \citet{schneider_pre-supernova_2021}, despite similar assumptions regarding core overshooting. These differences can be attributed to different choices for the wind mass-loss rate.
Much larger differences are found between our work and models computed with the KEPLER code. These models have core carbon abundances that are a factor of 1.7 lower than in our models. These differences probably stem from different physical assumptions, such as the choice for overshooting, but also from the value used for the uncertain carbon alpha-capture reaction rate. The models shown in \citep{patton_towards_2020} have a $^{12}\mathrm{C}(\alpha,\gamma)^{16}\mathrm{O}$ rate that is a factor of 1.2 higher than what we assume.

To obtain pre-supernova structures of binary-stripped stars at the onset of core collapse, we modeled donor stars in binary systems. These stars are often modeled as “naked” helium stars, with or without mass loss \citep[e.g.,][]{woosley_evolution_2019}.  It is expected that binary-stripped stars at low metallicity retain a substantial hydrogen-rich layer after mass transfer, sufficient to change their future evolution, and so for those cases helium stars are not a good approximation \citep{gotberg_ionizing_2017,yoon_type_2017,laplace_expansion_2020}. However, we are not aware of systematic work that has studied which results from the helium-star approximation significantly differ, in modeling supernova progenitors, from otherwise identical calculations using binary-stripped stars.   We encourage a careful investigation to determine with which parts of the parameter space, and for which questions, these stars can be adequately approximated with helium-star models.

\subsection{Wind mass loss}
\label{sec:discussion:winds}

Wind mass-loss rates remain a major uncertainty in stellar astrophysics, and are particularly poorly constrained for stripped stars in the mass range of our models \citep{yoon_type_2010,yoon_evolutionary_2015,gotberg_ionizing_2017}. Very few such stars have been observed so far, and the best observational constraint on the relevant mass-loss rates to date comes from a single system, the quasi Wolf-Rayet star in the system HD 45166 \citep[e.g.,][]{groh_qwr_2008}. Radiative transfer models have just begun to make mass-loss predictions for stripped stars in this mass range, and a large parameter space remains to be explored \citep{vink_winds_2017,sander_nature_2020}.

Wind mass loss plays a crucial role in the evolution, future interactions, and on the supernova properties of binary-stripped stars \citep{yoon_type_2010,yoon_evolutionary_2015,gilkis_effects_2019}. The extended gradient of the carbon and oxygen composition above the helium-depleted core of binary-stripped stars is a consequence of the receding outer edge of the convective core during central helium burning due to wind mass-loss \citep{langer_standard_1989,langer_wolf-rayet_1991,woosley_presupernova_1995}. 

We find a non-negligible effect of wind mass loss on the core structure of binary-stripped stars at solar metallicity. Appendix \ref{sec:appendix:winds} presents tests adopting a large range of possible wind mass-loss rates and their effect on the core structure of binary-stripped stars, spanning from low mass-loss rates predicted from the Monte Carlo radiative transfer models of \citealt{vink_winds_2017} to the wind mass-loss rates inferred from observations of WC and WO stars \citep{tramper_new_2016}. We confirm that our main conclusions would still be apparent for lower wind mass-loss rates than we assume, especially at the higher-mass end, though they would be less pronounced.

The wind mass-loss rate employed in our models is close to, but slightly smaller, than the wind mass loss inferred from HD 45166. \citet{tramper_new_2016} found that the wind mass-loss rate of WC and WO stars is higher than the mass-loss rate from \citet{nugis_mass-loss_2000} assumed for our default model. If binary-stripped stars have similar stellar winds to those inferred by \citet{tramper_new_2016}, the differences between single- and binary-stripped stars in the core composition and density structure could be even larger in reality than those we describe.

The vast majority of rapid population synthesis models rely on stellar structures for binary-stripped stars that were computed without mass loss \citep{hurley_comprehensive_2000, hurley_evolution_2002, pols_stellar_1998} and are based on calculations from single stars or naked helium stars only. In addition, these models are typically stopped at the end of core carbon burning. This means the core structures used in those rapid stellar population calculations are similar to those of single stars, and the core mass distribution is different from more detailed stellar models. This directly affects the number and properties of compact objects predicted from such calculations.

\subsection{Convection modeling and shell mergers}
\label{sec:discussion:shell_mergers}
Our calculations model convection with the MLT approximation (\citealt{bohm-vitense_uber_1958}, using the form from \citealt{cox_principles_1968}), with composition mixing treated as a diffusion process. These are common assumptions, even though they are imperfect approximations \citep{renzini_embarrassments_1987}, because MLT has not yet been replaced by a widely accepted practical alternative \citep[see, e.g.,][]{Arnett+321D}. This simplification in the modeling necessarily introduces uncertainties in some aspects of our results, especially during late phases in which mixing and burning occur on similar timescales \citep[e.g.,][]{Bazan+Arnett1998}. In particular, this leads to caution regarding the models in which material from the oxygen-rich layer is mixed down into the convective silicon-burning shell. The details of the predictions for such shell mergers may well change with a more sophisticated theory of convection (see also the discussion in \citealt{collins_properties_2018}, who found merging oxygen and neon shells in many of their single-star models). Nonetheless, mixing downward into convective burning shells has also been seen in multidimensional hydrodynamic models of late burning stages \citep[e.g.,][]{Bazan+Arnett1994,Bazan+Arnett1998,andrassy_3d_2020,yadav_large-scale_2020,mcneill_stochastic_2020}.

These shell mergers, if they occur in real stars as they do in our models, produce distinctly different distributions of alpha-chain elements in the oxygen-rich layer at the time of the explosion. A consequence of these shell mergers is a potentially observable abundance signature that allows for the occurrence of such shell mergers to be tested observationally. Potential implications of late-stage shell mergers for stellar nucleosynthesis have been studied by, for example, \citet{Rauscher+2002} and \citet{Ritter+2018}. \citet{dessart_radiative-transfer_2020} have also noted that such shell mergers in supernova progenitors would affect the interpretation of Ca/O abundance ratios from core-collapse supernovae. In our grid the oxygen-silicon shell mergers preferentially occur for reference core masses below 5 $\rm M_{\odot}$ and above 6.8 $\rm M_{\odot}$, which could produce an identifiable systematic pattern in observed stripped-envelope supernovae.

\section{Conclusions}
\label{sec:conclusion}
We have studied the pre-supernova structures of single-star models from 11 to 21\Msun and stars of the same initial mass that are donor stars in binary systems (binary-stripped stars). In the binary models, the stars transfer the majority of their hydrogen-rich envelopes to a companion right after the end of core hydrogen burning (early case B mass transfer). We computed models at solar metallicity and, to obtain accurate information for the composition profile at the onset of core collapse, used a large, fully coupled nuclear network of 128 isotopes after core oxygen burning. We focused on the systematic differences in the core density and composition structure at the moment of core collapse.
Overall, we confirm that binary-stripped stars in this mass range end their lives with systematically less massive helium cores than single stars with the same initial mass.

Our main findings regarding the core composition structure are:
\begin{itemize}
        \item Our composition diagrams confirm that the composition profiles are described by three main regions: (I) a helium-rich region; (II) an oxygen-rich region that contains high abundances of oxygen, neon, and magnesium; and (III) an iron-rich region that contains various isotopes of iron-group elements (see Figs. \ref{fig:end_comp_sb} and \ref{fig:end_comp}). These structures do not contain a clearly distinct carbon-rich layer, in contrast to simplified depictions of pre-supernova structures.
        \item Binary-stripped and single-star models have distinct chemical profiles at the onset of core collapse (see Fig. \ref{fig:end_comp_sb}), even for helium cores of similar mass. In agreement with previous studies, we find that stars stripped in binaries develop an extended gradient of carbon and oxygen at the edge of the helium-depleted core that does not exist in single stars. We show that this layer remains until core collapse, and is more extended in mass for higher initial stellar masses due to the mass dependence of winds (see Fig. \ref{fig:hedepl_composition}).
        \item We find that this extended chemical gradient is responsible for a systematically larger total mass of carbon in binary-stripped star models than comparable single-star models at the onset of core collapse (see Figs. \ref{fig:c12_o16_single_vs_stripped} and \ref{fig:core_overview}). In contrast, the total mass of oxygen remains similar (see Figs. \ref{fig:c12_o16_single_vs_stripped} and \ref{fig:burn_ev}).
        \item The difference in total carbon mass will probably affect the chemical yield predictions, pending possible alterations as the supernova shock propagates through the interior, which we have not modeled here (but are presented in \citealt{farmer_cosmic_2021}). 
        \item Shortly before core collapse, about half of our models experience a merging of the convective silicon-burning shell with the oxygen-rich layer, which results in an enrichment of heavier alpha elements (mainly $^{28}$Si, $^{32}$S, $^{36}$Ar, and $^{40}$Ca) in the outer oxygen-rich regions. Although their physical accuracy in one-dimensional models is unclear, similar effects are observed in multidimensional simulations. If shell mergers do occur in nature, these peculiar abundances could be observationally testable.

\end{itemize}
In addition, we investigate systematic differences in the core density structure and come to the following conclusions:
\begin{itemize}
        \item For the same reference helium core mass, stars stripped in binaries reach systematically larger helium core radii at the onset of core collapse than single stars (see Fig.~\ref{fig:he_core}).
        \item Systematic differences can be observed in the core density structure of single and binary-stripped stars at the onset of core collapse (see Fig. \ref{fig:profiles_cc}). In the helium-rich region, binary-stripped stars have higher final densities than single stars, and at the edge of the helium-rich region the density drops more sharply in the binary-stripped star models.
        The core density structures begin to diverge during core helium burning (see Figs. \ref{fig:profile_ev} and \ref{fig:burn_ev}).
        \item We find that, for the same initial mass, the compactness of binary-stripped stars is systematically lower than that of single stars. For the same reference core mass, the compactness is similar.
\end{itemize}
The differences in composition structures may lead to a systematically different creation of isotopes from the supernovae of single stars and binary-stripped stars. We expect the distinct density and shell-burning structures  to have a significant impact on the dynamics following core     collapse, on the propagation of the supernova shock through the stellar structure, and on the nature of the compact remnant. Multidimensional supernova simulations, based on stellar models similar to those presented here, are needed to quantify these effects and to compare them with supernova observations.

% --------------------------------- ACKNOWLEDGMENTS --------------------------
\begin{acknowledgements}
        We are grateful to the referee for a thorough and careful review, and for helpful suggestions, which have led to a significant improvement of the manuscript.
        We thank A. Burrows, L. van Son, T. Moriya, and M. Modjaz for useful discussions.
        The research was funded by the European Union's Horizon 2020 research and innovation program from the European Research Council (ERC, grant agreement No.\ 715063), and by the Netherlands Organisation for Scientific Research (NWO) as part of the Vidi research program BinWaves with project number 639.042.728. YG acknowledges the funding from the Alvin E.\ Nashman fellowship for Theoretical Astrophysics. RF is supported by the Netherlands Organisation for Scientific Research (NWO) through a top module 2 grant with project number 614.001.501 (PI de Mink). DV acknowledges funding from the U.S. Department of Energy Office of Science and the Office of Advanced Scientific Computing Research via the Scientific Discovery through Advanced Computing (SciDAC4) program and Grant DE-SC0018297 (subaward 00009650). \\
\end{acknowledgements}

% --------------------------------- BIBLIOGRAPHY -----------------------------
\bibliographystyle{aa} % style aa.bst
\bibliography{stripped_CCSNe,additional} % complete bibliography with additional entries

% --------------------------------- APPENDIX --------------------------------
\begin{appendix} %First appendix
\section{Impact of the nuclear network}
\label{sec:appendix:nuclear_network}
Toward the end of the evolution of massive stars, the core is composed by a multitude of synthesized isotopes, many of which are short-lived and neutron-rich. The properties of the stellar core are sensitive to weak interaction processes that can only be accounted for by tracking the many electron and positron captures, and $\beta$-decays from rare and unstable isotopes. This usually requires the quasi-statistical nuclear equilibrium approximation approach \citep{hix_silicon_1996} or to keep track of a large number of isotopes in the nuclear network \citep[e.g.,][]{farmer_variations_2016}. In this work, we present simulations where we simultaneously solve for the structure and nuclear burning using with a large nuclear network of 128 isotopes. \citet{farmer_variations_2016} showed that a nuclear network with at least 127 isotopes is required to obtain resolved core structures after the end of core oxygen burning. This is what we adopt in our models. For the prior evolution up to and including central oxygen burning, we employ an $\alpha$-chain nuclear network containing 21 isotopes, which is considered adequate to capture the energy generation rate during earlier nuclear burning phases. However, the accuracy of an $\alpha$-chain network decreases with the evolution, especially during core oxygen burning \citep[see, e.g.,][]{renzo:thesis}. 

\begin{figure}[ht!]% ----- tc /rhoc nuclear network comparison:   ----
        \centering
        \includegraphics[width=0.5\textwidth]{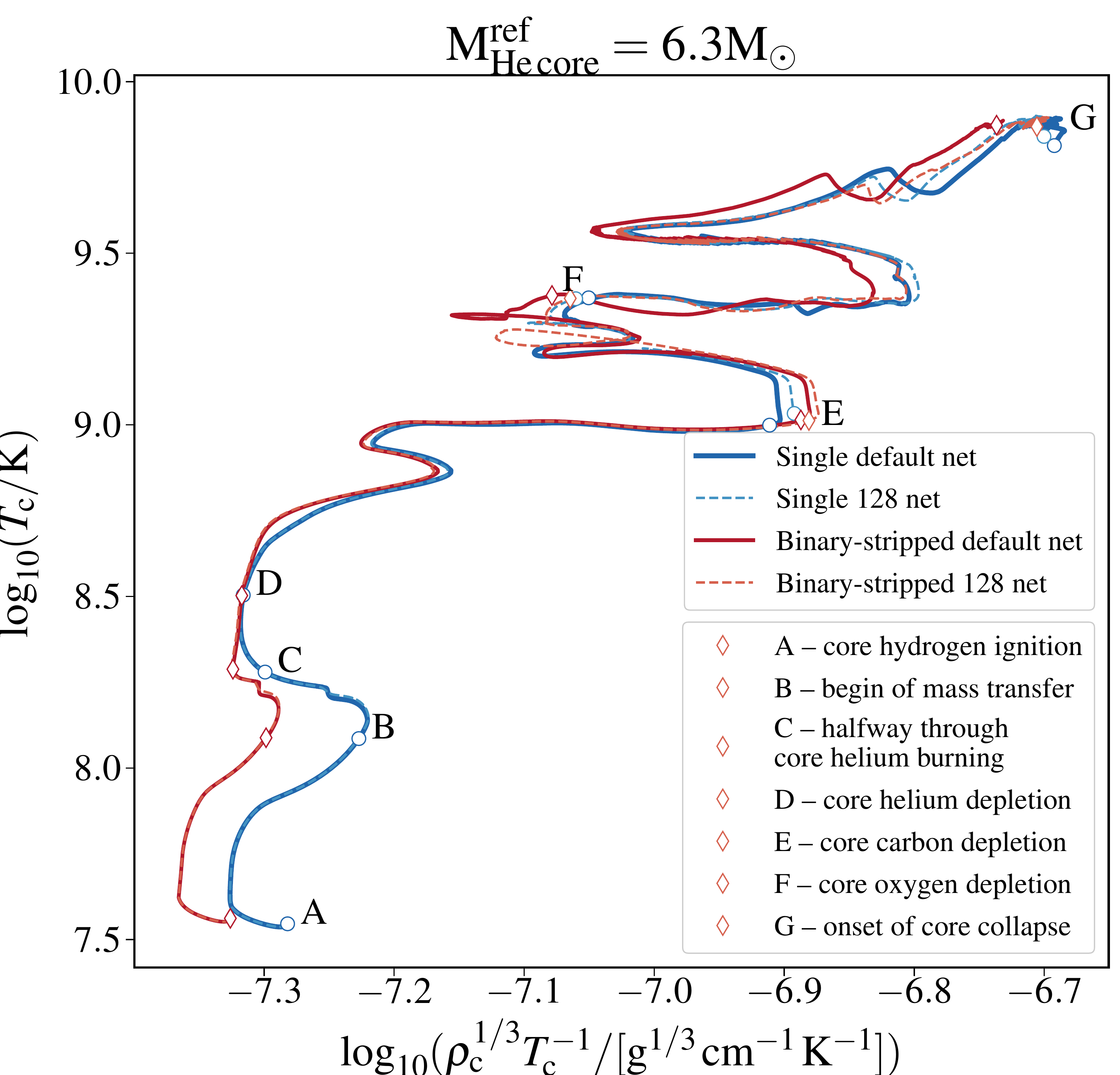}
        \caption{Comparison of stellar models on the central temperature, central density plane when using a large nuclear network with 128 isotopes throughout the evolution versus switching from a nuclear network with 21 isotopes to a nuclear network with 128 isotopes after core oxygen depletion. To better show the differences between the models, we have rescaled the horizontal axis by the dependence expected for homologous contraction of an ideal gas.}
        \label{fig:tc_rhoc_test128net}
\end{figure}

Here, we test the impact of computing the evolution of stars from birth to core oxygen depletion with a nuclear network that contains 128 isotopes compared to our default setup. We perform this test for two stellar models (1) the single-star model with an initial mass of 16\Msun (2) a binary-stripped model with an initial mass of 20\Msun. Both models reach almost exactly the same core mass of 6.3~\Msun at the end of core helium burning.
In Fig. \ref{fig:tc_rhoc_test128net}, we compare these models on the central density-central temperature space. Until central carbon depletion, the models have almost exactly the same evolution independently from the size of the nuclear networks. However, after this moment, differences start to appear. The models with a larger network reach higher densities. This is true for both the single and the binary-stripped star model. After this moment, differences in the binary-stripped model become more prominent.

The central temperatures and densities between the models do vary with the use of a larger nuclear network in the late evolutionary phases. However, the relative differences between the single and binary-stripped stellar models remain similar.

\begin{figure}[ht!]% ----- tc /rhoc nuclear network comparison:   ----
        \centering
        \includegraphics[width=0.5\textwidth]{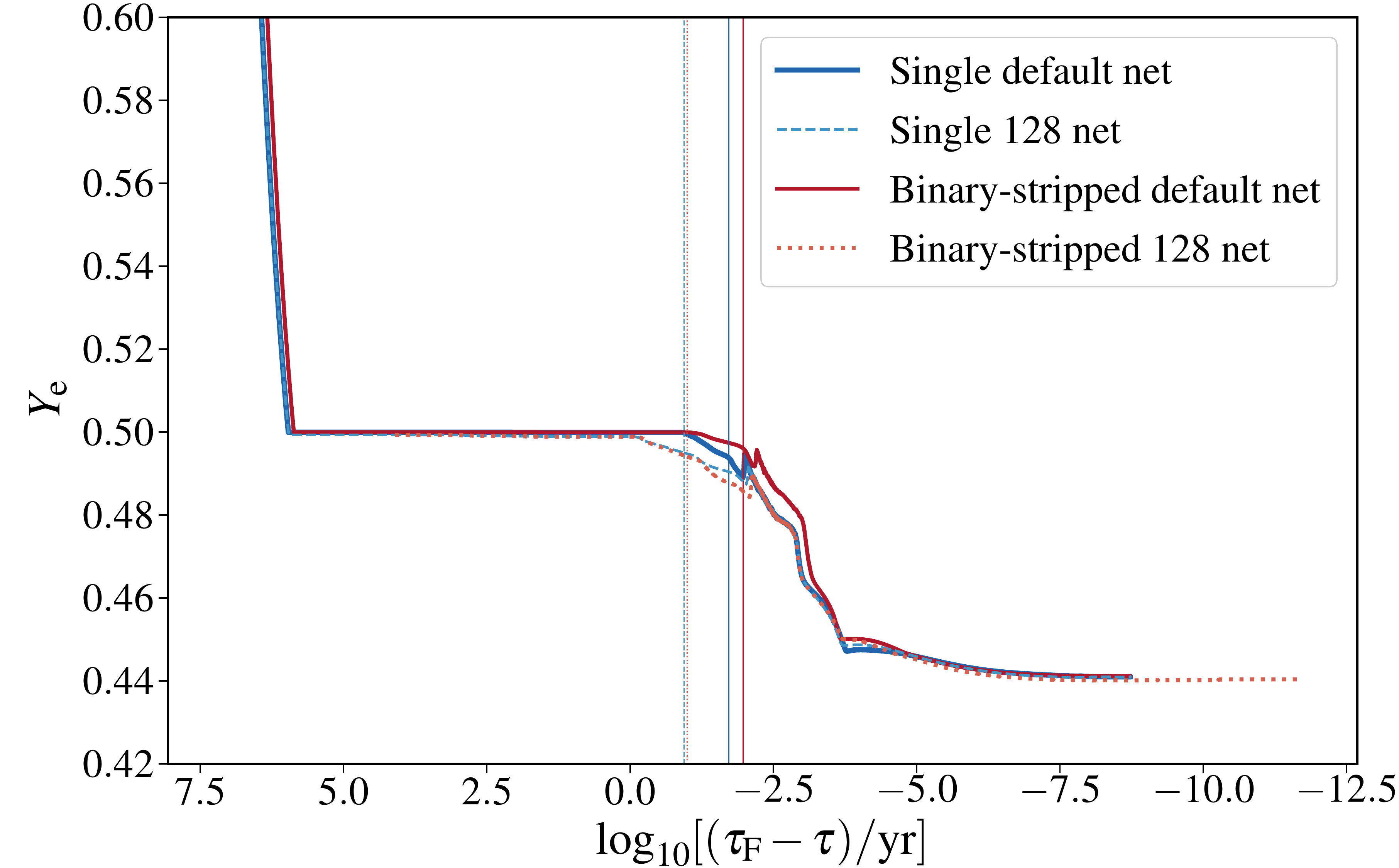}
        \caption{Comparison of the electron fraction, $Y_\mathrm{e}$, evolution when using a large nuclear network with 128 isotopes throughout the evolution versus switching from a nuclear network with 21 isotopes to a nuclear network with 128 isotopes after core oxygen depletion. Vertical lines mark the moment of core oxygen depletion for each model, which is when the switch between networks occurs for the models with the default network.}
        \label{fig:Ye_test128net}
\end{figure}

In Fig. \ref{fig:Ye_test128net} we show the evolution of the electron fraction $Y_\mathrm{e}$ for the same models with and without a large nuclear network from the beginning of the simulation. Generally, the evolution is similar between all models, and final values agree within 2\%. The largest difference can be observed after core oxygen depletion, where both the single and binary-stripped models with a large nuclear network from the start display smaller values than the alternate models.

 We conclude that models with larger nuclear networks throughout the evolution, though expensive, are desirable for the future, but we expect that our conclusions about the relative differences between single and binary-stripped star models are not significantly affected.

\section{Reference core mass}
\label{sec:appendix:helium_core_mass}
\begin{figure} % ----- Helium core mass as a function of initial mass
        \centering
        \includegraphics[width=0.5\textwidth]{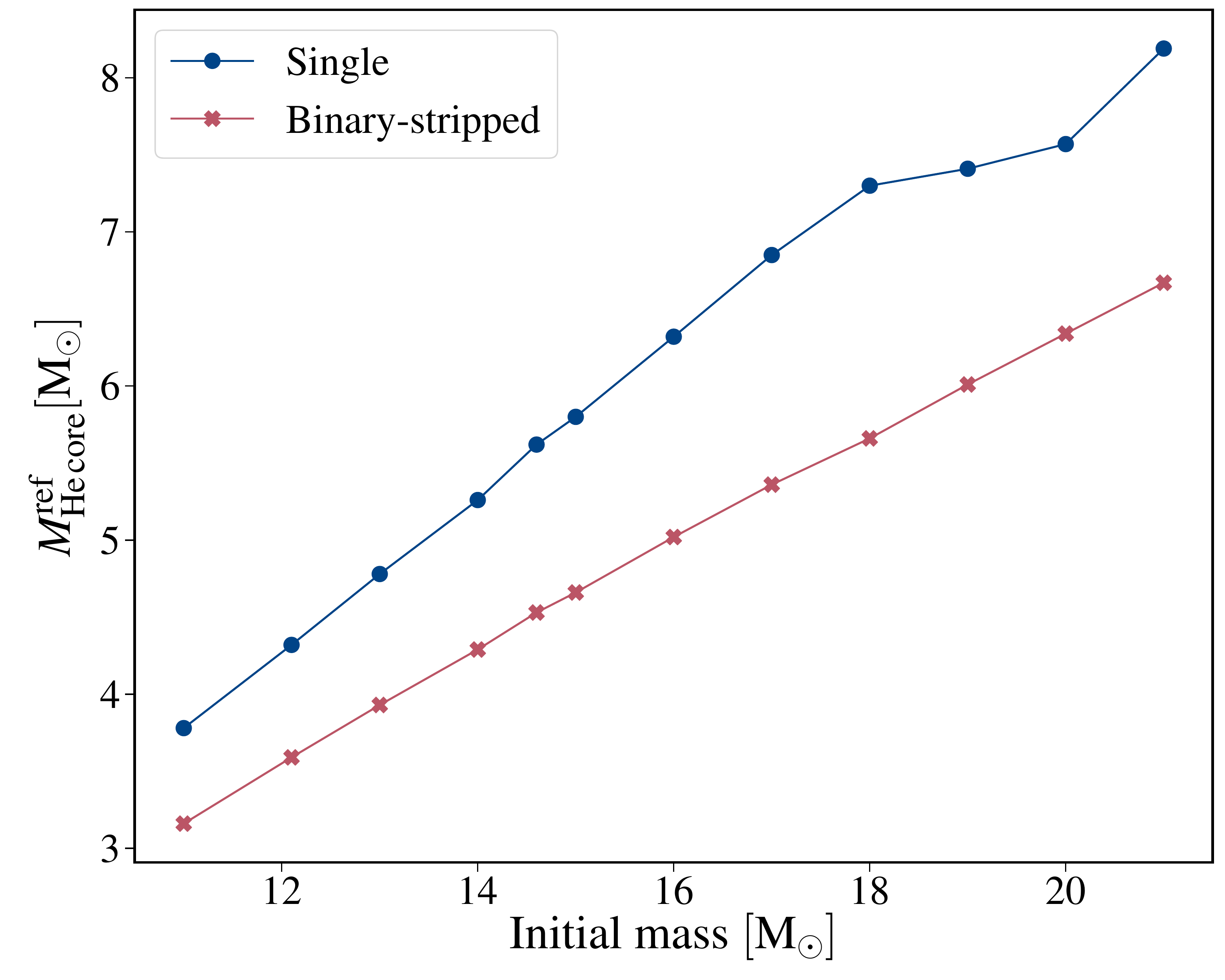}
        \caption{Helium core mass at core helium depletion as a function of initial mass for single stars and stars that were stripped due to Roche-lobe overflow in a binary system during the hydrogen-shell burning phase.}
        \label{fig:MHecore_vs_Mi}
\end{figure}
Giving an exact definition of the boundary between the core and the envelope of a star is not trivial and varies between different studies \citep{sukhbold_compactness_2014}. That is because, while a drop in density of several orders of magnitude, together with a larger change in composition, can typically be observed between the edge of the inner core and the envelope, there is a gradient in composition and density at this edge that make it difficult to define an exact boundary. In addition, details of semi-convection and core overshoot can have a large effect on the stellar core mass and on the core edge and complicate this definition even further \citep[e.g.,][]{schootemeijer_constraining_2019}.
For this study, we compare the late properties of stellar models as a function of their reference core mass, which we define as the mass of the helium core (i.e., the mass coordinate where the mass fraction of hydrogen decreases below 0.01) at the moment of core helium depletion (when the central mass fraction of helium decreases below $10^{-4}$). Here, we investigate the impact of this definition on our conclusions. We discuss the sensitivity to the evolutionary stage for which we define the reference core mass in Sect. \ref{sec:appendix:helium_core_mass:ev} and to the exact choice of the helium core mass boundary in Sect. \ref{sec:appendix:helium_core_mass:boundary}.

\subsection{Evolution of the helium core mass}
\label{sec:appendix:helium_core_mass:ev}
In Fig. \ref{fig:MHecore_vs_Mi}, we present the reference core mass as a function of initial mass for both single and binary models. We verify that donor stars in early case B mass-transfer binaries have systematically less massive cores at the end of their lives than single stars with the same initial mass \citep[cf.][]{podsiadlowski_presupernova_1992}. For both the single and binary-stripped models, the reference core mass increases as a function of initial mass. The highest-mass single-star models deviate from the linear trend due to mass loss.

\begin{figure*}[h] % ----- Evolution of the helium core mass as a function of time
        \centering
        \includegraphics[width=\textwidth]{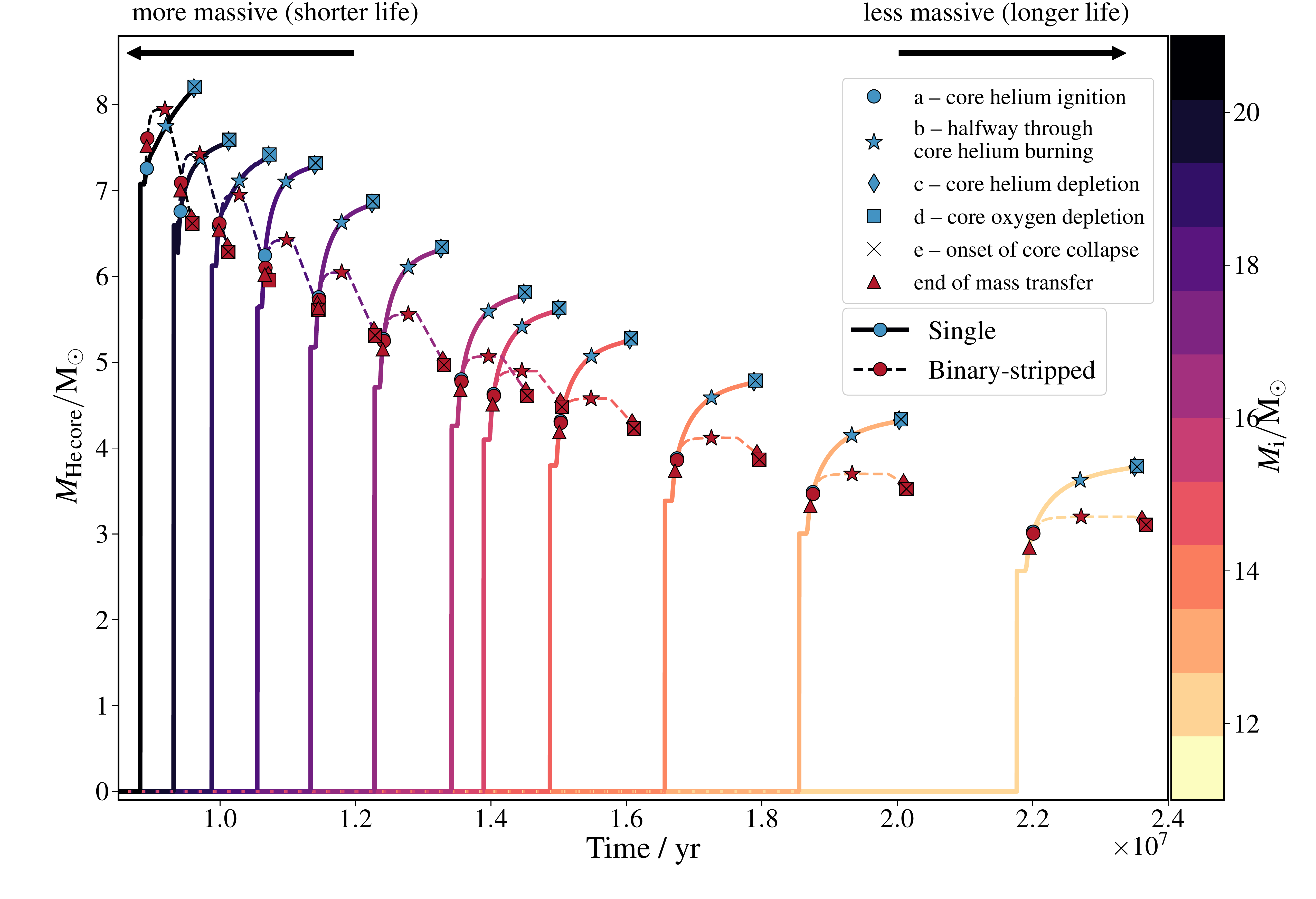}
        \caption{Evolution of the helium core mass of every model as a function of time. Full lines correspond to single stars and dashed lines to binary-stripped stars. Symbols indicate key evolutionary stages for the development of the helium core, and colors give the initial mass of the stars. Because massive stars live shorter lives, the more massive models are located on the left side and the less massive models on the right side of the plot.}
        \label{fig:ev_he_core_mass}
\end{figure*}

We present the evolution of the helium core mass as a function of time for all of the single (full lines) and binary-stripped (dashed lines) models in Fig. \ref{fig:ev_he_core_mass}. We indicate the initial mass of the models with colors and mark key evolutionary steps with symbols. All single-star models show the same trend of an increasing helium core mass with time. In contrast, the helium core mass of binary-stripped star models first increases, then reaches a plateau, and then decreases toward the end of their life. This is because single stars undergo hydrogen-shell burning, which leads to an increase of about 1\Msun in helium core mass from the beginning of core helium burning onward (circles in Fig. \ref{fig:ev_he_core_mass}) until core oxygen depletion (squares in Fig. \ref{fig:ev_he_core_mass}). 
In contrast, binary-stripped stars  at this metallicity stop hydrogen-shell burning shortly after mass transfer. Eventually, most of their hydrogen-rich envelope is removed due to the solar-metallicity wind mass-loss rate assumed here (we discuss the impact of the wind mass-loss rate in Appendix \ref{sec:appendix:winds}). For lower-metallicity stars, a hydrogen-rich layer is retained because the stripping process is less effective \citep{gotberg_ionizing_2017,yoon_type_2017,laplace_expansion_2020}. Halfway through core helium burning (central helium fraction < 0.5, indicated by star markers in Fig. \ref{fig:ev_he_core_mass}), the helium core mass of the binary-stripped stars reaches a plateau for all masses considered. For all the binary-stripped models, this happens after the end of mass transfer (triangle markers in Fig.~\ref{fig:ev_he_core_mass}). At this moment, hydrogen-shell burning stops, and only a small hydrogen layer remains on the stellar surface. Soon after, the helium core mass begins to decrease for all models due to the effect of winds that remove the outer layers. This effect is mass dependent \citep[cf.][]{langer_mass-dependent_1989}. For the stellar wind mass-loss rate we assume in this study, the decrease in mass due to wind mass loss in binary-stripped stars ranges from 0.1\Msun for the lowest-mass model (with an initial mass of 11\Msun) to 1.2\Msun for the highest-mass model (with an initial mass of 21\Msun). 

After core helium depletion, the lifetime of stars is so short that winds have a negligible impact on the helium core mass, as shown by the overlapping markers after this moment (diamonds, squares, and crosses in \ref{fig:ev_he_core_mass}). The strongest effect we observe is a decrease in the helium core mass by 0.1\Msun for the highest-mass binary-stripped star model.

\subsection{Helium core boundary}
\label{sec:appendix:helium_core_mass:boundary}
\begin{figure}[ht] % ----- Definition helium core mass in abundance plot ---
        \centering
        \includegraphics[width=0.5\textwidth]{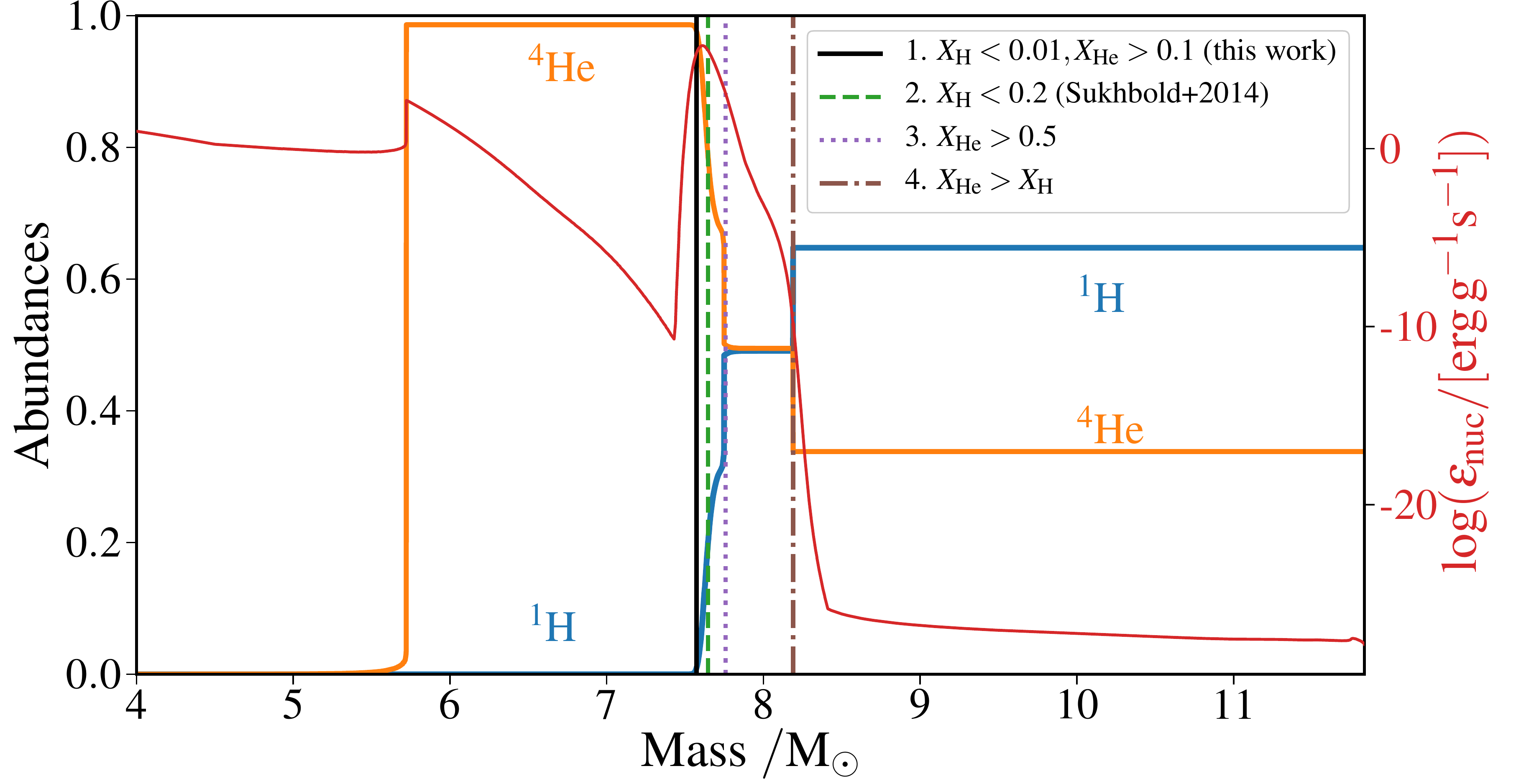}
        \caption{Mass fraction of $^{1}\mathrm{H}$ (blue line) and $^{4}\mathrm{He}$ (orange line) as a function of mass coordinate for an example single-star model with an initial mass of 20\Msun at core helium depletion. Vertical lines mark different choices of the helium core mass definition. In red, we indicate the specific nuclear energy generation rate throughout the model. The first peak in the nuclear energy generation rate around 5.8\Msun corresponds to helium-shell burning and the second peak around 7.6\Msun to hydrogen-shell burning.}
        \label{fig:he_core_mass_def}
\end{figure}

Various definitions are possible for the helium core mass, as illustrated in Fig.~\ref{fig:he_core_mass_def}. We discuss whether the choice of the core definition impacts our conclusions, in particular regarding structural differences between the helium cores of single and binary-stripped stars (see Sect. \ref{sec:endcomp:helium_core}).

In Fig.~\ref{fig:he_core_mass_def} we show the interior helium and hydrogen abundances of an example single-star model with an initial mass of 20\Msun at core helium depletion. Our definition of the helium core mass (1) is shown as the mass coordinate where the hydrogen mass fraction $X_{\mathrm{H}} < 0.01$ and the helium mass fraction $X_{\mathrm{He}} > 0.1$ and corresponds to the edge of the helium-rich region at 7.5\Msun. The definition employed in \citet{sukhbold_compactness_2014} is somewhat further out and corresponds to the mass where $X_{\mathrm{H}} < 0.2$ (2). The definition where $X_{\mathrm{He}} > 0.5$ leads to an even higher helium core mass and coincides with the beginning of a helium-enriched layer due to hydrogen-shell burning (corresponding to the peak of the nuclear energy generation rate around 7.6\Msun). Another possible definition is the mass coordinate where the mass fraction of hydrogen is strictly larger than the mass fraction of helium (4). This boundary is however set outside the hydrogen-burning shell and does therefore not follow the growth of the helium-rich core due to shell burning.

\begin{figure} % ----- Definition helium core mass in abundance plot ---
        \centering
        \includegraphics[width=0.5\textwidth]{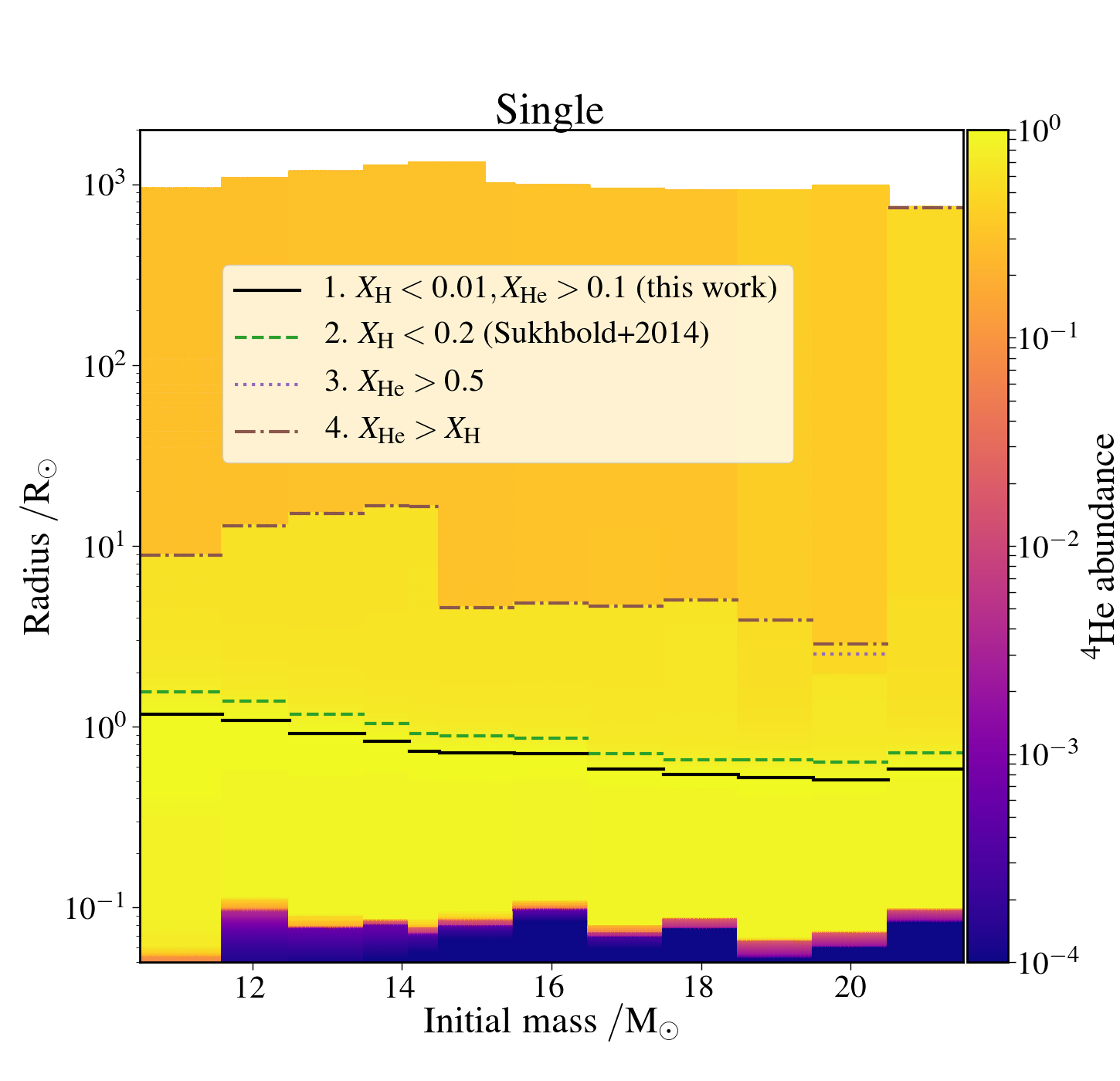}
        \includegraphics[width=0.5\textwidth]{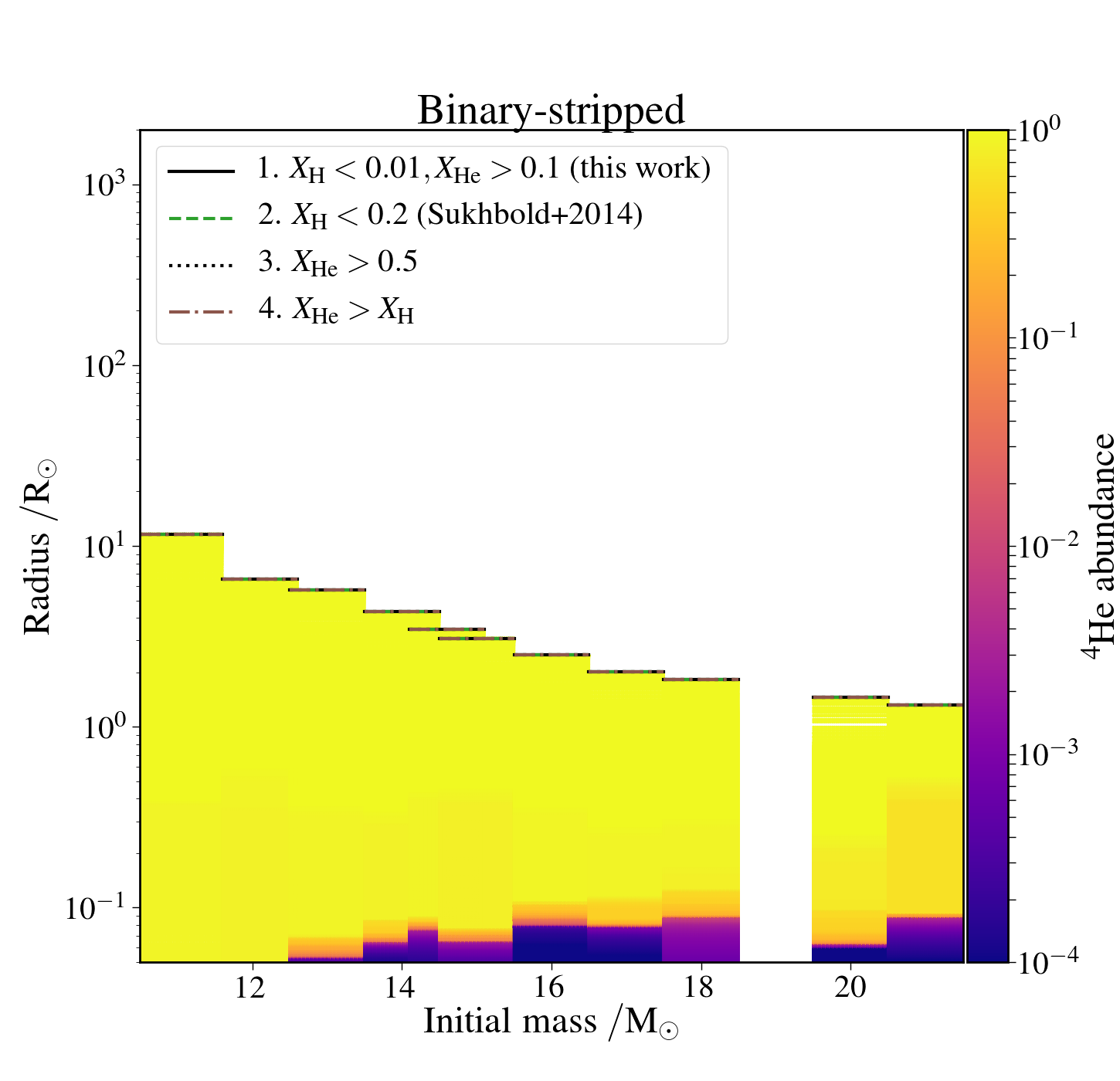}
        \caption{Radial coordinate as a function of the initial mass for the single-star models (\textit{top}) and the binary-stripped star models (\textit{bottom}) at the onset of core collapse. Colors indicate the mass fraction of $^{4}\mathrm{He}$ at each radial coordinate. Four different helium core mass definitions are indicated with horizontal lines.}
        \label{fig:r_he_core_def_cc_single}
\end{figure}

We show how the radius of the helium core changes when we change the definition of the helium core mass boundary in Fig. \ref{fig:r_he_core_def_cc_single}. The figure presents the radial coordinate of each stellar model (top: single, bottom: binary-stripped) as a function of the initial stellar mass. Every model in our grid is shown as a rectangle whose height corresponds to the surface radius of the model. The width of every rectangle is set as 1\Msun. We indicate the mass fraction of helium at each radius coordinate by color shading it. For the binary-stripped models at this metallicity, all helium core definitions correspond to the surface of the star and are thus identical. The choice of the helium core boundary is thus only relevant for the single-star models.

For the single-star models, the helium core radius changes depending on the core boundary chosen. With definition 3, the boundary of the helium core corresponds to the boundary between the helium-rich region and the hydrogen-rich envelope for all models, except for the highest-mass model. This is because for the highest-mass model, some helium has been mixed in the hydrogen-rich envelope. This means the core-mass definition is not consistent across all masses. Furthermore, this boundary lies above the hydrogen-burning shell and can therefore not account for the growth in core mass due to shell burning.
The other two definitions give similar trends to each other. Definition 2 from \citep{sukhbold_compactness_2014} results in systematically larger helium core radii, of about 0.5\Rsun larger than for definition 1. However, this difference is negligible compared with the difference to the helium core radii of the binary-stripped models. The helium core mass does not change significantly after core helium depletion, and our findings on the differences in helium core radii are robust against sensible variations in the core mass definition.
We conclude that the reference core mass we chose is robust for our purpose.

\section{All final composition diagrams}
\label{sec:appendix:full_set}
\begin{figure*} % ----- Evolution of the composition
        \centering
        \includegraphics[width=\textwidth]{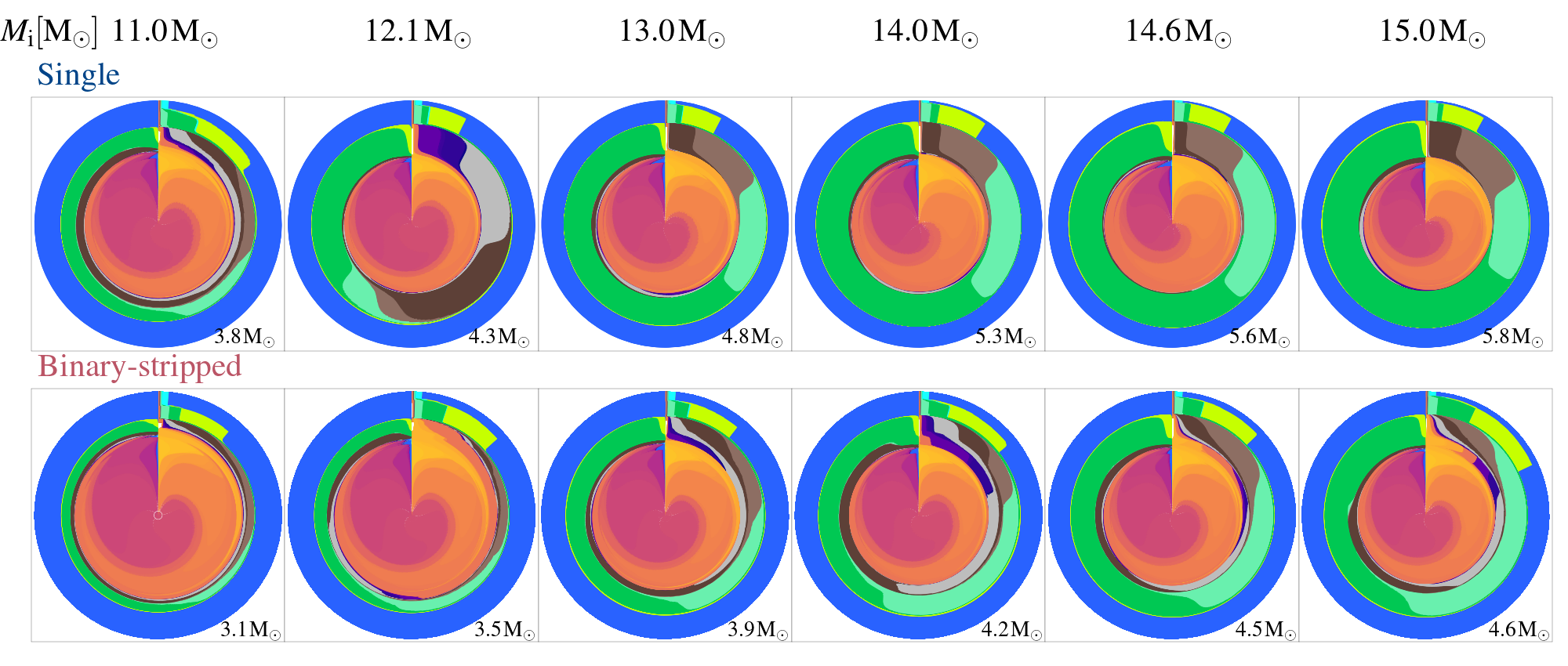}
        \includegraphics[width=\textwidth]{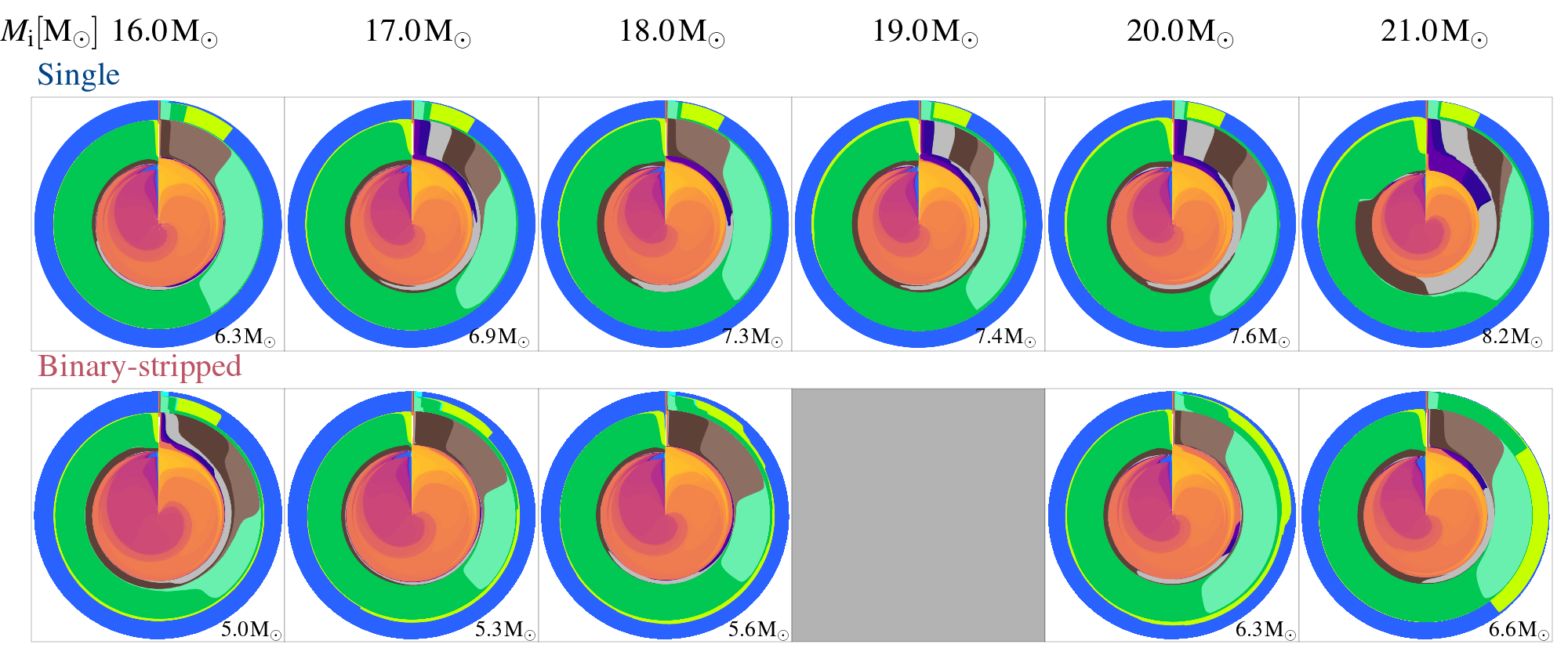}
        \caption{Composition diagrams showing the core composition inside the helium core at the onset of core collapse for the entire grid of models. The two pairs of rows show single and binary-stripped star models with the same initial masses. All models are scaled such that the radius of the diagram is proportional to the square root of the helium core mass (the helium core mass is indicated at the bottom of each diagram). The binary-stripped model with an initial mass of 19\Msun is not shown because it did not reach core collapse. Colors represent the same isotopes as in Fig. \ref{fig:end_comp_sb}.}
        \label{fig:end_comp}
\end{figure*}

In Sect. \ref{sec:endcomp} we discuss general trends in final chemical structure and compare single and binary-stripped star models with similar values of the reference core masses. In Fig. \ref{fig:end_comp}, we present composition diagrams at the moment of core collapse for the full set of models. The figure is divided into two sets of two rows that contain single (top) and binary-stripped (bottom) stellar models with the same initial masses. The binary-stripped model with an initial mass of 19\Msun is not shown because it did not reach core collapse due to numerical issues. Below each model we indicate its reference core mass.  

\section{Effect of winds on the cores of binary-stripped stars}
\label{sec:appendix:winds}
How much mass massive stars lose to winds is one of the biggest open question in stellar astrophysics \citep{smith_mass_2014}. The choice of stellar winds assumed for stellar evolution models greatly impacts the evolution of massive stars and in particular their core structure \citep{renzo_systematic_2017}. Stellar winds of binary-stripped stars are among of the most unconstrained \citep{gotberg_ionizing_2017,vink_winds_2017} and can have a large impact on their evolution, future interactions, and on their supernova properties \citep{yoon_type_2010,yoon_evolutionary_2015,gilkis_effects_2019}, though efforts are being made to better constrain these with new-generation stellar atmosphere models \citep[e.g.,][]{sander_nature_2020}. Here, we investigate the effect of changing the adopted wind mass-loss rate for binary-stripped stars on the core properties. To this end, we simulated two example models with initial masses of 11 and 20\Msun, at the extremes of the mass range we chose, and varied the wind mass-loss rate by a constant factor after the end of mass-transfer. 

We varied the wind mass-loss rate from a factor of ten smaller to a factor of three higher than the default wind mass-loss rate. These values are motivated by the range of mass-loss rates inferred from binary-stripped stars both through observations and from latest theoretical studies \citep{yoon_type_2010,yoon_evolutionary_2015,vink_winds_2017}. In Table~\ref{table:winds}, we give values for the surface properties and wind mass-loss rates halfway through core helium burning (when the central mass fraction of helium decreases below 0.5). The values for the lowest wind mass-loss rates (wind factors of 0.10 and 0.33) are similar to the solar-metallicity wind mass-loss rate computed by \citet{vink_winds_2017} through Monte Carlo radiative transfer models. Our lowest-mass model with an initial mass of 11 \Msun can be compared to observations of the quasi Wolf-Rayet star in the system HD 45166, which has a wind mass-loss rate of $\log_{10}\dot{M}/(M_{\odot}\,\mathrm{yr}^{-1})=-6.66$ and a luminosity of $\log_{10}(L/\Lsun)=3.75\pm0.08$ \citep{groh_qwr_2008}. Though the luminosity of this star, and therefore its mass, is smaller than the luminosity of the model, its wind mass-loss rate is similar to the mass-loss rates we obtain when increasing the wind mass-loss rate by a factor of between 1.25 or 2. For the binary-stripped star model with an initial mass of 20\Msun, the values for the highest wind mass-loss rates (wind factors 2 and 3) are similar to the wind mass-loss rates derived from observations of WC and WO stars \citep[$\log_{10}\dot{M}/(M_{\odot}\,\mathrm{yr}^{-1})=-5$ for $\log_{10}(L/\Lsun)=5$,][]{tramper_new_2016}.

\begin{table}
        \caption{Effect of a modified wind mass loss factor after mass transfer on the stellar properties of binary-stripped stars}     
       \begin{tabular}{ccccc}
\toprule\midrule
Wind factor & $M_{*}$ & $\log_{10}$ & $\log_{10}$ & $\log_{10}$ \\
 & ($M_{\odot}$) & ($T_{\mathrm{eff}}/\mathrm{K}$) & ($L/L_{\odot}$) & ($\dot{M}/[M_{\odot}\,\mathrm{yr}^{-1}]$) \\
\midrule
\multicolumn{4}{l}{\textbf{Binary-stripped star model: $M_{i} = 11\Msun$}} \\
0.10 & 3.50 & 4.84 & 4.19 & -7.79 \\
0.33 & 3.46 & 4.85 & 4.17 & -7.28 \\
0.50 & 3.43 & 4.85 & 4.15 & -7.11 \\
0.75 & 3.39 & 4.86 & 4.14 & -6.94 \\
1.00 & 3.36 & 4.86 & 4.12 & -6.82 \\
1.25 & 3.32 & 4.87 & 4.11 & -6.73 \\
2.00 & 3.22 & 4.89 & 4.08 & -6.52 \\
3.00 & 3.08 & 4.92 & 4.03 & -6.27\\
\midrule
\multicolumn{4}{l}{\textbf{Binary-stripped star model: $M_{i} = 20\Msun$}} \\
0.10 & 8.10 & 4.92 & 5.04 & -6.68 \\
0.33 & 7.94 & 4.95 & 5.02 & -6.17 \\
0.50 & 7.84 & 4.97 & 5.00 & -5.98 \\
0.75 & 7.68 & 4.99 & 4.99 & -5.80 \\
1.00 & 7.52 & 5.02 & 4.97 & -5.62 \\
1.25 & 7.34 & 5.04 & 4.95 & -5.46 \\
2.00 & 6.81 & 5.03 & 4.88 & -5.34 \\
3.00 & 6.22 & 5.02 & 4.80 & -5.27 \\
\bottomrule   
\end{tabular}

       {\textbf{Notes.} The values are taken halfway through central helium burning (defined as the moment when $X_{\mathrm{He}}=0.5$).}
        \label{table:winds}
\end{table}

\begin{figure} % ----- Profiles winds ---
        \centering
        \includegraphics[width=0.5\textwidth]{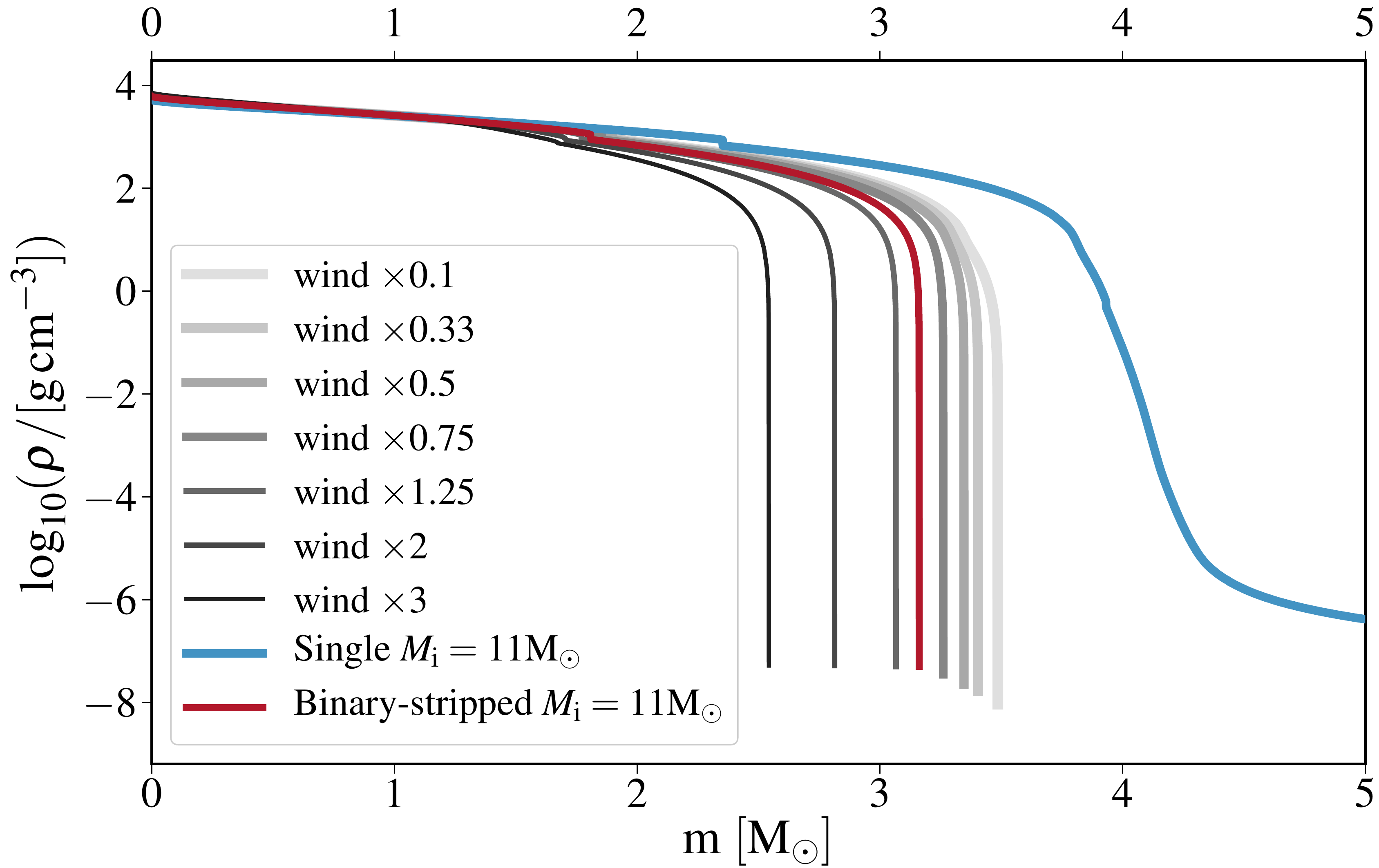}
        \includegraphics[width=0.5\textwidth]{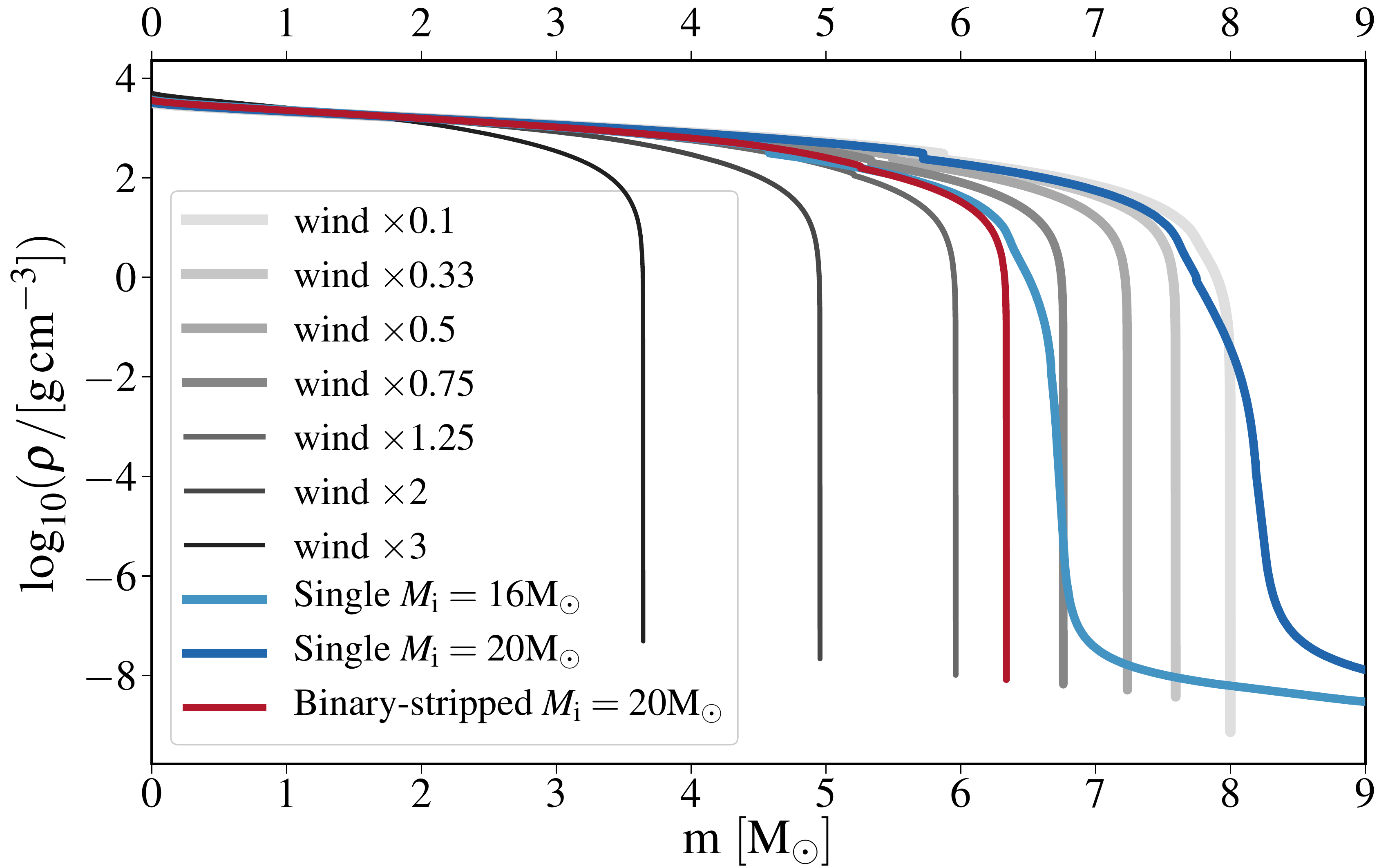}
        \caption{Impact of wind variations on the density profiles of binary-stripped stars. We show the profiles at the moment of core helium depletion for a single (blue) and binary-stripped (red) star model with the same initial mass of 11\Msun (\textit{top panel}) and 20\Msun (\textit{bottom panel}). Gray lines show variations in the density profile for the binary-stripped star when we vary the wind mass-loss rate after Roche-lobe overflow by the factor indicated in the legend.}
        \label{fig:profile_winds}
\end{figure}

Varying the wind mass-loss rate factor has a systematic impact on the stellar structure. The total mass of the binary-stripped star, and with it its density profile in the outer layers at core helium depletion, strongly depends on the wind mass-loss rate, as shown in Fig. \ref{fig:profile_winds}. In the top panel of Fig. \ref{fig:profile_winds}, we show the models with an initial mass of 11\Msun, and in the bottom panel, the models with an initial mass of 20\Msun.
Independently of the initial mass, we find that models with increased wind mass-loss rates have more centrally concentrated density profiles, a lower-mass core and a lower total mass. In contrast, models with decreased wind mass-loss rates have higher total masses and achieve smaller densities at the surface. The models with the smallest wind mass-loss rates have density profiles with a similar shape as for the single-star profiles because they retain a hydrogen-rich layer above the helium core. The 20\Msun binary-stripped model (bottom panel of Fig. \ref{fig:profile_winds}) with the lowest wind mass-loss rate achieves a helium core mass that is larger than that of the single-star model with the same initial mass. 

\begin{figure} % ----- central mass fractions ---
        \centering
        \includegraphics[width=0.5\textwidth]{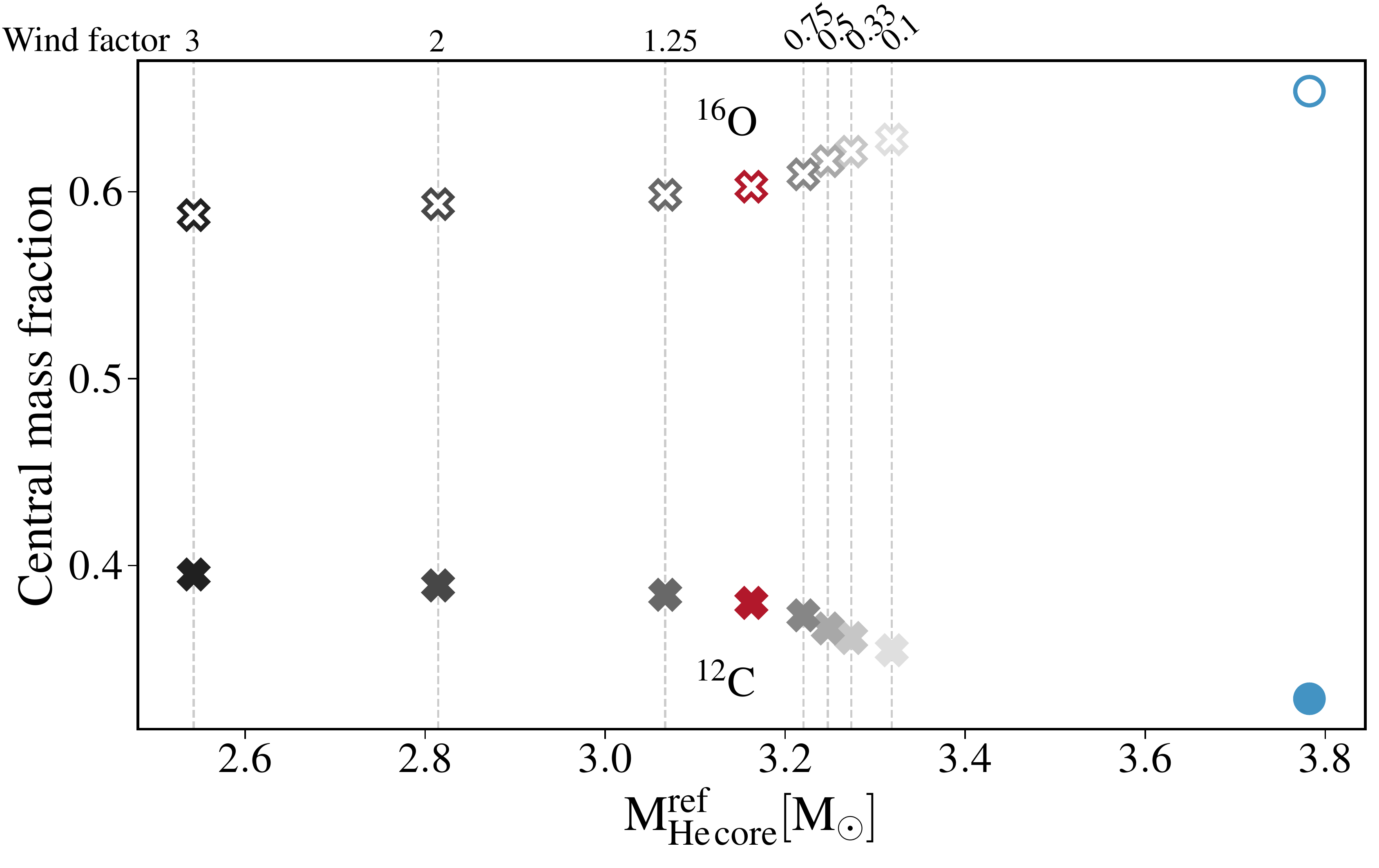}
        \includegraphics[width=0.5\textwidth]{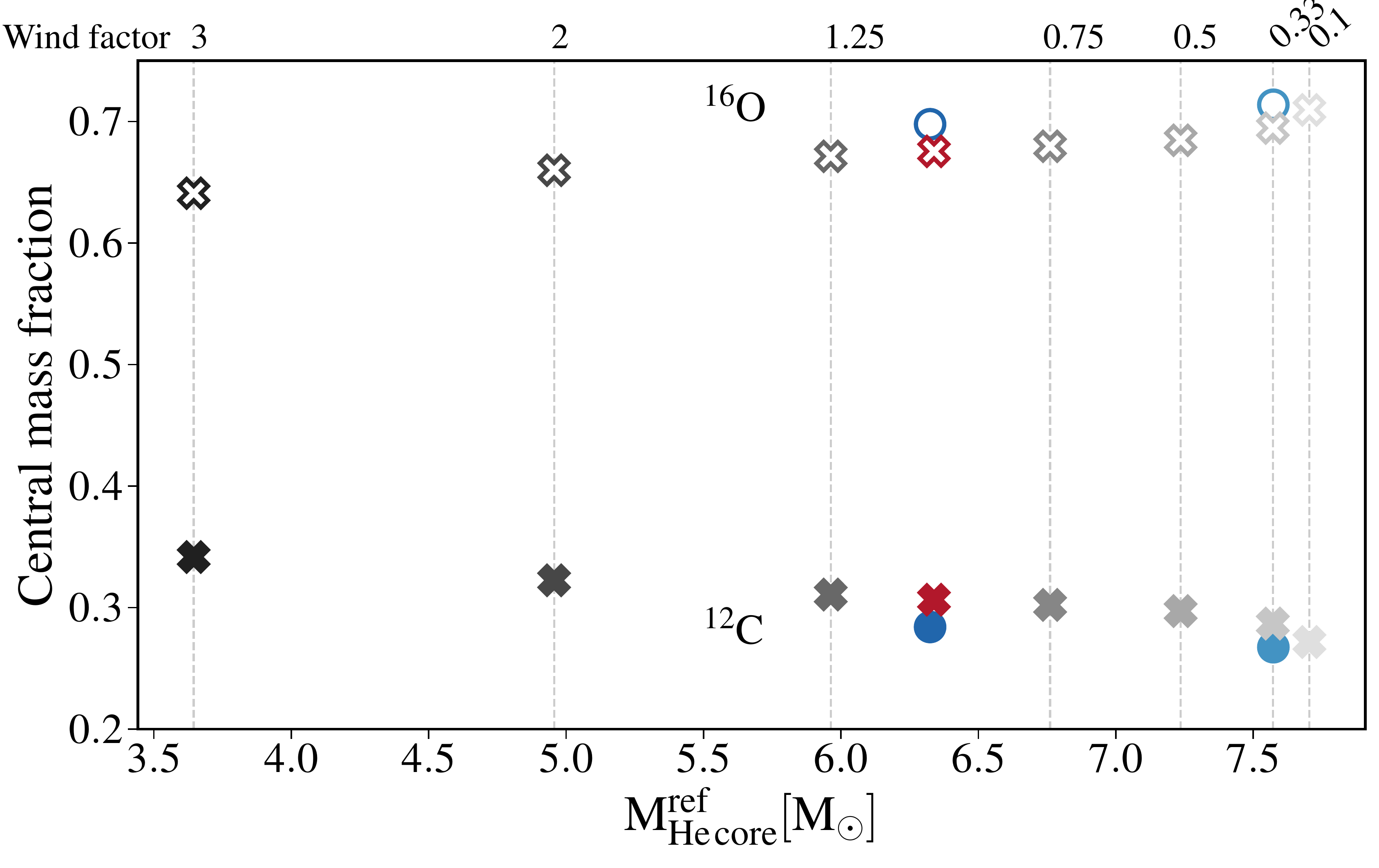}
        \caption{Impact of binary-stripped star wind variations on the central mass fractions at the moment of core helium depletion. Open and filled crosses indicate the central mass fractions of $^{16}$O and $^{12}$C, respectively. Gray crosses show variations for binary-stripped star models when we decrease the strength of the winds after Roche-lobe overflow by a constant factor, indicated on top of the vertical lines. For reference, we show a single (light blue) and binary-stripped (red) star model with the same initial mass of 11\Msun (\textit{top panel}) and 20\Msun (\textit{bottom panel}). In the bottom panel, we also compare the binary-stripped star models to a single-star model with the same reference core mass of 6.3\Msun with an initial mass of 16\Msun.}
        \label{fig:cmf_winds}
\end{figure}

In Fig. \ref{fig:cmf_winds} we show that the central mass fraction of $^{12}$C and $^{16}$O at the moment of core helium depletion is also systematically affected by a changing wind mass-loss rate. The higher the wind mass loss, the lower (higher) the central mass fraction of $^{12}$C ($^{16}$O). For models with low wind mass-loss rates, we find a more pronounced effect on the central mass fraction. 

\begin{figure*} % ----- Composition profiles  ---
        \centering
        \includegraphics[width=\textwidth]{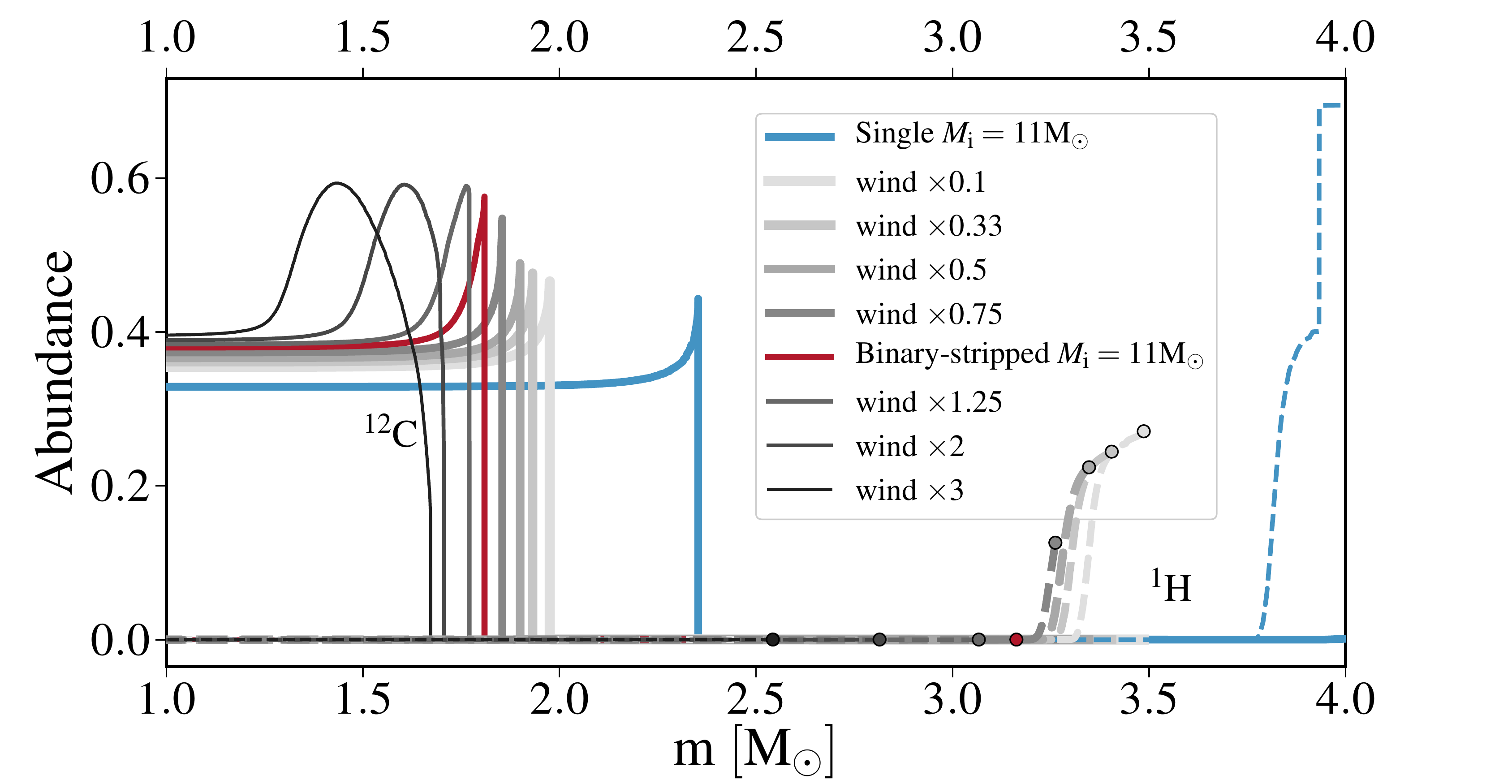}
        \includegraphics[width=\textwidth]{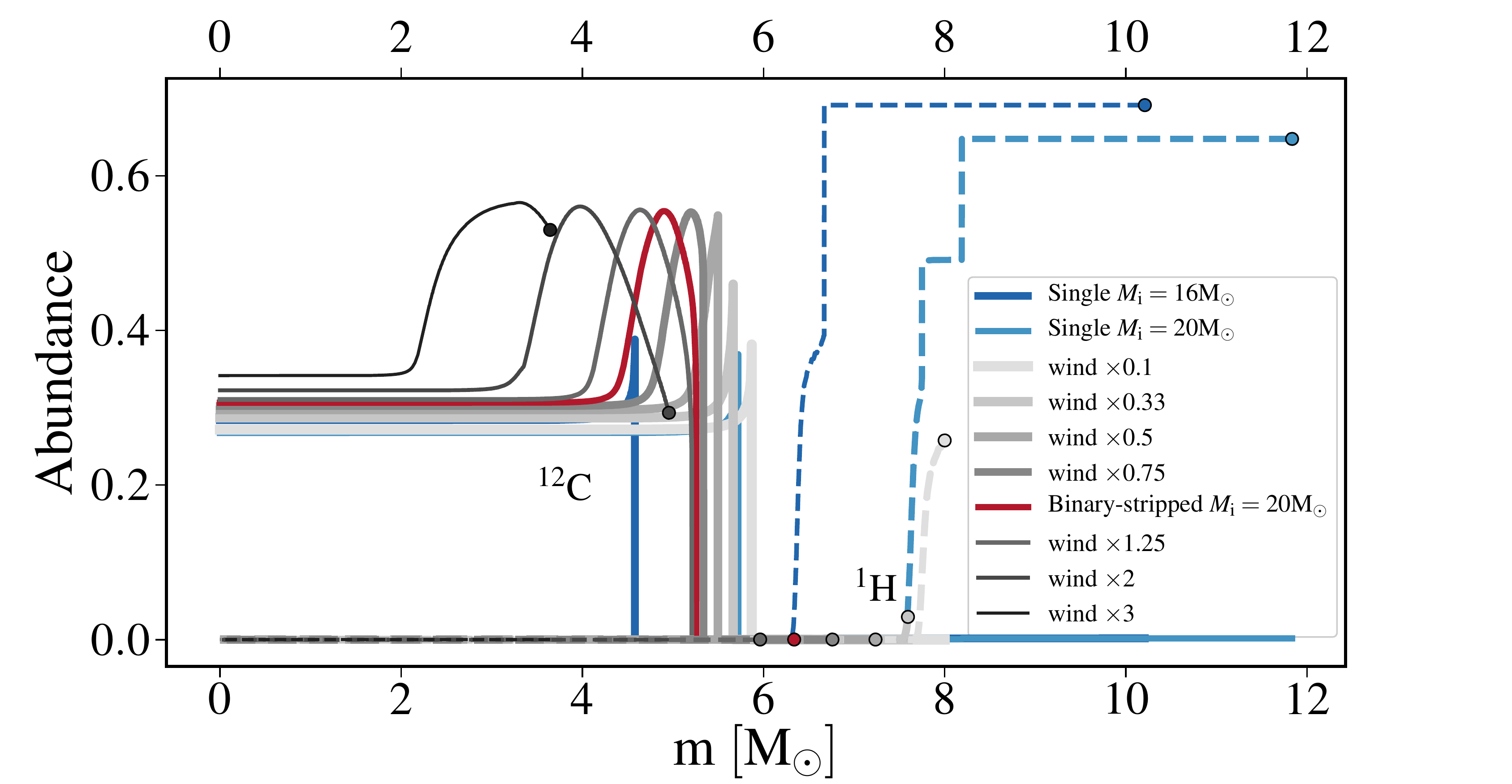}
        \caption{Impact of binary-stripped star wind variations on the composition profiles at the moment of core helium depletion as a function of the mass coordinate. We show our reference single (blue) and binary-stripped (red) star models with the same initial masses of 11\Msun (\textit{top panel}) and 20\Msun(\textit{bottom panel}). In gray, we show how the profiles of the binary-stripped star models change when we vary the wind mass-loss rates by the factor indicated in the legend. Dashed and full lines indicate the mass fractions of $^{1}$H and $^{12}$C, respectively. The mass extent of the carbon-rich layer increases with higher wind mass-loss rates. The surface of the stellar models is indicated with a colored circle. In the top panel, we limit the mass range for clarity and do not show the entire hydrogen layer of the single-star model.}
        \label{fig:comp_winds_11}
\end{figure*}

The wind mass-loss rate affects the extent of the carbon-rich gradient above the helium-depleted core, as we show in the composition profiles in Fig.~\ref{fig:comp_winds_11}. Models with high wind mass-loss rates have a more extended carbon-rich layer. This can be understood as the consequence of a more pronounced retreat in mass of the convective core during central helium burning (see also \ref{sec:origins:co_gradient}), which leaves helium burning products behind. Even for lower wind mass-loss rates, the carbon-rich layer is still present in the binary-stripped stars and more extended than for single-star models. Models with reduced wind mass-loss rates retain a hydrogen-rich layer (as shown by the dashed lines in Fig.~\ref{fig:comp_winds_11}). 

For the example binary-stripped star model with an initial mass of 20\Msun, the effect of winds is more pronounced than for the 11\Msun model, due to the indirect mass dependence of winds. In the two models with the highest wind mass-loss rates (a factor of two and three larger, respectively), the winds are so strong that even the helium-rich layer is removed, and the carbon-rich layer reaches the surface of the star (indicated with a circle in the bottom panel of Fig.~\ref{fig:comp_winds_11}). Observationally, such stars would likely be observed as carbon- or oxygen-rich Wolf-Rayet stars (type WC/WO) and a supernova resulting from such stars would most likely be of type Ic.

We conclude that variations in the wind mass-loss rate have a systematic effect on the core structure of binary-stripped stars. The extent of the carbon-rich layer above the helium-depleted core strongly depends on the wind mass-loss rate. However, the mass extent of the carbon-rich layer and the central fraction of carbon remains larger for binary-stripped star models than for single-star models with the same initial mass, even for the lowest wind mass-loss rates (a factor of ten weaker, and similar to the wind mass loss inferred by \citealt{vink_winds_2017}).


\section{Shell mergers}
\label{sec:appendix:shell_merger}
About half of the models in our grid experience the merging of silicon-burning shells with oxygen-rich layers \citep[e.g.,][]{couch_revival_2013,yoshida_one-_2019,andrassy_3d_2020,fields_development_2020}. They occur in both single and binary-stripped models. In this section we discuss how these shell mergers arise and how the composition is modified. To this end, we focus on an example model, the single-star model with an initial mass of 11\Msun.
\begin{figure}[h]% ----- Example shell merger: composition structure evolution ----
        \centering
        \includegraphics[width=0.5\textwidth]{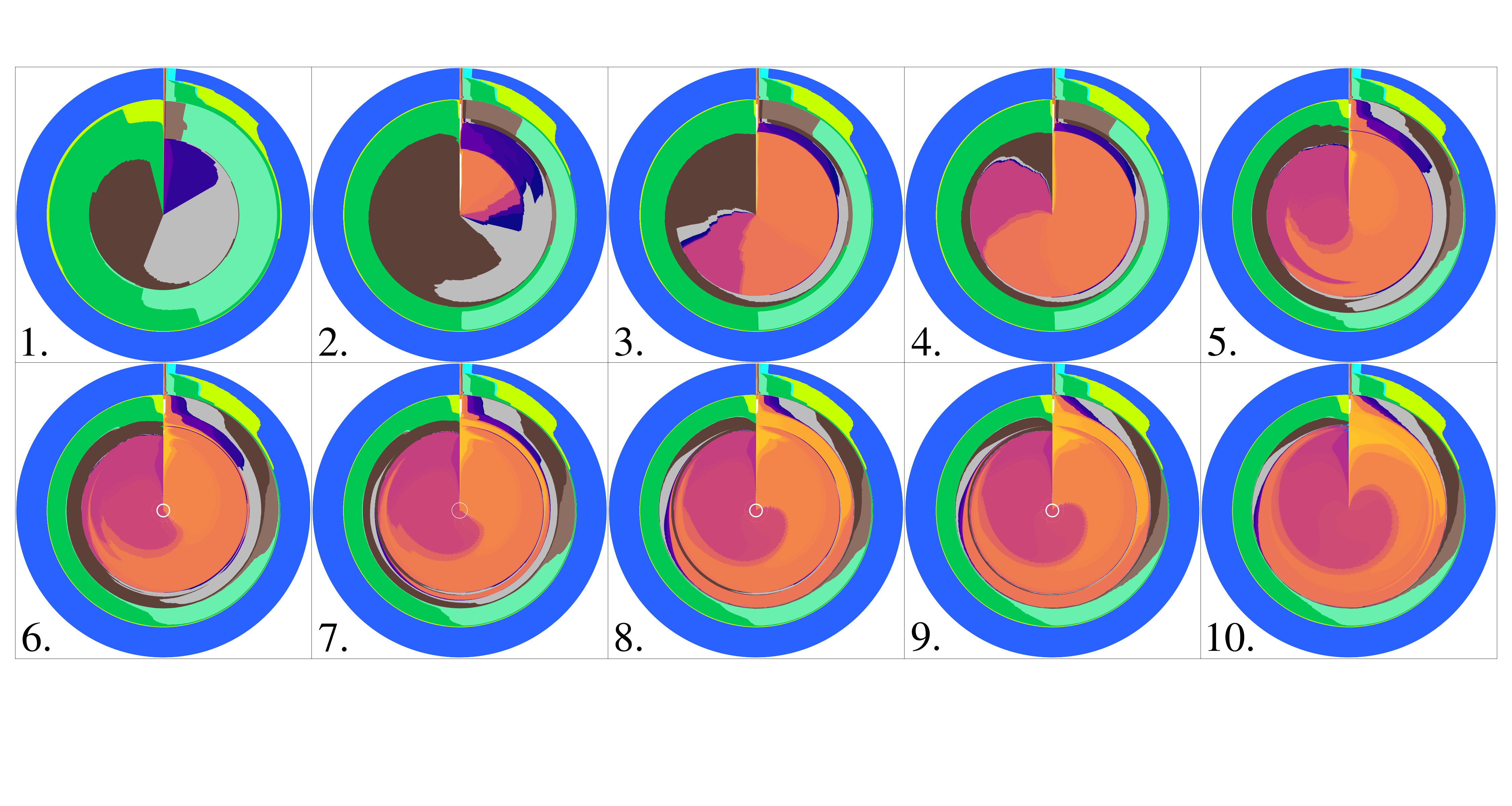}%
        \vspace{-1cm}
        \caption{Evolution of the interior composition structure of an example single-star model with an initial mass of 11\Msun after core oxygen depletion. From diagrams 1 to 3, core silicon burning occurs, leading to the creation of $^{54}$Fe. The silicon-burning shell of the star merges with the oxygen-rich layers above, between points 4 and 5, leading to the creation of $^{28}$Si, $^{32}$S, $^{36}$Ar, $^{40}$Ca, and $^{54}$Fe in the oxygen-rich layer. From points 6 to 9, the outer layers of the silicon-rich region turn into iron-group elements.}
        \label{fig:comp_shell_merger}
\end{figure}

In Fig. \ref{fig:comp_shell_merger} we show the evolution of the composition structure at selected points in the evolution. The star burns silicon in a convective core (labels 1 to 3 in Fig. \ref{fig:comp_shell_merger}), before igniting silicon in a convective shell (label 4 in Fig. \ref{fig:comp_shell_merger}). Shortly after the ignition of this shell, the burning shell moves upward in mass coordinate. It then reaches the oxygen-rich region above and becomes mixed in (labels 4 to 5 in Fig. \ref{fig:comp_shell_merger}). The merger leads to a more luminous and extended burning of silicon. Because silicon now burns in a different composition environment, different branching rates of the burning reactions occur. The ingestion of oxygen helps boost alpha-chain reactions and leads to an increased abundance of $^{28}$Si, $^{32}$S, $^{36}$Ar, $^{40}$Ca, and $^{54}$Fe in the oxygen-rich layer. From point 6 to 9, the silicon-rich outer layers turn into iron-group elements.

\section{Kippenhahn diagrams}
\label{sec:appendix:kipp}
For completeness, we show the Kippenhahn diagrams for our single and binary-stripped star models with initial masses of 11\Msun in this section.
\begin{figure}[!ht] % ----- Kippenhahn diagrams ---
        \centering
        \includegraphics[width=0.5\textwidth]{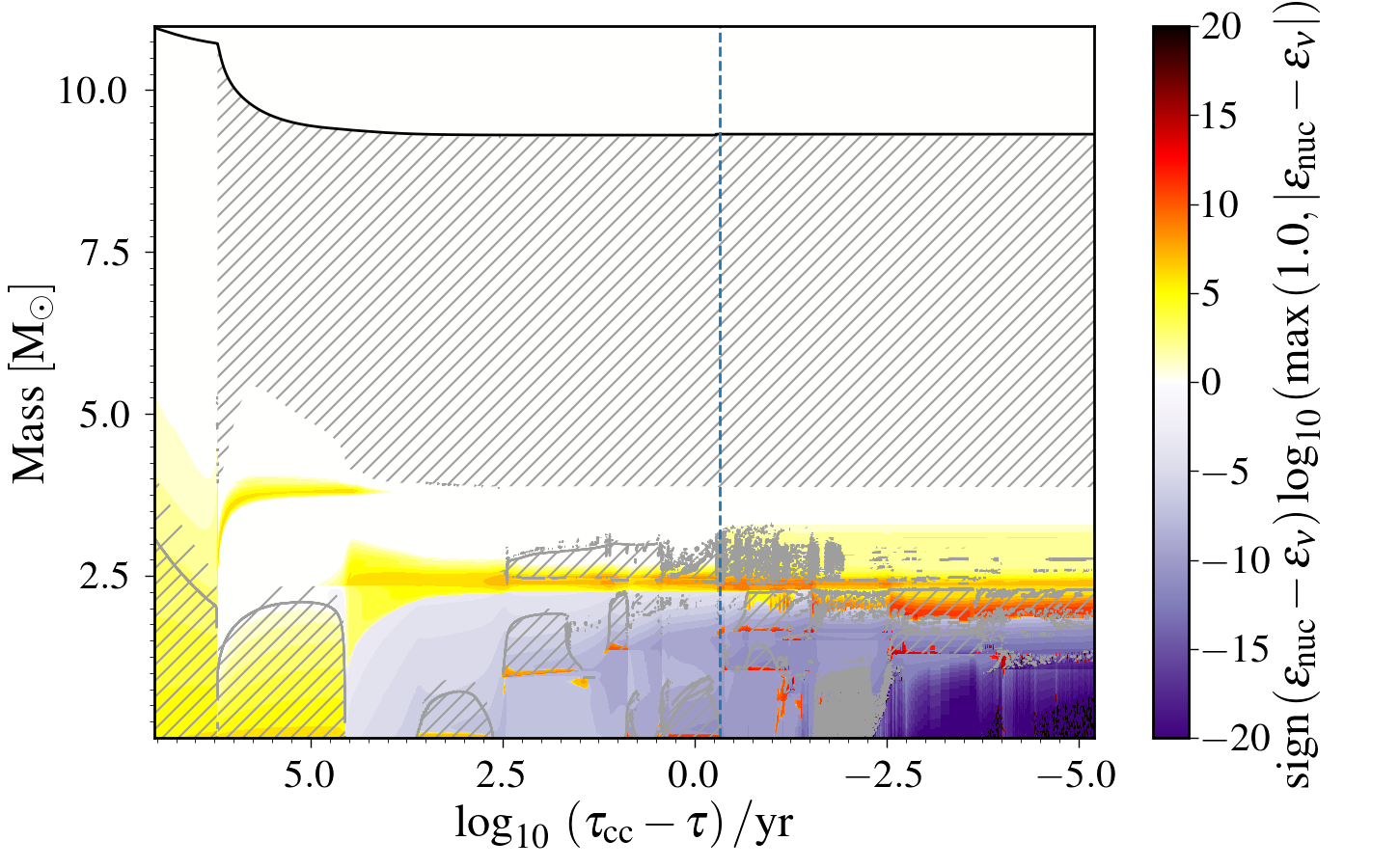}
        \includegraphics[width=0.5\textwidth]{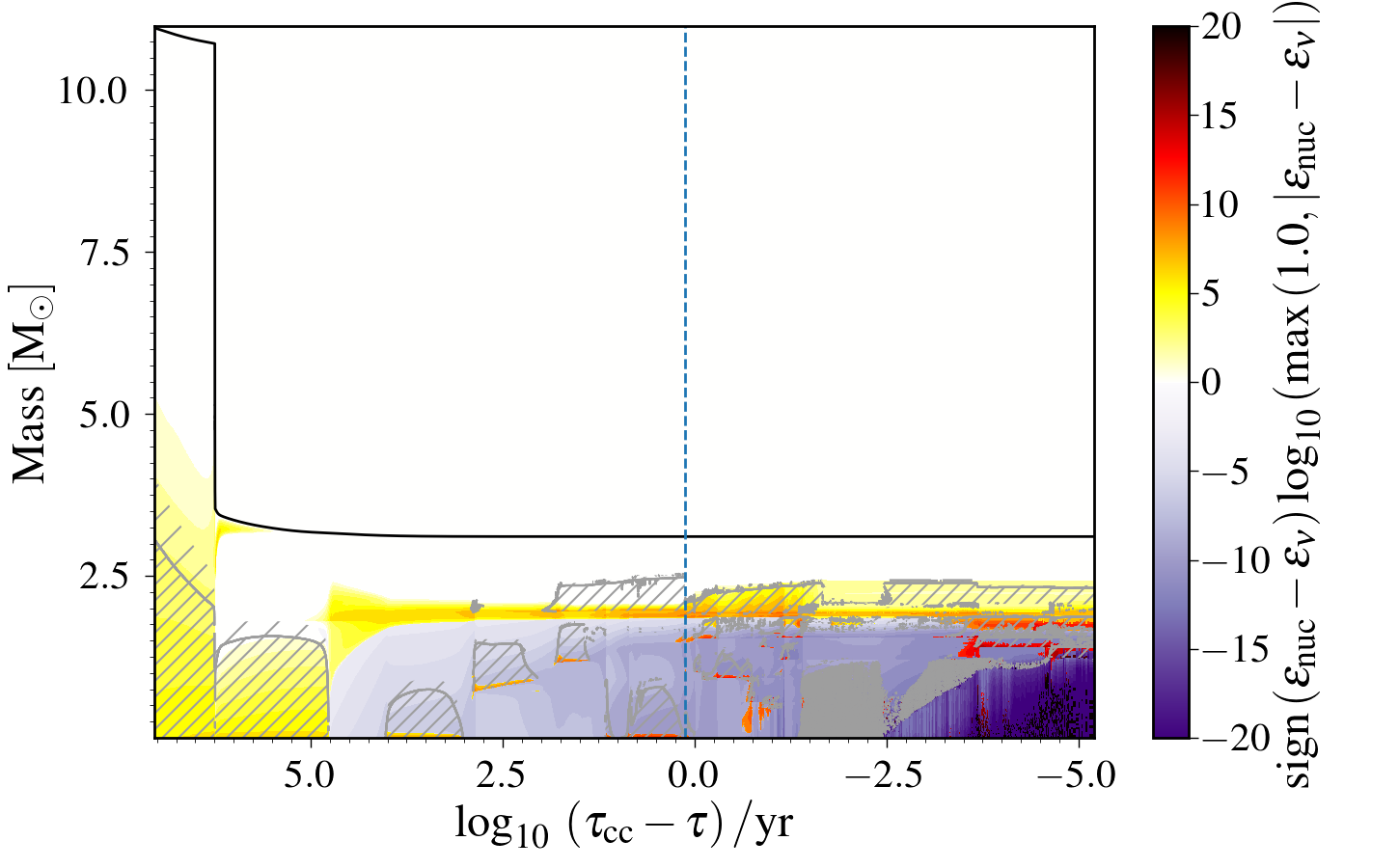}
        \caption{Kippenhahn diagrams. These figures show the evolution of the interior stellar structures of the single (top) and binary-stripped star (bottom) models with initial masses of 11\Msun as a function of time until core collapse. Single and double hatching indicates overshooting and convection, respectively. The colors mark the difference between the specific energy generation rate, $\epsilon_{\mathrm{nuc}}$, and the specific energy lost due to neutrino emission, $\epsilon_{\mathrm{\nu}}$. For each model, a vertical dashed blue line marks the moment when we switch to a nuclear network of 128 isotopes.}
        \label{fig:kipp}
\end{figure}
\end{appendix}

%%%% End of manuscript
\end{document}